\documentclass[nohyperref]{article}

\usepackage{microtype}
\usepackage{graphicx}
\usepackage{booktabs} 
\usepackage{diagbox}
\usepackage{bm}
\usepackage{hyperref}

\usepackage[accepted]{icml2022}

\usepackage{graphicx}
\usepackage{amsmath}
\usepackage{amssymb}
\usepackage{mathtools}
\usepackage{amsthm}
\usepackage{float}
\usepackage{subfig}
\usepackage{newfloat}
\usepackage{wrapfig}
\usepackage[capitalize,noabbrev]{cleveref}

\usepackage{algpseudocode}
\usepackage[vlined,ruled,linesnumbered]{algorithm2e}
\theoremstyle{plain}

\theoremstyle{definition}

\theoremstyle{remark}

\usepackage{xspace}

\newcommand{\alg}{PMIC\xspace}
\newcommand{\dnb}{Du-PCB\xspace}
\newcommand{\mie}{Du-MIE\xspace}

\usepackage[textsize=tiny]{todonotes}

\icmltitlerunning{PMIC: Improving Multi-Agent Reinforcement Learning with Progressive Mutual Information Collaboration}

\begin{document}

\twocolumn[
\icmltitle{PMIC: Improving Multi-Agent Reinforcement \\Learning with Progressive Mutual Information Collaboration}



\icmlsetsymbol{equal}{*}

\begin{icmlauthorlist}
\icmlauthor{Pengyi Li}{yyy}
\icmlauthor{Hongyao Tang}{yyy}
\icmlauthor{Tianpei Yang}{yyy,comp}
\icmlauthor{Xiaotian Hao}{yyy}
\icmlauthor{Tong Sang}{yyy}
\icmlauthor{Yan Zheng}{yyy}
\icmlauthor{Jianye Hao}{yyy}
\icmlauthor{Matthew E.Taylor}{comp}
\icmlauthor{Wenyuan Tao}{yyy}
\icmlauthor{Zhen Wang}{sch}
\icmlauthor{Fazl Barez}{fb}
\end{icmlauthorlist}

\icmlaffiliation{yyy}{College of Intelligence and Computing, Tianjin University, China}
\icmlaffiliation{comp}{University of Alberta, Canada}
\icmlaffiliation{sch}{Northwestern Polytechnical University, China}
\icmlaffiliation{fb}{The University of Edinburgh, U.K}
\icmlcorrespondingauthor{Yan Zheng}{yanzheng@tju.edu.cn}
\icmlcorrespondingauthor{Jianye Hao}{jianye.hao@tju.edu.cn}

\icmlkeywords{Multi-Agent Reinforcement Learning, Multi-Agent Collaboration, Deep Reinforcement Learning, Mutual Information}

\vskip 0.3in
]



\printAffiliationsAndNotice{}  

\begin{abstract}
Learning to collaborate is critical in Multi-Agent Reinforcement Learning (MARL). Previous works promote collaboration by maximizing the correlation of agents’ behaviors, which is typically characterized by Mutual Information (MI) in different forms.
However, we reveal sub-optimal collaborative behaviors also emerge with strong correlations,
and simply maximizing the MI can, surprisingly, \emph{hinder} the learning towards better collaboration.
To address this issue,
we propose a novel MARL framework, called Progressive Mutual Information Collaboration (PMIC), for more effective MI-driven collaboration.
PMIC uses a new collaboration criterion measured by the MI between global states and joint actions.
Based on this criterion, the key idea of PMIC is maximizing the MI associated with superior collaborative behaviors and minimizing the MI associated with inferior ones. 
The two MI objectives play complementary roles
by facilitating better collaborations while avoiding falling into sub-optimal ones. 
Experiments on a wide range of MARL benchmarks show the superior performance of PMIC compared with other algorithms. Our code is open-source and available at \url{https://github.com/yeshenpy/PMIC}.
\end{abstract}

\section{Introduction}
\label{intro}
With the potential to solve complex real-world problems, Multi-Agent Reinforcement Learning (MARL) has attracted much attention in recent years~\citep{lyu2022deeper,DBLP:conf/nips/YangWTHMMLLCHFZ21,DBLP:journals/jcst/ZhengHZMH20,wang2019action,2019jaamas} and has been applied to many practical domains like Game AI~\citep{peng2017multiagent}, Robotics Control~\citep{matignon2012coordinated}, Transportation~\citep{li2019efficient}. 
However, efficiently achieving collaboration and learning optimal policies still remains challenging in MARL~\citep{liu2020multi,wen2019probabilistic,DBLP:journals/corr/abs-2109-06668,DBLP:conf/pricai/ZhengMHZ18,DBLP:conf/nips/ZhengMHZYF18}. 

Centralized Training with Decentralized Execution (CTDE)~\citep{rashid2018qmix,sunehag2017value} is a popular MARL paradigm, adopted to promote collaboration among agents.
During centralized training, agents are granted access to other agents’ information and possibly the global state, 
while during decentralized execution, agents make decisions independently based on their individual policies. 
There are many CTDE-based MARL algorithms being proposed, including MADDPG~\citep{lowe2017multi}, MASAC~\citep{kim2020maximum}, VDN~\citep{sunehag2017value}, and QMIX~\citep{rashid2018qmix}.
However, although global information is incorporated during centralized CTDE training, optimizing the decentralized policies of multiple agents only through reward signals is often inefficient, especially when the reward signals are stochastic or sparse --- \emph{ therefore, additional mechanisms are often critical to facilitating effective collaboration.} A complementing branch of works proposes to leverage the correlation or influence of agents~\citep{08647,jaques2019social,xie2020learning,liu2020multi,merhej2021lief}.
The intuition behind these works is that if agents make decisions that account for their influence on the behaviors of other agents, the problem of non-stationarity could be mitigated, and thus agents are more likely to achieve collaboration.
Several works~\citep{chen2019signal,mahajan2019maven,kim2020maximum} proposed to maximize the correlation of agents' behaviors to promote collaboration, which commonly quantifies the correlation of agents' behaviors by the mutual information (MI). Unfortunately, these previous works overlook the fact that agents with a high degree of collaboration may not necessarily generate high rewards.
In complex environments, there exist multiple types of collaborations differing in return when agents achieve them --- \emph{simply maximizing the MI of agents’ behaviors cannot guarantee high-quality collaboration} because agents in sub-optimal collaborations can also have a high degree of correlation. 
Furthermore, maximizing MI exacerbates the problem: an agent can easily overfit its strategy to the behaviors of other agents~\citep{zhang2019efficient, lanctot2017unified}.

To solve the problem, this paper proposes a novel framework, called \textbf{P}rogressive \textbf{M}utual \textbf{I}nformation \textbf{C}ollaboration (\alg). In PMIC, a new collaboration criterion is measured by the MI between global states and joint actions, freeing us from relying on additional global input (existing in previous works) and addressing the scalability issue. Based on the new criterion, PMIC uses two main components to promote agents' collaboration. The first component is the \textit{{D}ual { Pr}ogressive C{o}llaboration {B}uffer} (\dnb), which includes a positive and a negative buffer to dynamically maintain data about superior and inferior collaborations, respectively. The second component is the \textit{{Du}al {M}utual {I}nformation {E}stimator} (\mie), which employs two MI neural estimators to estimate the MI of global states and joint actions for transitions in \dnb. 
In particular, one estimator is trained on the positive buffer to provide the lower bound of MI and the other is trained on the negative buffer to provide the upper bound of MI. By maximizing the lower bound and minimizing the upper bound, the agents can progressively break the current sub-optimal collaboration and learn to achieve better ones, which thus promotes an efficient and stable learning process.
Importantly, \alg is general and can be easily combined with existing MARL algorithms.
Our experiments show that \alg significantly accelerates existing MARL algorithms, outperforming other baseline algorithms on a wide range of MARL benchmarks.

In summary, our contributions are threefold: First, we reveal that simply maximizing the correlation of agents without distinguishing what kind of behaviors are expected can hinder the learning towards better collaborations.
Secondly, we propose a novel framework \alg to solve the problem, including a new collaboration criterion and a progressive MI estimation designed by \mie and \dnb. Last but not least, we build \alg on many MARL algorithms such as MADDPG~\citep{lowe2017multi}, MASAC~\citep{kim2020maximum}, RODE~\cite{wang2020rode} and show their superior performance by comparing them with other competitive methods on a wide range of MARL benchmarks.

\begin{figure*}[htb]
\begin{center}
\subfloat[A motivating example]{
\centering
\includegraphics[width=0.19\linewidth]{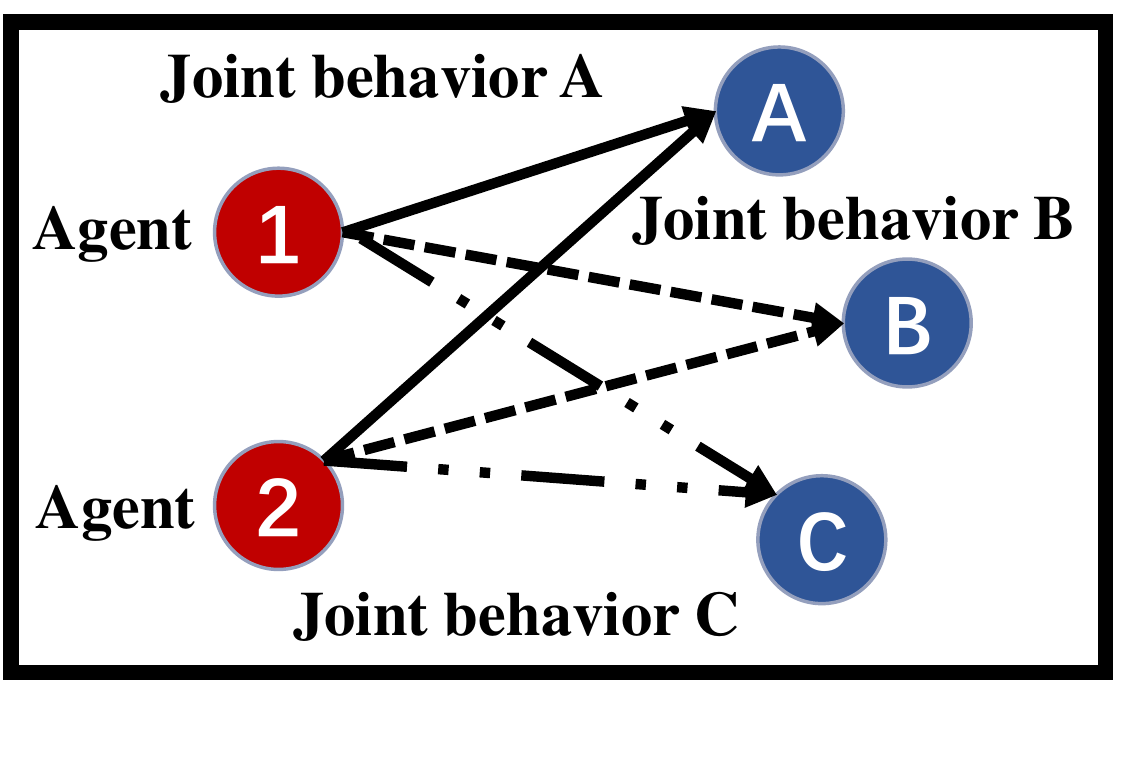}
\label{(a)}}
\subfloat[Reward matrix]{
\centering
\includegraphics[width=0.19\linewidth]{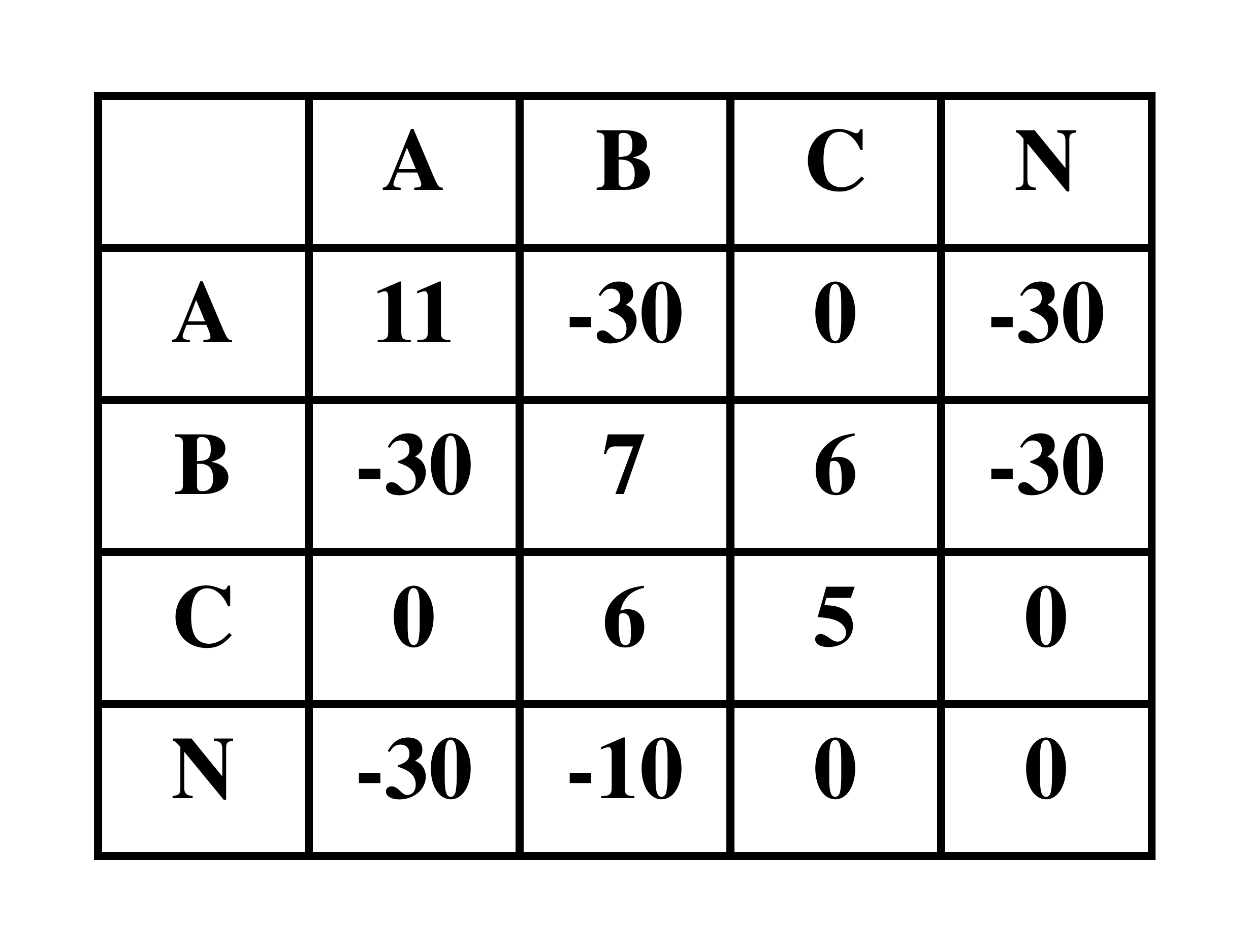}
\label{(b)}}
\subfloat[Behavior distribution]{
\centering
\includegraphics[width=0.19\linewidth]{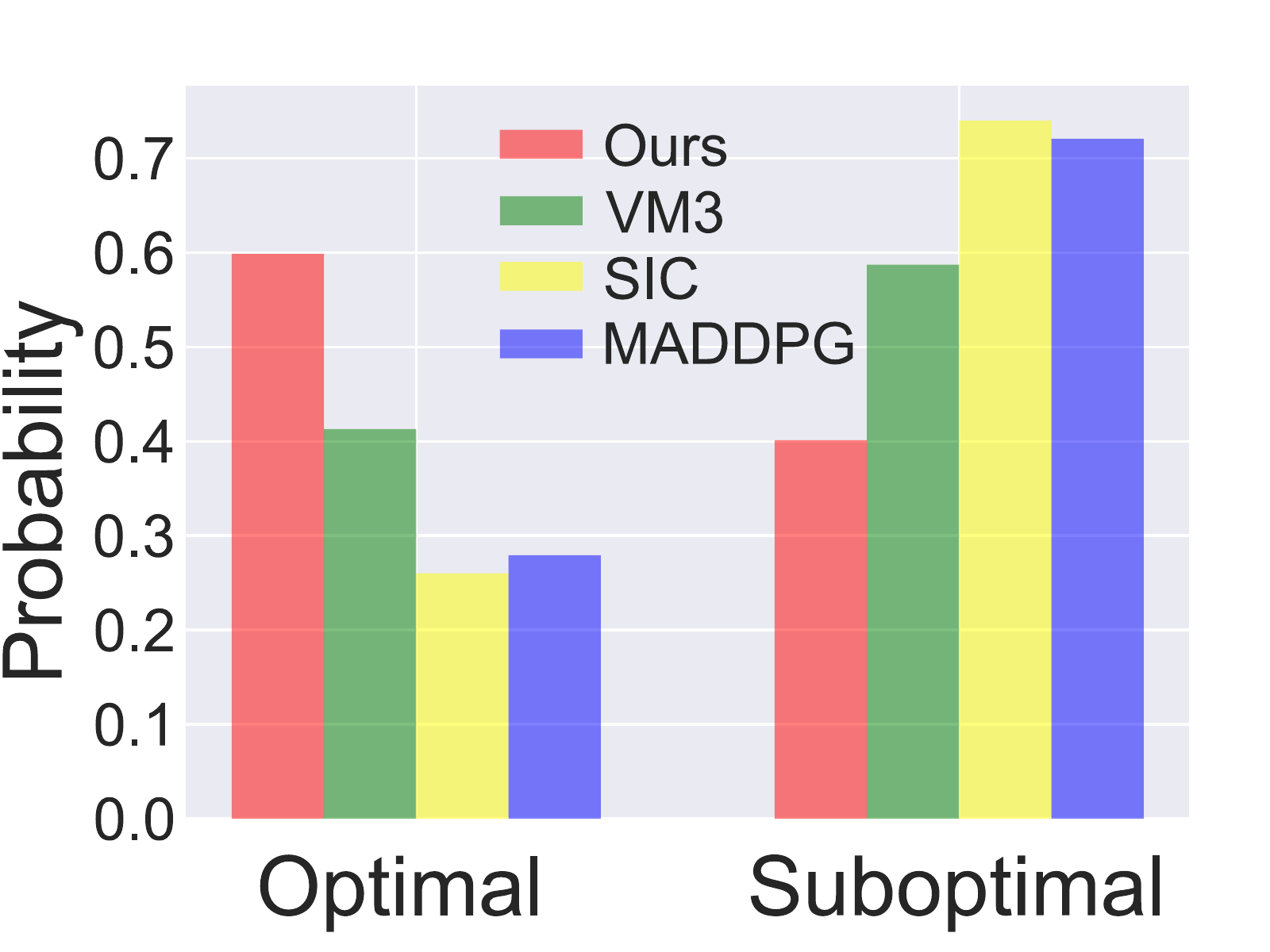}\label{(c)}
}
\subfloat[Degree of correlation]{
\centering
\includegraphics[width=0.19\linewidth]{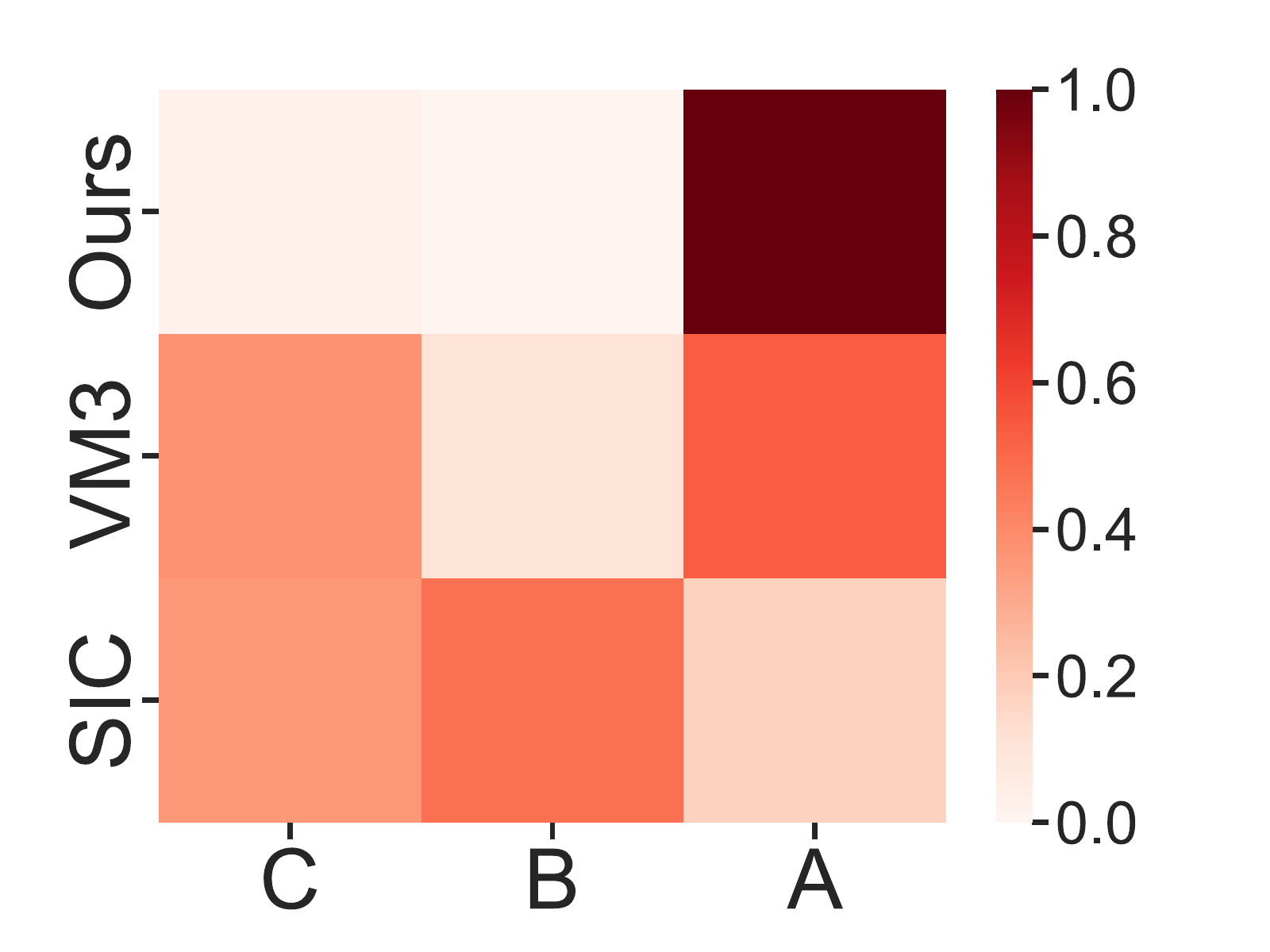}
\label{(d)}
}
\subfloat[Learning performance]{
\centering
\includegraphics[width=0.19\linewidth]{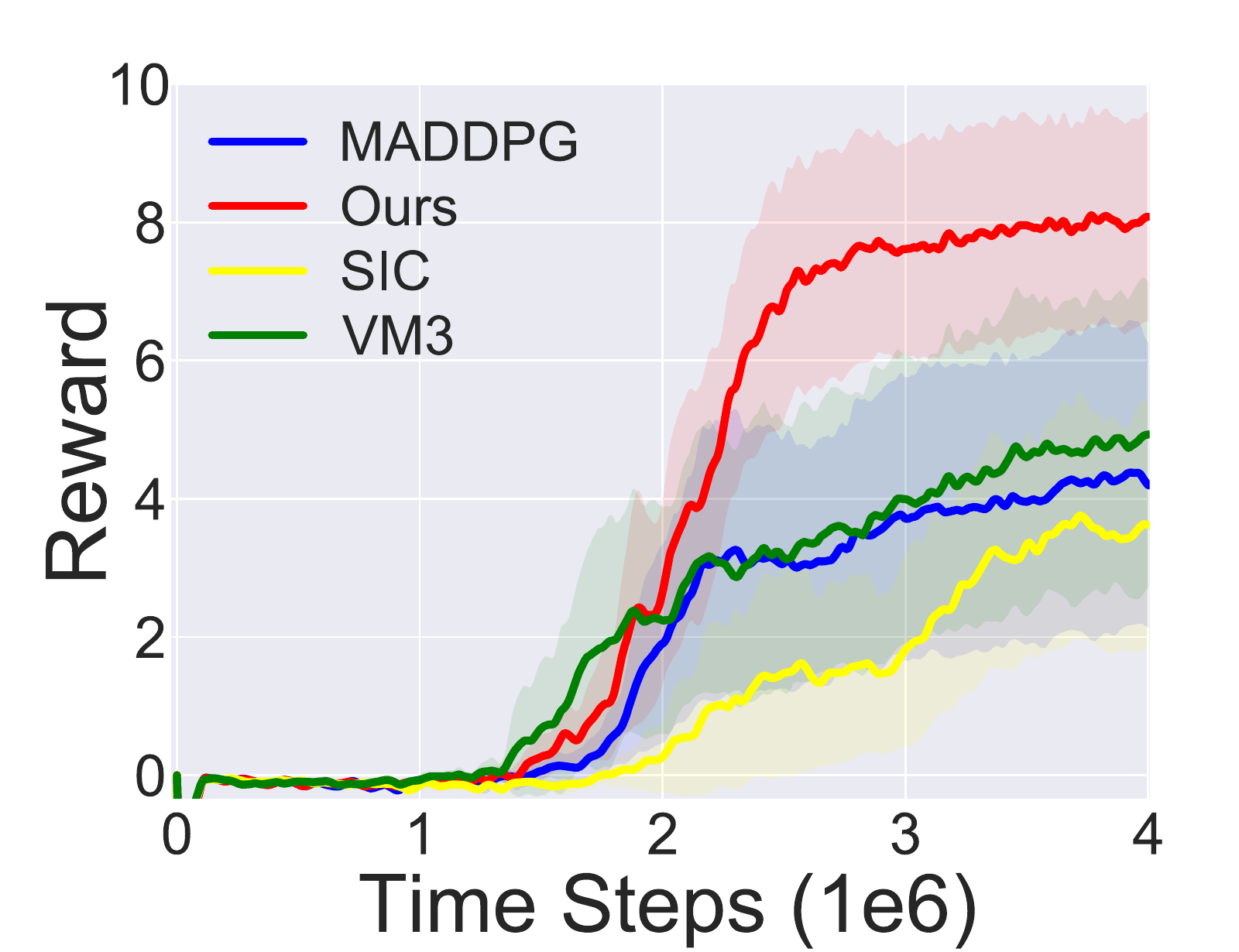}
\label{(e)}
}\hspace{-0.2cm}
\end{center}
\caption{\textit{(a)} A motivating example: \textit{Wildlife Rescue}~\citep{lowe2017multi}. The optimal joint behavior here is to rescue target A collaboratively, while other joint behaviors lead to sub-optimal collaborations. 
\textit{(b)} The reward matrix for the agents capturing different targets at the end of the game, where `N' means an agent does not catch any target.
\textit{(c)} The probability of optimal and sub-optimal joint behaviors learned by different algorithms, averaged over
10k (1k x 10 seeds) episodes.
\textit{(d)} The degree of correlation is measured by different algorithms for different joint behaviors.
\textit{(e)} The performance of episodic rewards averaged over 10 seeds with 95\% confidence regions for different algorithms.
}
\label{Collaboration pattern}
\end{figure*}

\section{Background}\label{related work}

This section presents the necessary background to understand \alg and its relationship to existing works.

\subsection{Preliminaries}
We consider a fully cooperative multi-agent task where a team of agents are situated in a stochastic, partially observable environment, it can be modeled as a
\textit{decentralised partially observable Markov decision process} (Dec-POMDP) \citep{oliehoek2016concise}, which can be defined as a tuple: $\left\langle\mathcal{N}, \mathcal{S},\mathcal{U}, \mathcal{O}, \mathcal{T}, \mathcal{R}, \gamma \right\rangle$. 
Here $\mathcal{N} = \{1,...,N\}$ denotes the set of $N$ agents. 
In Dec-POMDP, the full state of the environment $s_t \in \mathcal{S}$ cannot be observed by agents at each time step $t$. Each agent $i\in \mathcal{N}$ can only observe its individual observation $o_t^i$ determined by observation function $\mathcal{O}(s_t,i)$, each agent $i$ uses a stochastic policy $\pi_{i}$ to choose actions $u_{t}^{i} \sim \pi_i(\cdot|o_t^{i})$, yielding the joint action $u_t = \{u_{t}^{i}\}_{i=1}^N \in \mathcal{U}$. 
After executing $u_{t}$ in state $s_{t}$, the environment transits to the next state $s_{t+1}$ according to transition function $\mathcal{T}(s_t,u_t)$ and agents receive a common reward $r_t$ from $\mathcal{R}(s_t,u_t)$, with a discount factor $\gamma \in [0,1)$. 
We denote the joint policy as $\pi = (\pi_1,\pi_2,...,\pi_N) \in \Pi$, where $\Pi$ is the joint policy space.
In cooperative MARL, the collaborative team aims to find a joint policy to maximize the total expected discounted return, denoted by $J(\pi) = \mathbb{E}_{\pi}\left[\sum_{t=0}^{\infty} \gamma^{t} r_{t}\right]$.

\subsection{Related Work}

\textbf{Centralized training \& decentralized
execution} (CTDE) has been a major paradigm in
recent efforts in MARL.
For example, MADDPG ~\citep{lowe2017multi} uses a centralized critic to train decentralized policies. 
VDN~\citep{sunehag2017value},  QMIX~\citep{rashid2018qmix}, MAAC~\citep{iqbal2019actor}, COMIX, and FAC-MADDPG~\citep{de2020deep} achieve CTDE through value function factorisation.

\textbf{MI-based collaboration MARL:}
Many existing algorithms explicitly maximize the correlation or influence of agents to facilitate collaboration, where the correlation or influence is often quantified by the MI of the agent's behavior. For example, Signal Instructed Coordination (SIC) ~\citep{chen2019signal} takes a holistic view and facilitates collaboration by increasing the MI of the agent's behavior and the joint policy. Specifically, SIC extracts the information of the joint policy into the latent variables $z$ (sampled from a predefined distribution), which are then used as the input of agents' policy networks. SIC then maximizes the MI of each agent's behaviors and the latent variables $z$ to improve the correlation of agents. Based on the latent variables, the agents can know what kind of joint policy the whole team is executing and what kind of action should be selected. 
Multi-agent Variational Exploration (MAVEN) ~\citep{mahajan2019maven} shares a similar idea with SIC, except that MAVEN extracts the latent variables about joint policy information from the initial global state and maximizes the mutual information of future trajectories and the latent variables. However, one common drawback of both methods is the shared latent variable required during decentralized execution violates the CTDE paradigm. This makes algorithms fail in some real-world deployment scenarios where global communication is not available. 
EITI~\citep{wang2019influencebased} leverages MI to capture the influence between one agent’s current actions/states and the other agents' transitions in grid environments. SI~\cite{jaques2019social} proposes a social influence intrinsic reward measured by the MI between any two agents' actions to achieve coordination in sequential social dilemmas.
SI-MOA~\cite{jaques2019social} extends this idea to CTDE by modeling the other agents' actions and promotes coordination by the MI between one agent’s current action and the other agent' next action. VM3-AC~\cite{kim2020maximum} also tries to extend SI to CTDE. VM3-AC modifies policy iteration based on the MI of any two agents' current action and introduces additional input to explicitly represent the relation of agents' policies, and VM3 achieves better performance than previous methods.

Unfortunately, these methods may fall into the trap of blindly maximizing MI, resulting in a high degree of correlation, but failing to achieve high-performing policies. The next section will explain this failure mode, while Section~\ref{sec:methodology} will provide a solution to avoid the problem.

\section{Why Can MI-based Collaboration Fail?}
\label{motivation1}


\begin{figure}[t]
\centering
\vspace{-0.3cm}
\includegraphics[width=0.98\linewidth]{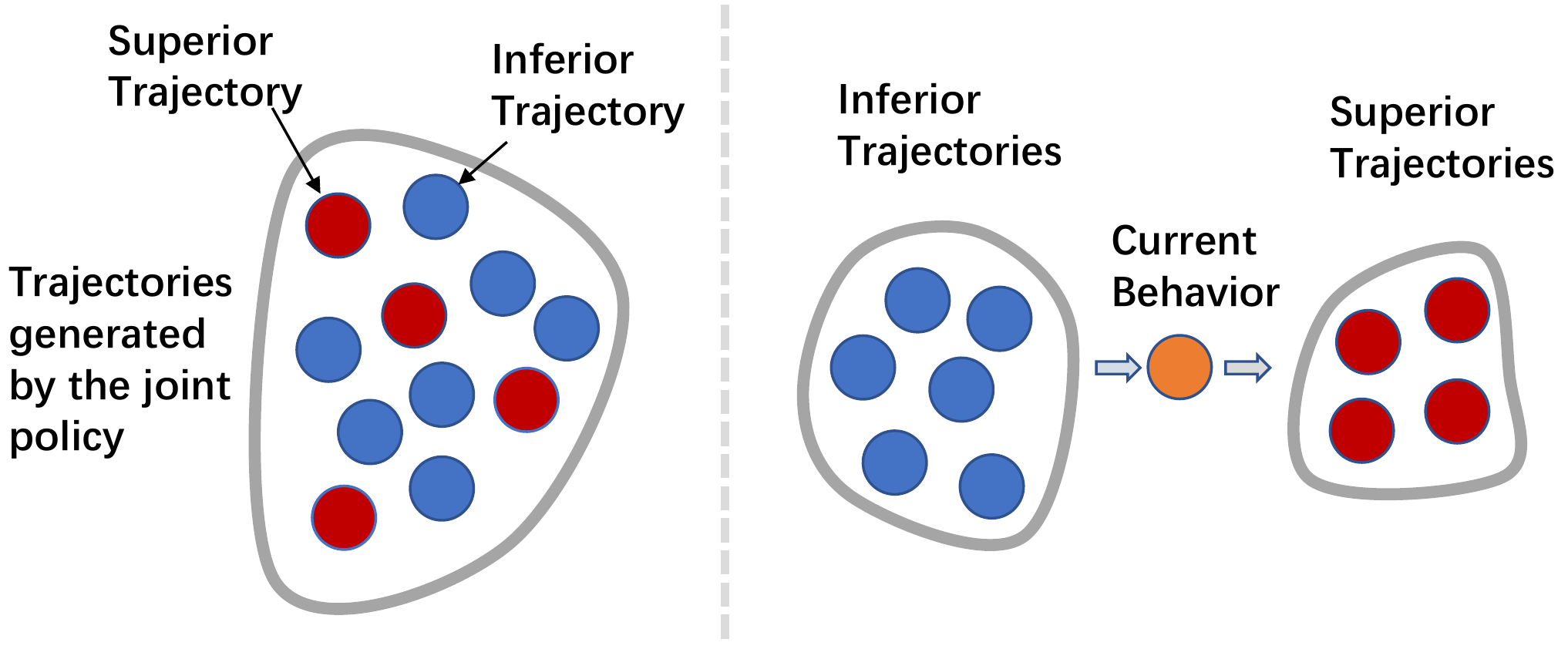}
\vspace{-0.2cm}
\caption{\textit{Left}: Trajectories collected during the learning process contain a mix-up of different joint behaviors.
\textit{Right}: An ideal learning process that identifies the distinctions and performs progressive improvement.
}
\label{figure figure}
\vspace{-0.3cm}
\end{figure} 

This section motivates our approach, showing why blindly enhancing the correlation of agents' behaviors can lead to agents falling into sub-optimal collaborations.
Figure \ref{(a)} illustrates a motivating example where two agents need to collaborate to rescue three targets (i.e., A, B, and C), and receive a team reward (Figure~\ref{(b)}).
The joint behaviors of the two agents are various in this game, and figure \ref{(a)} presents three kinds of joint behaviors. Apparently, the expected joint behavior is that both agents collaborate to rescue target A, since this achieves the highest reward.


Due to the stochasticity of the environment and the learning dynamics of agents' policies,
trajectories of different behaviors are collected by the two agents' joint policy, as shown in the left of Figure \ref{figure figure}.
These trajectories mix up many joint behaviors of a high collaboration degree, which have distinct outcomes (e.g., rescue B or C).
Intuitively, to achieve an ideal learning process as depicted in the right of Figure \ref{figure figure}, agents not only 1) need to enhance the correlation of their joint behaviors to form collaborations,
but also 2) need to be capable of escaping from a sub-optimal collaboration to reach a better one.
The contradiction between the two objectives indicates that the correlation should be enhanced and loosened in an adaptive manner.

Existing MARL methods with MI-based collaboration only focus on the first point, maximizing MI while neglecting the second.
In principle, this 
can be problematic because the consistent enhancement of behavioral correlation can prevent learning other joint behaviors.
One natural solution is to enhance the correlation of agents in superior trajectories and reduce the correlation in inferior ones.
Following this idea, in our PMIC framework, we maintain the superior and inferior trajectories separately. PMIC then maximizes the MI associated with the superior trajectories and minimizes the MI associated with the inferior trajectories.
Thus, agents learn to form superior joint behavior and avoid inferior ones.


\begin{figure*}[t]
\centering
\includegraphics[width=\linewidth]{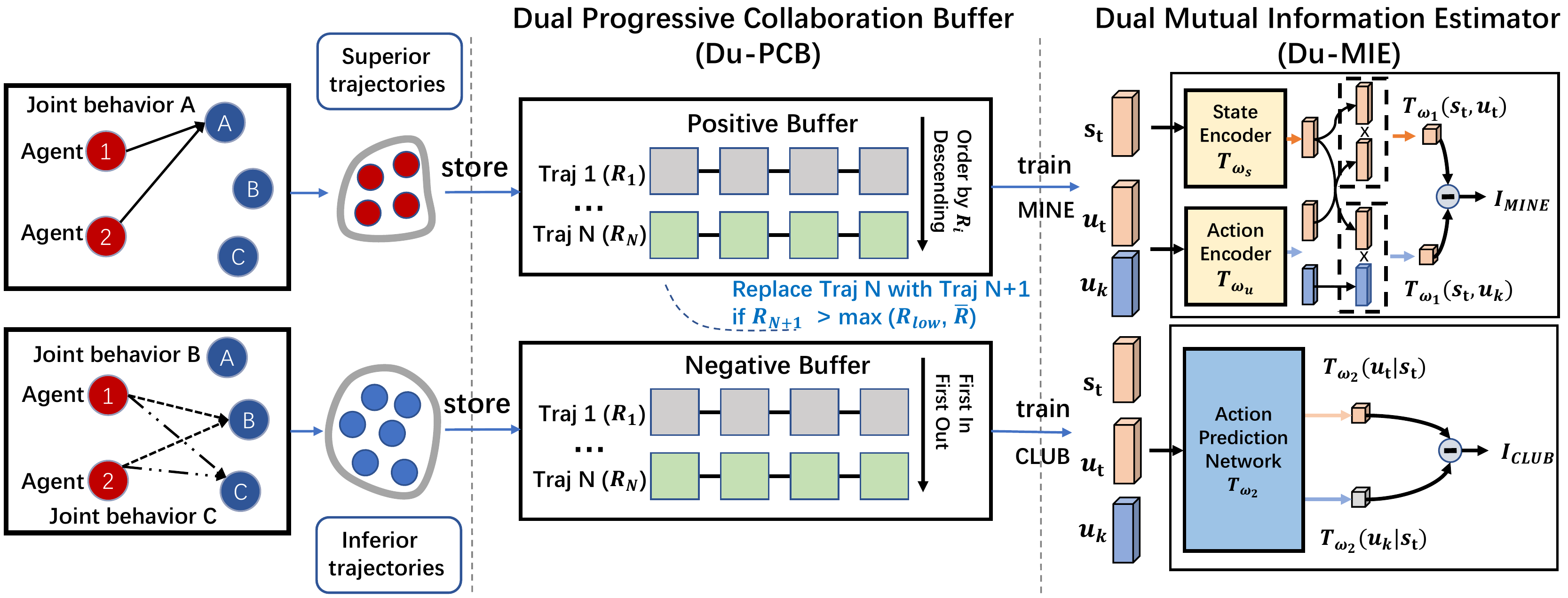}
\caption{An overall illustration of Progressive Mutual Information Estimation,
consisting of two main components:
1) 
\dnb maintains superior and inferior trajectories separately in a progressive manner
and 2) 
\mie estimates the collaboration criterion $I(s;u)$ from Equation \ref{e55} using $I_{\text{MINE}}$ and $I_{\text{CLUB}}$. 
}
\label{321}
\end{figure*}

We verify our analysis by empirically evaluating the policies achieved by different methods in our motivating example.
VM3~\citep{kim2020maximum} and SIC~\citep{chen2019signal} are representative methods that use MI maximizing approaches and MADDPG serves as a basic reference. 
Figure \ref{(c)} shows the probability of converging to the optimal and sub-optimal joint behaviors.
We can see that the methods maximizing the correlation of agents indiscriminately (i.e., SIC and VM3) have a greater probability of falling into the sub-optimal joint behaviors than PMIC (our algorithm).
In turn, PMIC achieves higher rewards during the learning process (Figure \ref{(e)}).
Furthermore, Figure \ref{(d)} shows the degrees of correlation over three joint behaviors measured by different methods with corresponding MI forms.
The results show that 
sub-optimal joint behaviors (rescuing B and C) can also gain high degrees of correlation (even higher than rescuing A) in VM3 and SIC. In contrast, the degree of correlation for the optimal joint behavior in PMIC is significantly higher than in other methods. This further explains the results shown in Figures \ref{(c)} and \ref{(e)}.
Overall, these empirical results in our motivating example
demonstrate that only maximizing MI can make agents fall into sub-optimal collaboration behaviors and prevent them from learning the optimal ones,
revealing the necessity of breaking the correlation over sub-optimal behaviors during the learning process.
We detail our methodology and deeper experimental studies in the following sections.

\section{Progressive Mutual Information Collaboration for MARL}
\label{sec:methodology}

This section introduces our framework, Progressive Mutual Information Collaboration (\alg)
to improve cooperative MARL based on our discovery in the previous section. 
The key idea of \alg is to identify superior and inferior collaboration behaviors during the learning process, and encourage agents to achieve superior behaviors while avoiding sticking to inferior ones.
This dual guidance of joint policy learning is imposed in a progressive manner as the learning proceeds.
We first propose a new MI-based collaboration criterion (Sec.~\ref{subsection:new_collaboration_criterion}) for measuring the collaboration degree of agents.
Second, we introduce our approach to realize the dual collaboration guidance by progressive MI estimation (Sec.~\ref{subsection:progressive_mi_estimation}).
Third, we show how to integrate PMIC into general MARL algorithms (Sec.~\ref{subsection:marl_with_pmic}).


\subsection{A New Collaboration Criterion}
\label{subsection:new_collaboration_criterion}

Previous works measure the correlation of agents using MI in different forms. However, they often suffer from at least one of the following limitations. First, measuring correlation with the MI of any two agents' actions 
~\citep{jaques2019social,kim2020maximum} can be computationally infeasible with the increase in the number of agents (i.e., the scalability issue). 
Second, other methods~\citep{mahajan2019maven,chen2019signal} leverage the MI of additional shared latent variables and the joint policy (or trajectories), 
which violates the CTDE paradigm and makes the methods fail in some real-world deployment scenarios 
when global communication is not available during execution.

To resolve these problems, we propose a new criterion to measure the degree of multiagent collaboration, which is defined as the mutual information between global state $s$ and joint action $u$, which is
formulated as follows:
\begin{equation}
\begin{aligned}
 I(s;u) & = H(u) - H(u\mid s) \\ 
        & = H(u) -H(u_i\mid s) - H(u_{-i}\mid u_{i},s), 
\end{aligned}
\label{e55}
\end{equation}
where $H(\cdot)$ and $H(\cdot|\cdot)$ denote the entropy and conditional entropy respectively,
$u_{i}$ is the action of any agent $i$, and $u_{-i}$ is the joint action of all agents except $i$.
Note that $I(s;u)$ can be decomposed into three distinct terms: 
(1) $H(u)$ describes the ability to explore various behaviors of all agents (via joint actions), which could help generate diverse trajectories and avoid policy collapse when maximized; 
(2) $H(u_i|s)$ measures the behavioral uncertainty of agent $i$, which encourages the agent to behave deterministically given global state $s$ when minimized;
(3) $H(u_{-i}|u_i, s)$ measures the uncertainty of agent $i$ about the actions of other agents, which implicitly characterizes the correlation between agents' behavior and will drive agents to coherent joint behaviors when minimized.
Overall, $I(s;u)$ can serve as a quantitative measure of collaboration,
which can be optimized to incentivize agents to enhance or break different joint behaviors.

Compared with the aforementioned MI-based criteria,
our collaboration criterion obeys the CTDE paradigm because it does not incorporate extra latent variables.
Moreover, the MI measurement of our criterion
is free of calculating the MI for all possible two agents,
thus it does not suffer from the scalability issue as the number of agents increases.
In the following subsection, we introduce how $I(s;u)$ is estimated and used as collaboration guidance.



\subsection{Progressive Mutual Information Estimation}
\label{subsection:progressive_mi_estimation}
We now introduce dual collaboration guidance by progressive estimation of the collaboration criterion $I(s;u)$ to
help agents learn to achieve better collaboration and avoid sub-optimal collaborations.
This is achieved by two components, which are illustrated in Figure~\ref{321}.
The first component, Dual Progressive Collaboration Buffer (\dnb), stores superior and inferior trajectories in separate buffers,
which correspond to joint behaviors to achieve and avoid, respectively.
The second component, Dual Mutual Information Estimator (\mie),
provides MI estimates of $I(s;u)$ as quantitative signals of dual collaboration guidance. We detail the two components below.

\textbf{Dual Progressive Collaboration Buffer} (\dnb) consists of a positive buffer $D^{+}$ and a negative buffer $D^{-}$ to store superior and inferior trajectories respectively.
To identify the superior trajectories, we use the average return $\bar{R}$ of the most recent $M$ episodes as a measurement.
We denote the trajectory with the lowest return in the positive buffer as $R_{\text{low}}$.
For each episode $k$, we store a trajectory with return $R_k$ if $R_k > \max(R_{\text{low}},\bar{R})$ until $D^{+}$ is full; then \dnb overwrites the trajectories with return $R_{\text{low}}$.
This ensures the quality of trajectories stored in $D^{+}$ monotonically increases during the learning process.
In contrast, trajectories with returns $R_k \le \max(R_{\text{low}},\bar{R})$ are stored in $D^{-}$ in a First-In-First-Out (FIFO) manner, since most recent inferior behaviors are more needed to be avoided.
By this means, \dnb maintains the trajectories of both superior and inferior joint behaviors
progressively according to the policy learned at present.

\textbf{Dual Mutual Information Estimator} (\mie) will be used to estimate the collaboration criterion $I(s;u)$ described in Section~\ref{subsection:new_collaboration_criterion}, based on the trajectories stored in \dnb.
MINE \citep{belghazi2018mutual} will estimate the lower bound of $I(s;u)$ for maximization and CLUB \citep{cheng2020club} will estimate the upper bound of $I(s;u)$ for minimization, based on the positive buffer $D^{+}$ and negative buffer $D^{-}$, respectively. 
To be concrete, MINE approximates the lower bound of $I(s;u)$ based on samples in $D^{+}$ as follows:
\begin{equation}
\begin{aligned}
& I(s;u) \ge I_{\text{MINE}}(s;u) = \sup_{\omega_1 \in \Omega}  \\& 
\underbrace{\mathbb{E}_{\mathbb{P}_{\mathcal{S} \mathcal{U}}} \left[-sp\left(-T_{\omega_1}(s_t, u_t)\right)\right]  - \mathbb{E}_{\mathbb{P}_{\mathcal{S}} \otimes \mathbb{P}_{\mathcal{U}}} \left[ sp\left(T_{\omega_1}(s_t, u_k)\right) \right]}_{-\mathcal{L}(\omega_1)},
\end{aligned}
\label{e7}
\end{equation}
where $\mathbb{P}_{\mathcal{S}\mathcal{U}}$ is state-action joint distribution, 
$\mathbb{P}_{\mathcal{S}}$ and $\mathbb{P}_{\mathcal{U}}$ are the marginals.
The samples can be obtained by sampling $s_t,u_t$ pairs jointly, $u_k$ solely from $D^{+}$.
$T_{\omega_1}$ is a neural network with parameters ${\omega_1} \in \Omega$ that outputs a scalar and soft-plus function $sp(z) = \log(1 + \exp(z))$. 
By contrast, CLUB approximates the upper bound of $I(s;u)$:
\begin{equation}
\begin{aligned}
& I(s;u) \le I_{\text{CLUB}}(s;u) = 
\\& \underbrace{\mathbb{E}_{\mathbb{P}_{\mathcal{S} \mathcal{U}}} \left[ \log T_{\omega_2}\left(u_t \mid s_t\right) \right]}_{-\mathcal{L}(\omega_2)}  - \mathbb{E}_{\mathbb{P}_{\mathcal{S}} \otimes \mathbb{P}_{\mathcal{U}}} \left[ \log T_{\omega_2}\left(u_k \mid s_t\right) \right],
\label{upper bound}
\end{aligned}
\end{equation}
where $T_{\omega_2}$ is a neural network with parameters $\omega_2$ that approximates the conditional distribution.
The joint and marginals are similarly sampled as in MINE, but are based on the negative buffer $D^{-}$.

The training losses of MINE and CLUB neural estimators are $\mathcal{L}(\omega_1)$ and $\mathcal{L}(\omega_2)$, as defined in Equations~\ref{e7} and \ref{upper bound}.
Thus, the total loss of \mie is:
\begin{equation}
\begin{aligned}
    \mathcal{L}_{\text{\mie}}(\omega) = \mathcal{L}(\omega_1) + \mathcal{L}(\omega_2),
\end{aligned}
\label{e11}
\end{equation}
where $\omega = (\omega_1, \omega_2)$.
The architecture of MINE and CLUB are illustrated in Figure \ref{321}.
The complete process of training is detailed in Appendix \ref{MINE CLUB}.
After training, given state and joint-action samples, we can use MINE and CLUB to estimate the upper and lower bounds of MI.

Since MINE is trained 
following Equation \eqref{e7} based on the samples from the positive buffer,
only trajectories that resemble the joint behavior of superior collaboration have large MI estimates calculated by MINE;
it is similar to CLUB.
Therefore, given the interaction samples collected by the joint policy of current agents,
MINE and CLUB can 
provide effective signals in guiding agents' behaviors towards or away from superior and inferior ones, progressively.
In the following subsection, we show how PMIC functions with MARL for more efficient learning.


\subsection{Integration of PMIC and MARL}
\label{subsection:marl_with_pmic}

With the MI estimations introduced in the previous subsection,
now we aim to ﬁnd the joint policy that 
maximizes the expected discounted return and follows the progressive collaboration guidance by optimizing the MI estimates from \mie.
In particular,
we propose a new objective function for PMIC-MARL that combines the two types of MI estimates (as additional per-step rewards) with the conventional objective $J(\pi)$:
\begin{equation}
\begin{aligned}
J^{\text{PMIC}}(\pi) =\mathbb{E}_{s,u \sim \pi}
\left[
\sum_{t=0}^{\infty} \gamma^{t}(r_{t} + r_{t}^{\text{\alg}})
\right],
\label{etotal}
\end{aligned}
\end{equation}
where $r_{t}^{\text{\alg}} = \alpha I_{\text{MINE}}(s_{t};u_{t}) -\beta I_{\text{CLUB}}(s_t;u_t))$, and
$\alpha,\beta$ are the hyperparameters that weight the impact of MI guidance. 



In principle, PMIC-MARL is a general framework that can be implemented with different MARL algorithms. 
We use PMIC-MADDPG for a representative demonstration, building upon MADDPG~\citep{lowe2017multi}.
We also implement PMIC-RODE, built on RODE~\citep{wang2020rode}, and study it in our experiments. 
The pseudo-code of \alg-MADDPG is shown in Algorithm~\ref{algo2}.
In each episode, agents interact with the environment and store the trajectory samples into
experience replay $D$ (Lines 6-9).
The trajectory is added to the positive buffer or the negative buffers in \dnb according to its return (Lines 10-13). 
Every $k$ steps,
\mie is trained with the samples from \dnb (Line 14) to reflect current superior and inferior joint behaviors.
Lastly,
\alg-MADDPG updates the centralized critic and the actors according to $J^{\text{PMIC}}(\pi)$ (Line 15),
by minimizing the loss functions $\mathcal{L}_{Q}$ and $\mathcal{L}_{\pi}$, deﬁned below:
\begin{equation}
\begin{aligned}
& \mathcal{L}_Q(\phi)=  \mathbb{E}_{s_t,u_t,r_t,s_{t+1} \sim \mathcal{D}}\left[\left(\hat{y}-Q_{\phi}\left(s_t, u_t\right)\right)^{2}\right]; \\& \mathcal{L}_{\pi}(\theta) =\mathbb{E}_{s_t \sim \mathcal{D}}[-Q_{\phi}(s_t, \pi_{\theta}(\cdot|s_t))], 
\end{aligned}
\label{e12}
\end{equation}
where $\hat{y}=r_t+r_{t}^{\text{\alg}} + \gamma Q_{\phi^{\prime}}(s_{t+1},\pi_{\theta^{\prime}}({s_{t+1}}))$,
$\theta = (\theta_1, ..., \theta_n)$
and $\phi^{\prime}, \theta^{\prime}$ are the parameters of corresponding target networks.

Overall, we provide the technical details of PMIC, as well as how to combine PMIC with MARL algorithms, which we then evaluate empirically in the following section.


\begin{algorithm}[t!]
\small
\caption{\alg-MADDPG}
\label{algo2}
    \textbf{Input:} the update frequency $k$ for \mie, maximum episode length $T$, hyperparameters $\alpha$ and $\beta$ to balance the effects of maximizing and minimizing MI.\\
    
     \textbf{Initialize} the critic network $\phi$, $n$ actor networks $\theta_1 ... \theta_n$ and corresponding target networks $\phi^\prime$, ${\theta_1}^\prime, \dots, {\theta_n}^\prime$.\\
     \textbf{Initialize} \mie parameterized by $\omega_1$ and $\omega_2$.\\
     \textbf{Initialize} \dnb and experience replay buffer $\mathcal{D}$
    
    \Repeat{reaching maximum training steps} {
        \For{$t = 1,..., T$}{
            Execute joint actions \textbf{$u_t$} via collecting $u_{t}^{i} \sim  \pi_{\theta_i}({o_{t}^{i}})$.
            \\Receive $o_{t+1}=$ $\{o_{t+1}^{i}\}_{i=1}^{n}$ and team reward $r_t$.
        }
        Store trajectory $\nu = \{o_{t}, u_t, o_{t+1}, r_t \}_{t=1}^{T}$ to $D$\\
        
        \eIf{$R_{\nu} > \max(R_{\text{low}},\bar{R})$}{
            Add $\nu$ to the positive buffer
        }{
            Add $\nu$ to the negative buffer
        }
        
        Update \mie with \dnb every $k$ steps  \Comment{see Eq.~\ref{e11}}\\
        
        Update the actors and critic networks \Comment{see Eq.~\ref{e12}} 
    }
    
        
\end{algorithm}

\section{Experiments}
This section empirically evaluates PMIC on multiple multiagent tasks
to answer the following research questions (RQs):

\noindent\textbf{RQ1 (Performance)} 
Can \alg effectively achieve collaboration and outperform related baselines? Is \alg a generic  framework?\\
\noindent\textbf{RQ2 (Superiority of Dual MI)} Is maximizing and minimizing
$I(s;u)$ necessary? Is $I(s;u)$ effective for performance improvements?\\
\noindent\textbf{RQ3 (Necessity of \dnb)} Does \dnb significantly improve performance over a normal replay buffer?

\subsection{Benchmarks \& Baselines}
For a comprehensive comparative study, 
we evaluate our algorithms on both discrete and continuous action spaces. 
For the continuous action space, we consider the Multi-Agent Particle Environment (MPE) and the Multi-Agent MuJoCo benchmark, comparing \alg-MADDPG with six advanced algorithms as baselines: SIC-MADDPG \citep{chen2019signal}, VM3-AC \citep{kim2020maximum}, MASAC \citep{kim2020maximum}, MADDPG \citep{lowe2017multi}, FacMADDPG \citep{de2020deep}, and COMIX \citep{de2020deep}, where FacMADDPG and COMIX are the state-of-the-art (SOTA) algorithms in Multi-Agent MuJoCo benchmark~\citep{de2020deep}. 
The StarCraft II micromanagement (SMAC) benchmark~\citep{samvelyan2019starcraft} is also considered, which has high complexity of control and requires learning policies in large discrete action space, and we compare our \alg-RODE algorithm with the current SOTA algorithm, RODE~\citep{wang2020rode}. More experiments to integrate \alg with MASAC and QMIX~\citep{rashid2018qmix} are provided in Appendix \ref{more exp}.
The environment description is provided in Appendix \ref{env details}. 

For all baseline algorithms, we use the official code if available or reproduce it according to the original papers. Hyperparameters have been ﬁne-tuned in all environments. For a fair comparison, we use the same structures to avoid the influence of different structures on the results.
Further implementation details are in Appendix \ref{Hyperparameters}.

\subsection{Performance (RQ1)}
\label{ex2}
\begin{figure}[t]
\centering
\includegraphics[width=0.49\linewidth]{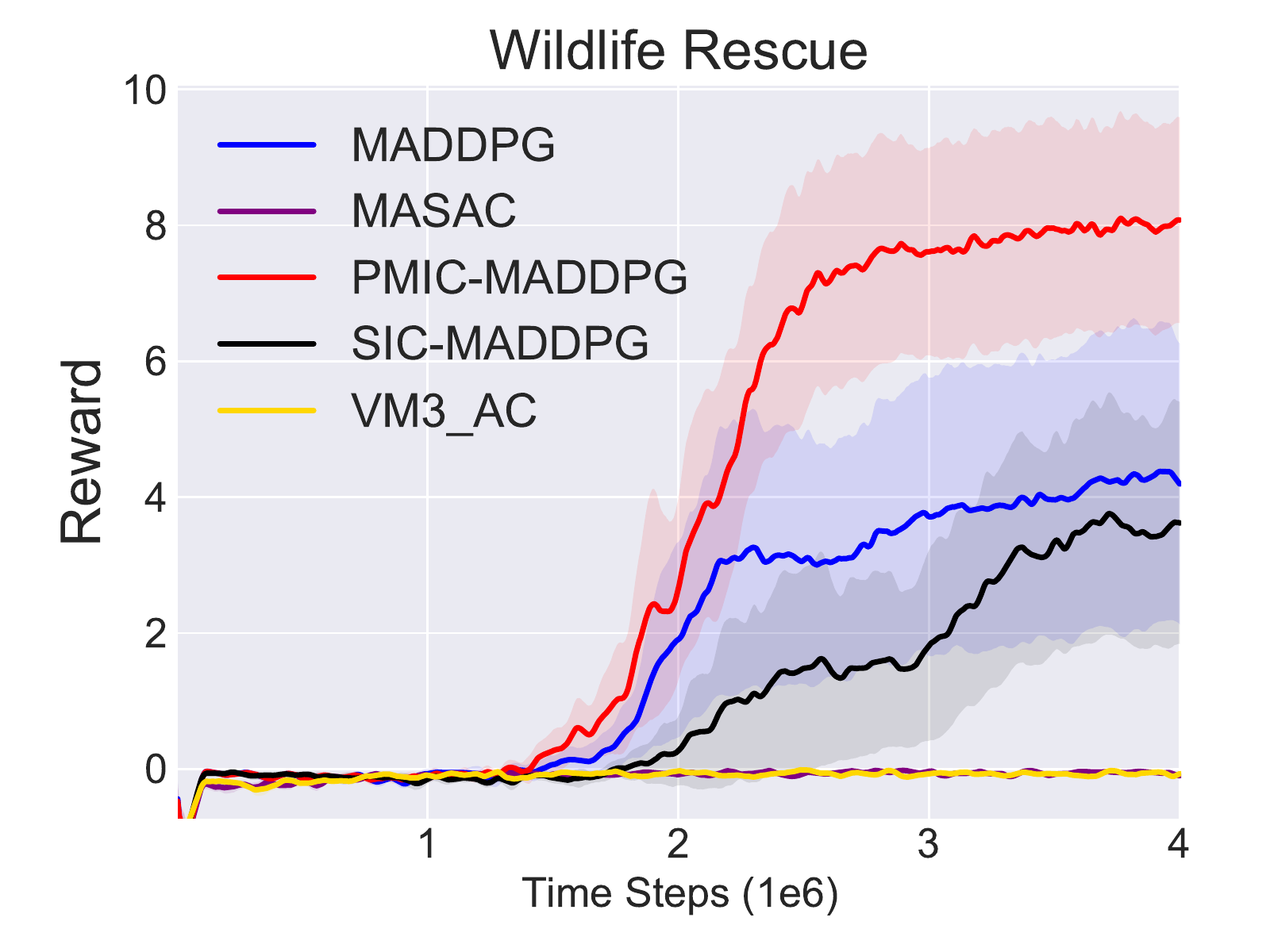}
\includegraphics[width=0.49\linewidth]{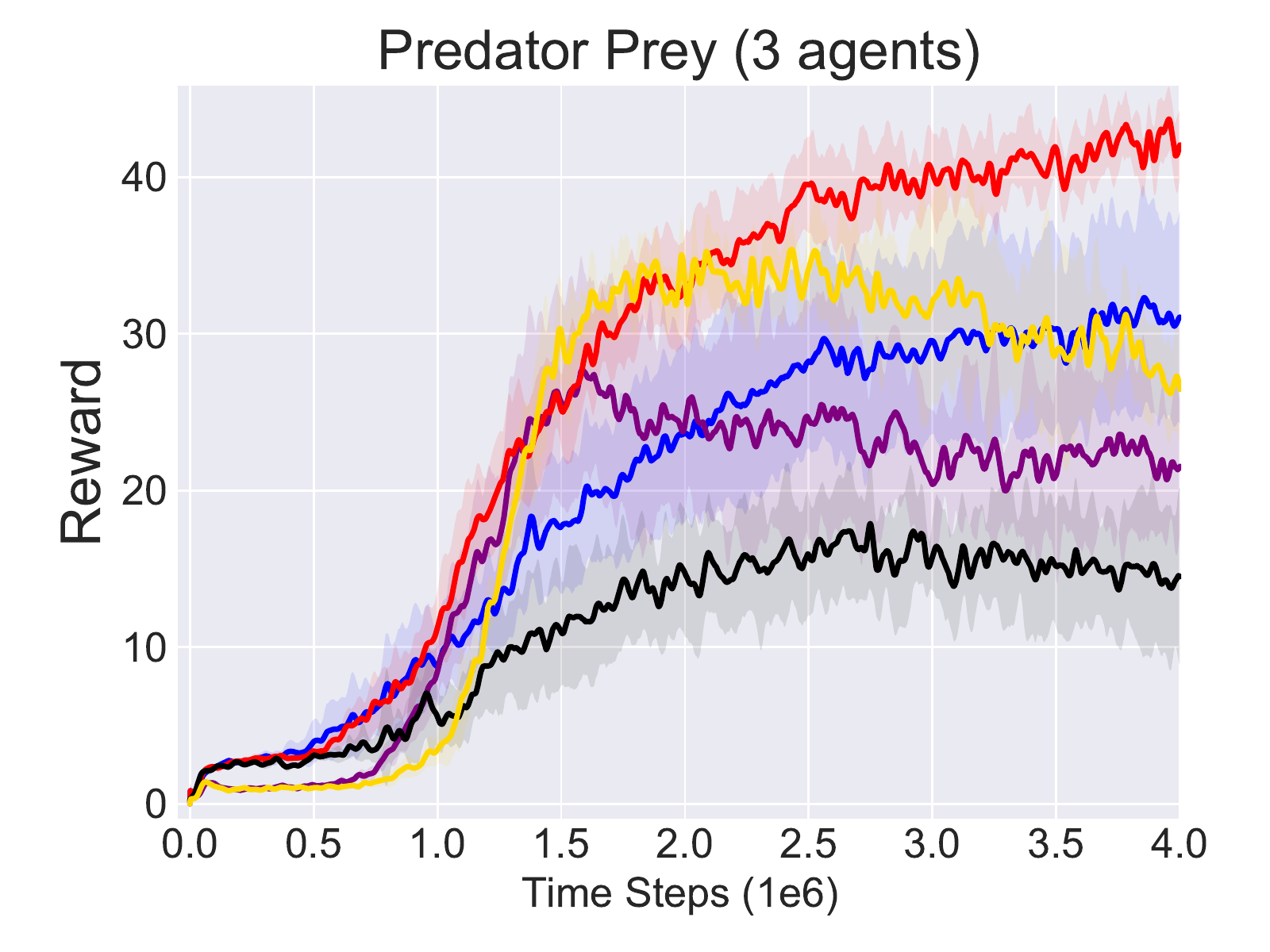}
\includegraphics[width=0.49\linewidth]{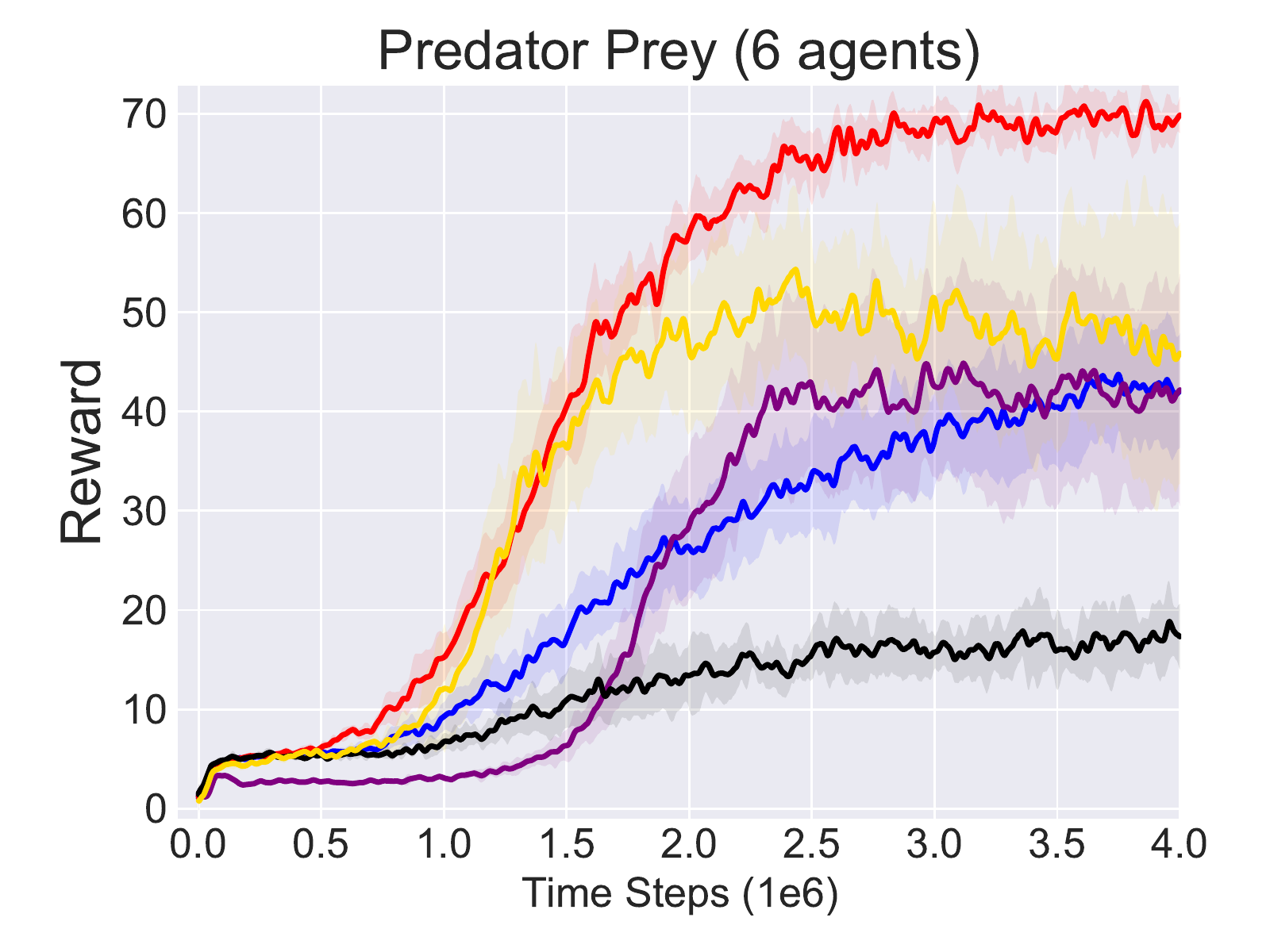}
\includegraphics[width=0.49\linewidth]{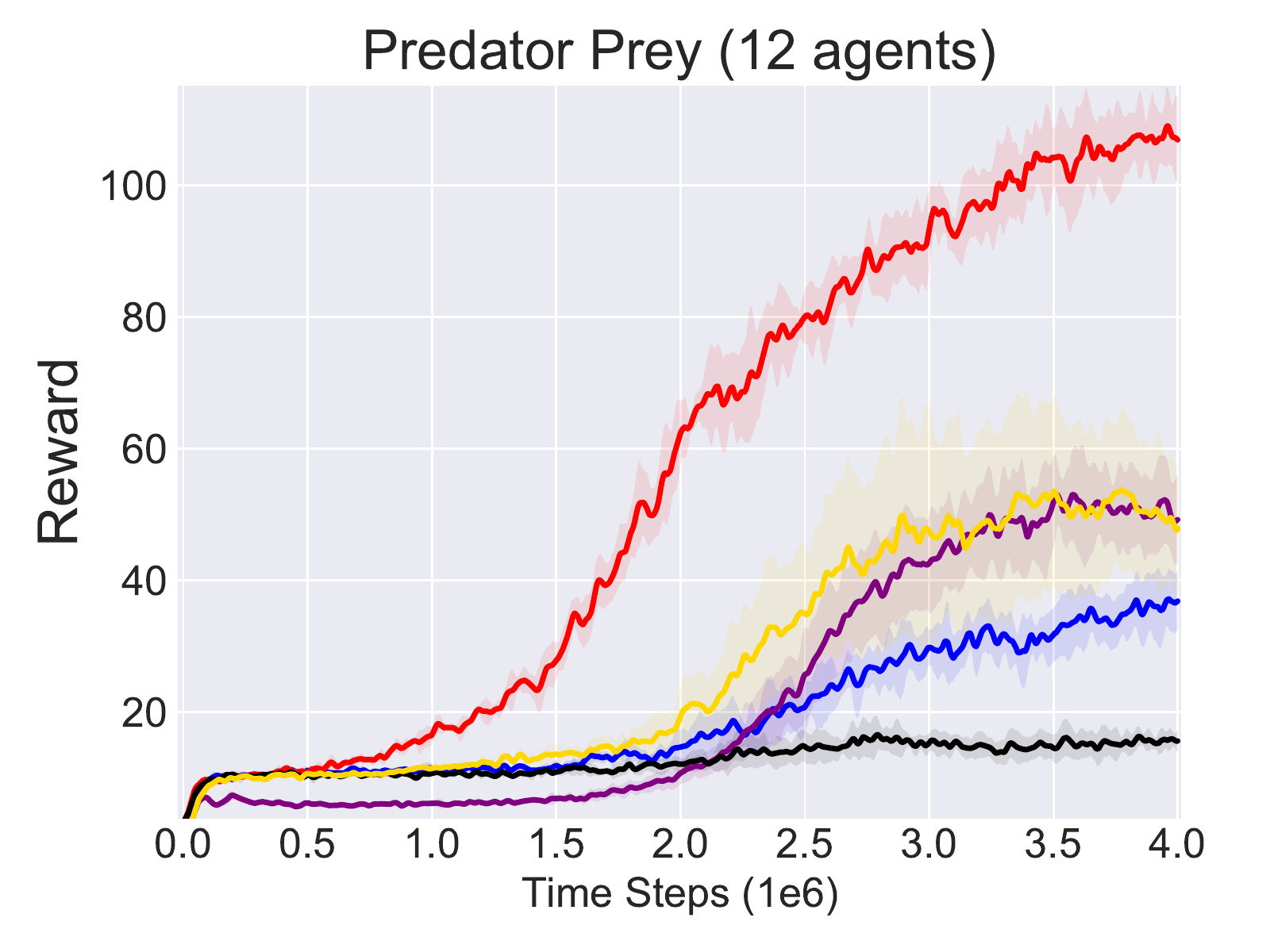}
\includegraphics[width=0.49\linewidth]{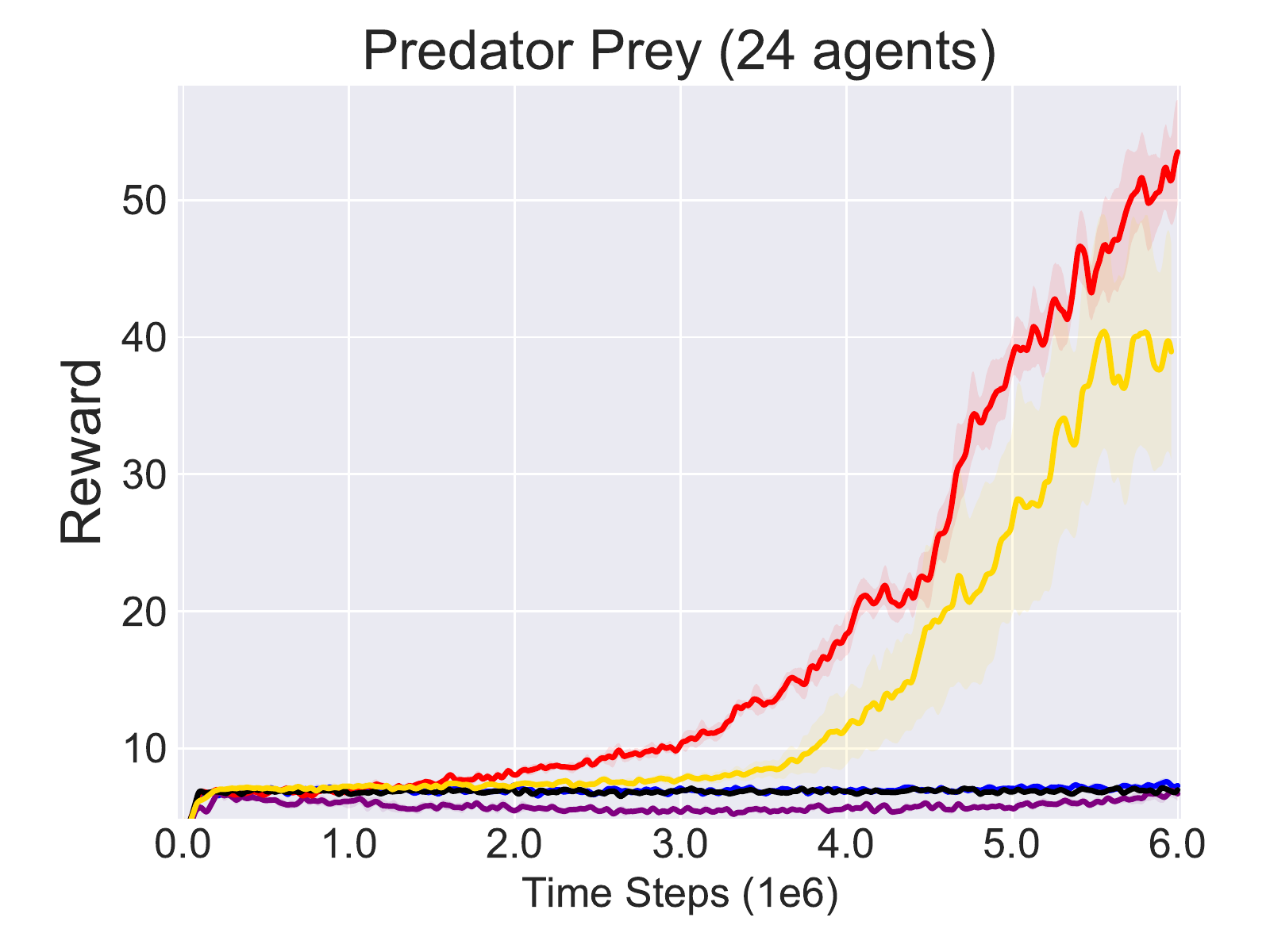}
\includegraphics[width=0.49\linewidth]{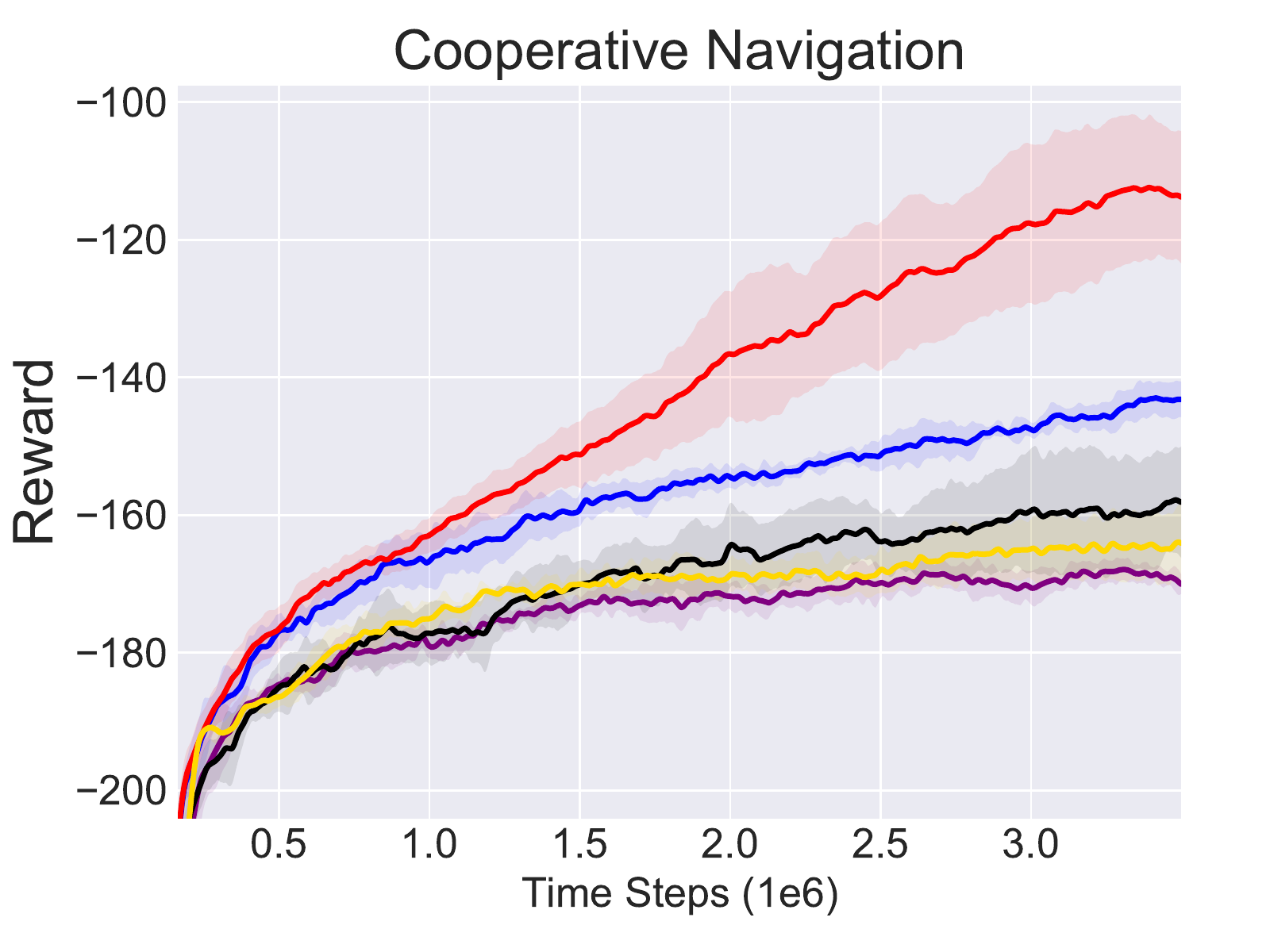}
\vspace{-0.2cm}
\caption{Comparisons of averaged return on MPE.}
\label{domain1}
\vspace{-0.2cm}
\end{figure}
We first evaluate the performance on 6 environments of MPE with 10 different random seeds: Wildlife Rescue, Cooperative Navigation, and Partial Observation Cooperative Predator Prey with 3, 6, 12, and 24 predators (where the agents control predators and the policy of prey is fixed).
The results in Figure~\ref{domain1} show that \alg-MADDPG outperforms other methods across all tasks. 
Both VM3-AC and SIC-MADDPG receive lower rewards than \alg-MADDPG, indicating that simply maximizing MI does not guarantee high-performing collaboration behaviors. 
\alg can also deliver significant performance gains as the number of agents increases, while some of the other algorithms cannot learn to collaborate effectively. For example, on Predator Prey with 24 agents, only \alg-MADDPG and VM3-AC can learn collaborative behaviors (but VM3-AC requires three times more time than \alg-MADDPG). For Wildlife Rescue, results show that the MASAC-related algorithms (e.g., MASAC and VM3-AC) can not make agents capture any wildlife, which is due to underestimation of using double Q-learning. To remove the influence of irrelevant factors, in the motivating example, we use single Q-learning instead. Details of the experiments are in Appendix \ref{ver MASAC}.


\begin{figure}[t]
\centering
\includegraphics[width=0.49\linewidth]{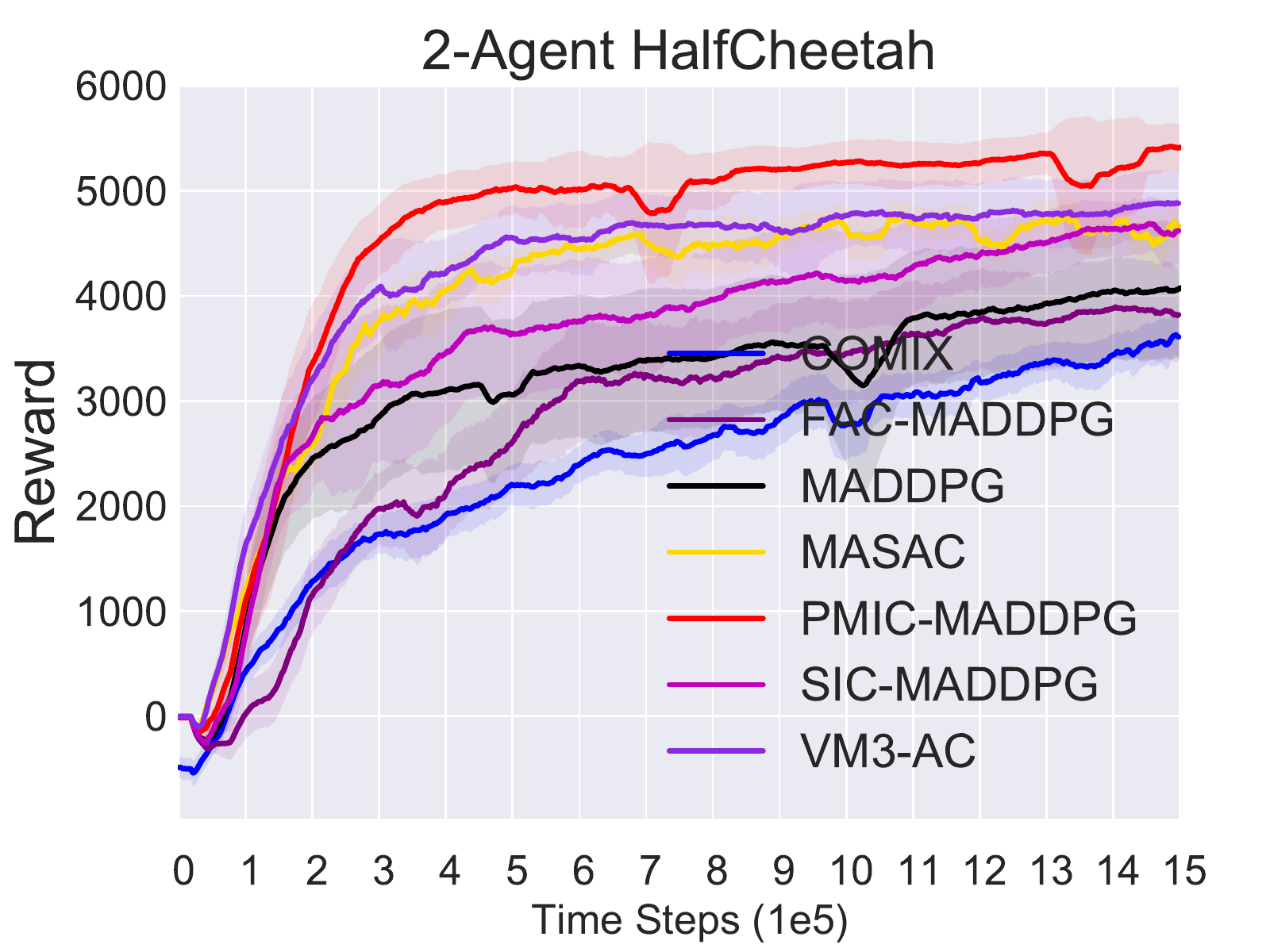}
\includegraphics[width=0.49\linewidth]{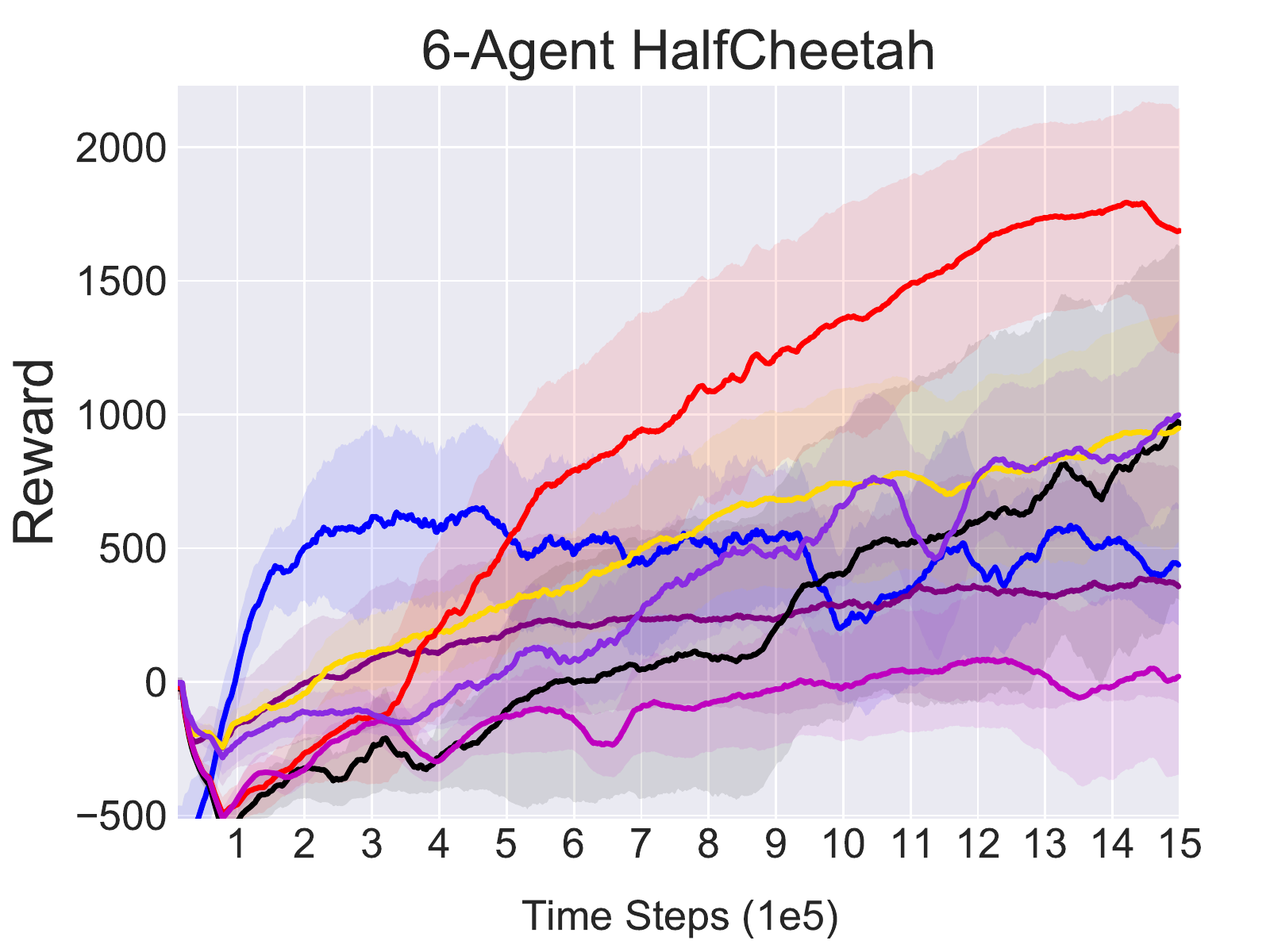}
\caption{Comparisons of averaged return on MA-MuJoCo.}
\label{domain2}
\end{figure}


We further evaluate \alg on 2 tasks of Multi-Agent MuJoCo benchmark with 10 different random seeds. In these tasks, agents need to cooperate in robot control and different agents control different joints of the robot. In our experimental setting, agents do not share information, which is the most difficult setting in Multi-Agent MuJoCo.
The results in Figure~\ref{domain2} show that \alg-MADDPG outperforms other baselines (COMIX, Fac-MADPPG, SIC-MADDPG, VM3AC, and MASAC), which demonstrates the benefits of \alg in challenging continuous control tasks.
We perform significance analysis (t-test) and prove that \alg-MADDPG achieves statistically significant advantages over other baseline algorithms.

\begin{figure}[t]
\vspace{-0.1cm}
\setlength{\belowcaptionskip}{-0.15cm}
\centering
\includegraphics[width=0.49\linewidth]{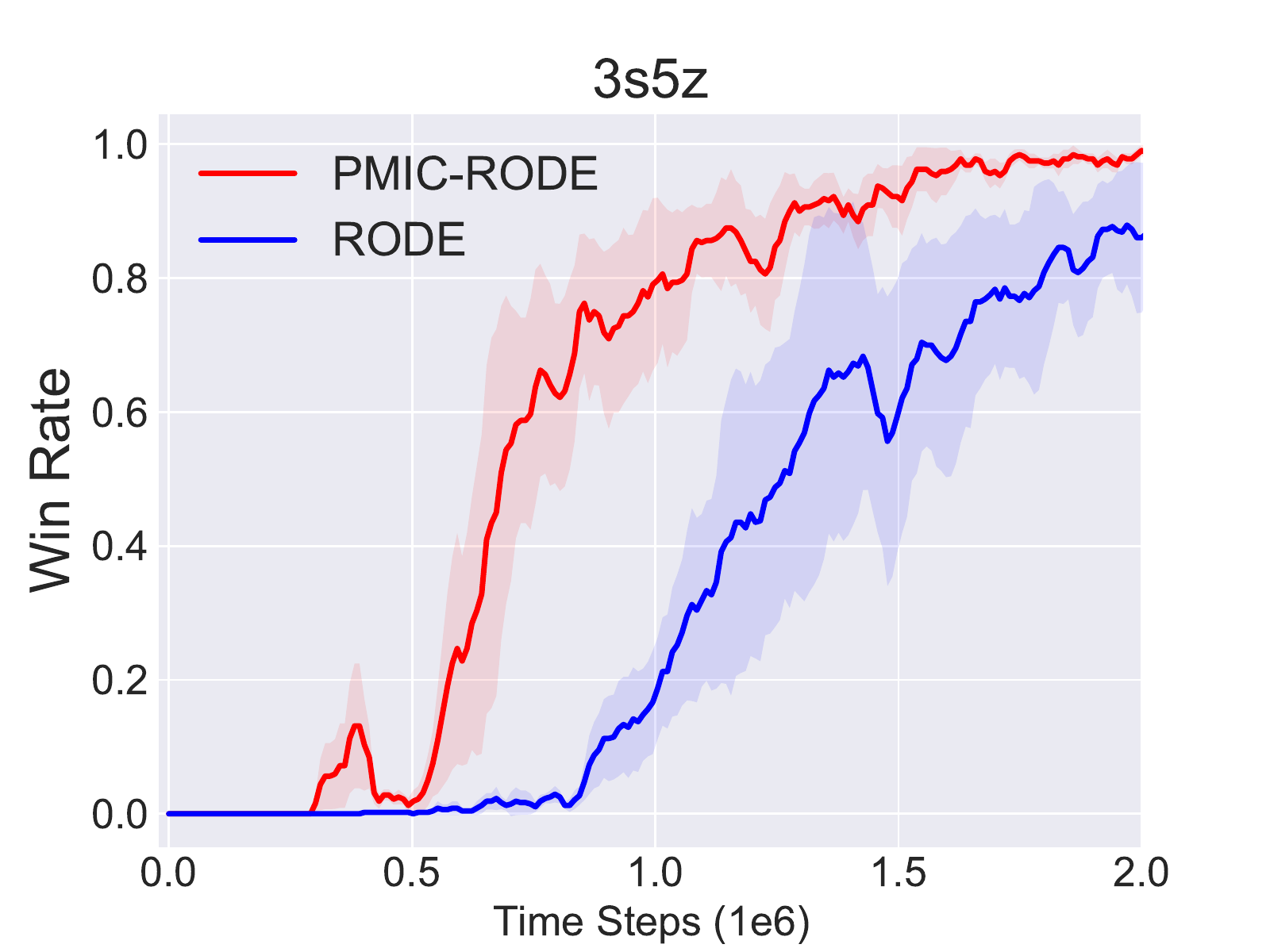}
\includegraphics[width=0.49\linewidth]{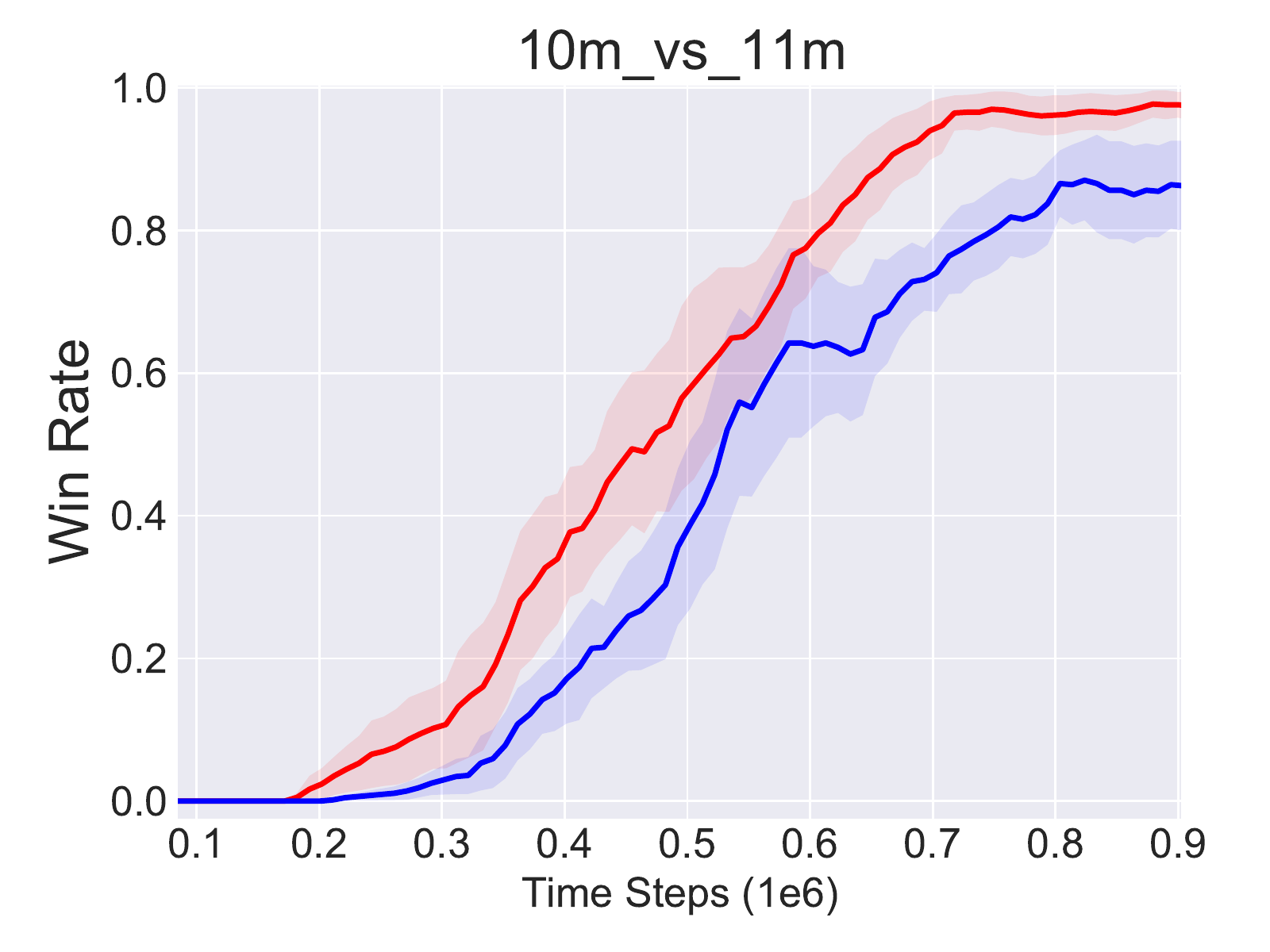}
\includegraphics[width=0.49\linewidth]{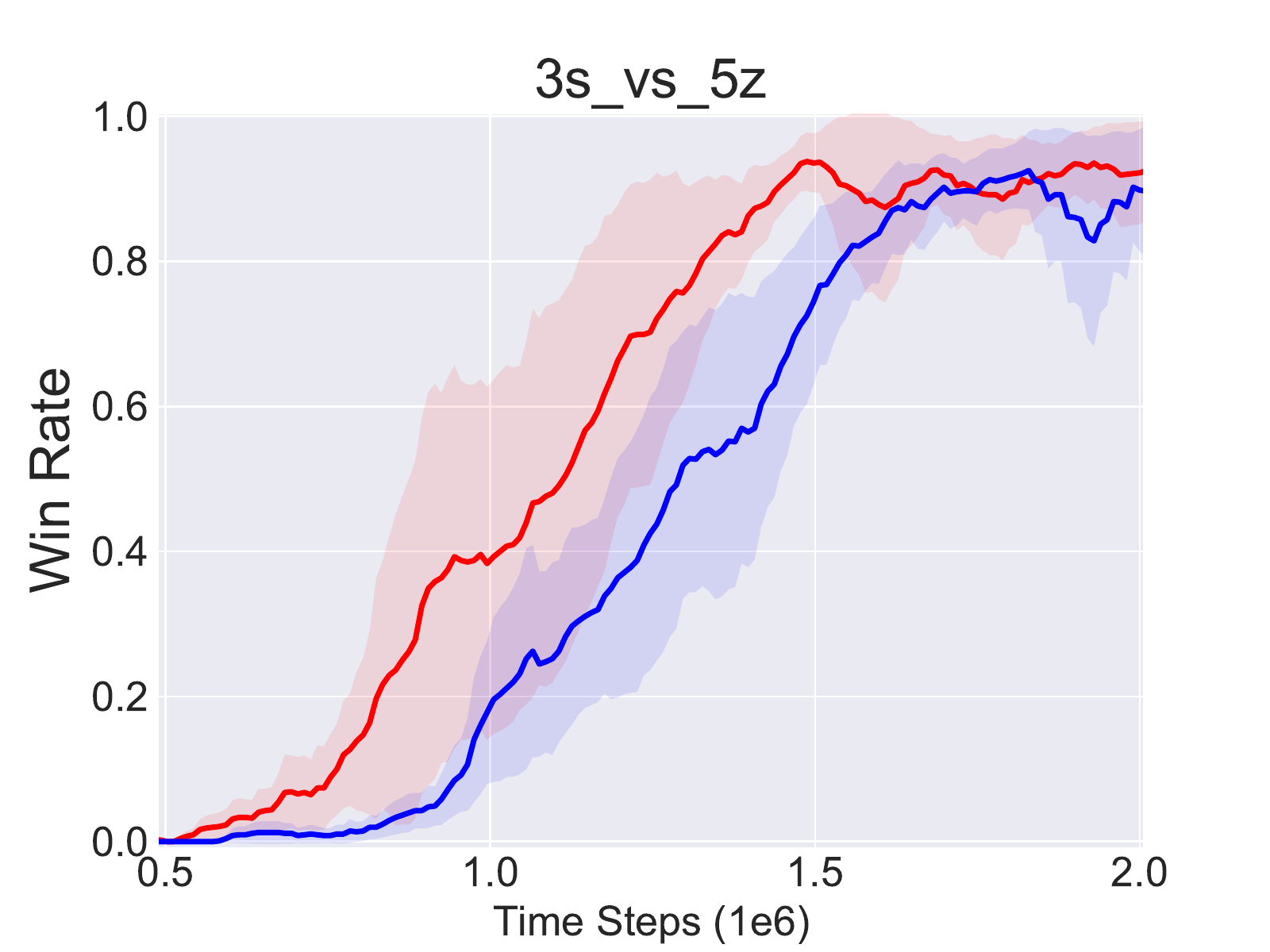}
\includegraphics[width=0.49\linewidth]{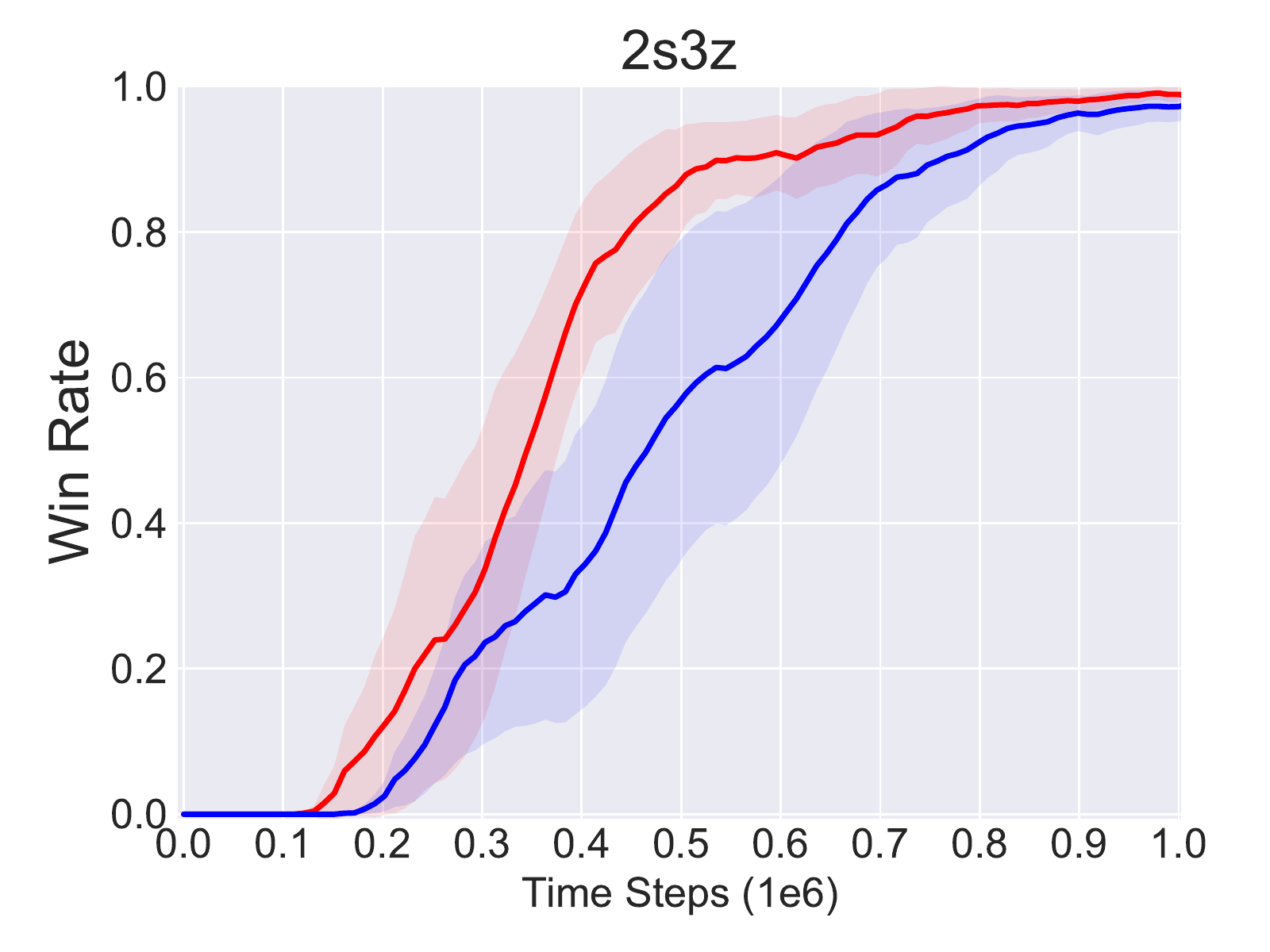}
\includegraphics[width=0.49\linewidth]{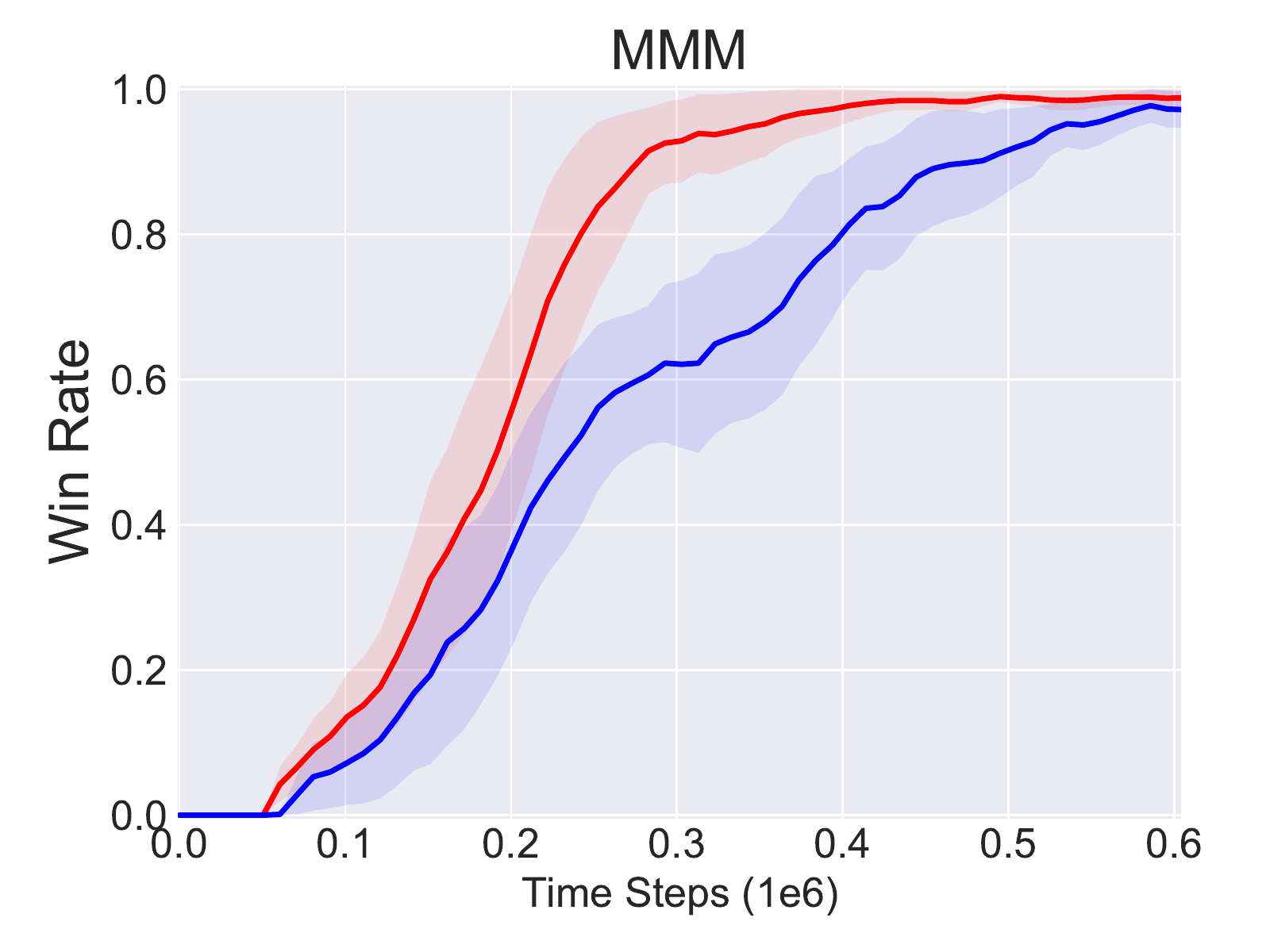}
\includegraphics[width=0.49\linewidth]{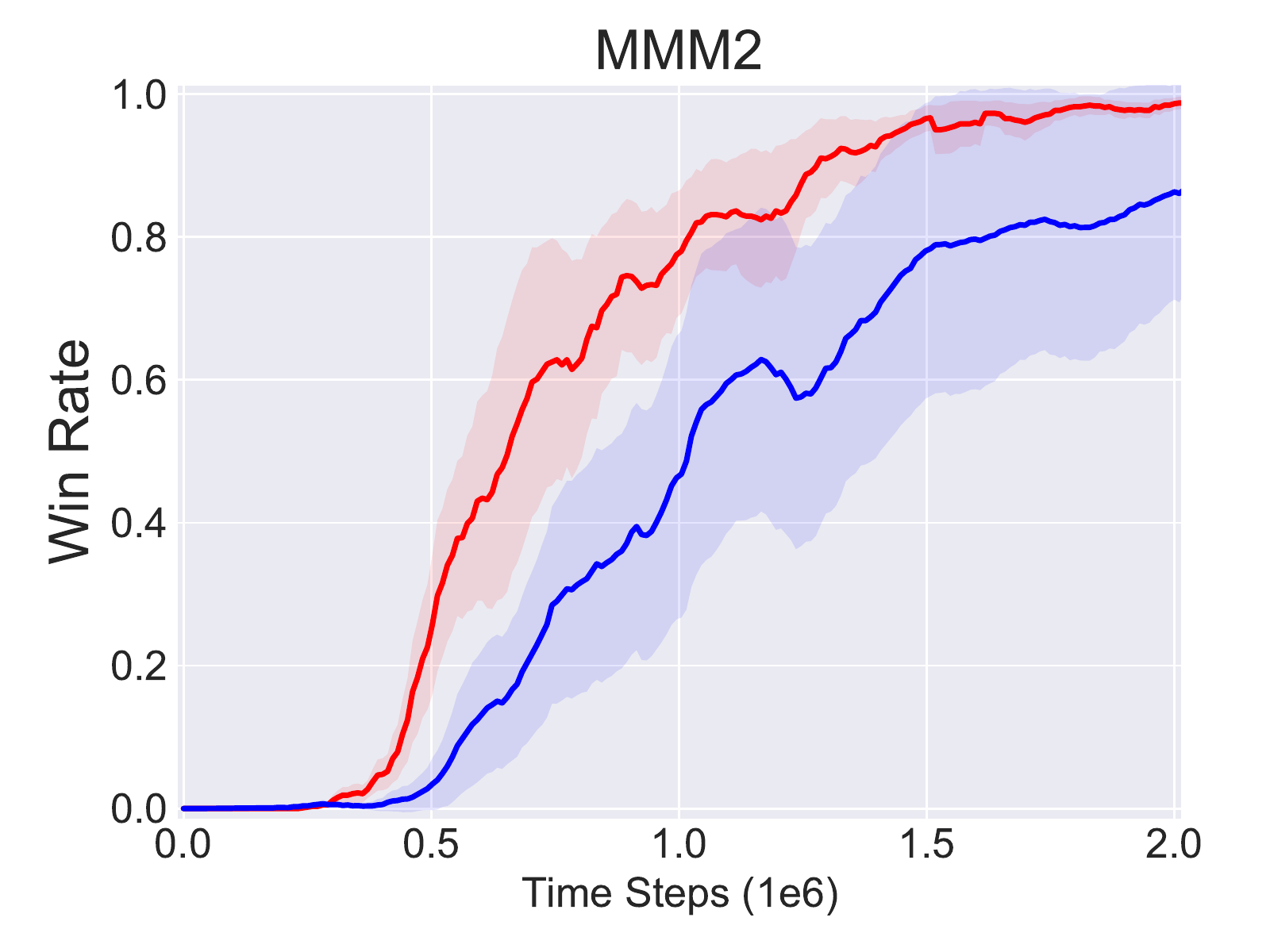}
\vspace{-0.2cm}
\caption{Comparisons of averaged test win rate on SMAC.}
\label{domain3}
\end{figure}
For SMAC, we evaluate \alg on 6 maps with 5 different random seeds. Note that RODE is the SOTA algorithm on SMAC.
The results in Figure~\ref{domain3} show that \alg can provide an improvement over RODE, reaching convergence faster and achieving higher performance. 

In summary, these experiments show that \alg is an effective framework that can be integrated with multiple algorithms and can provide significant improvements in both continuous and discrete action space tasks. More experiments, such as \alg-MASAC and \alg-QMIX, are included in Appendix \ref{more exp}.

\subsection{Superiority of Dual MI (RQ2)}
\label{RQ2}

\begin{figure}[t]
\setlength{\belowcaptionskip}{-0.15cm}
\centering
\includegraphics[width=0.49\linewidth]{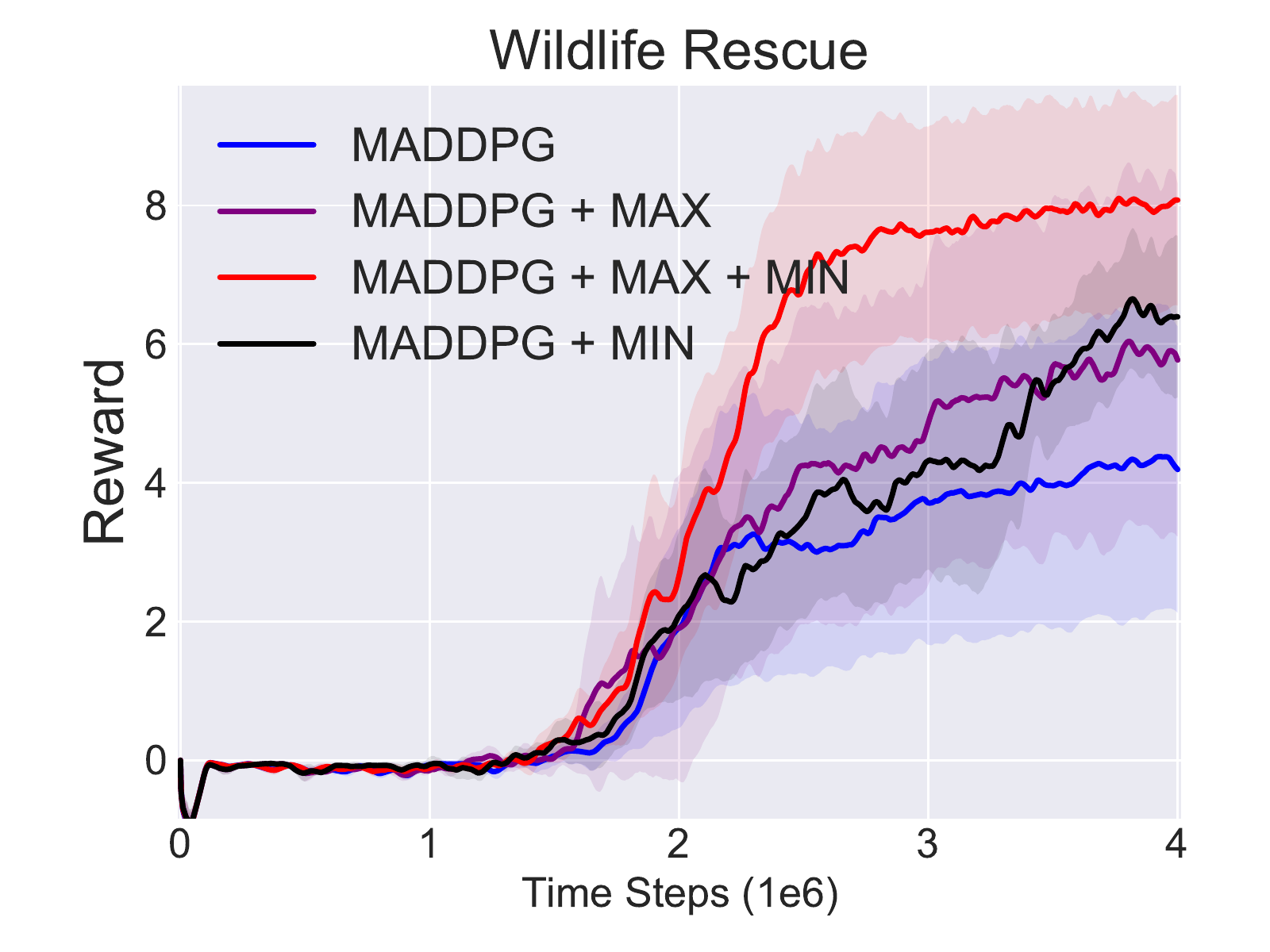}
\includegraphics[width=0.49\linewidth]{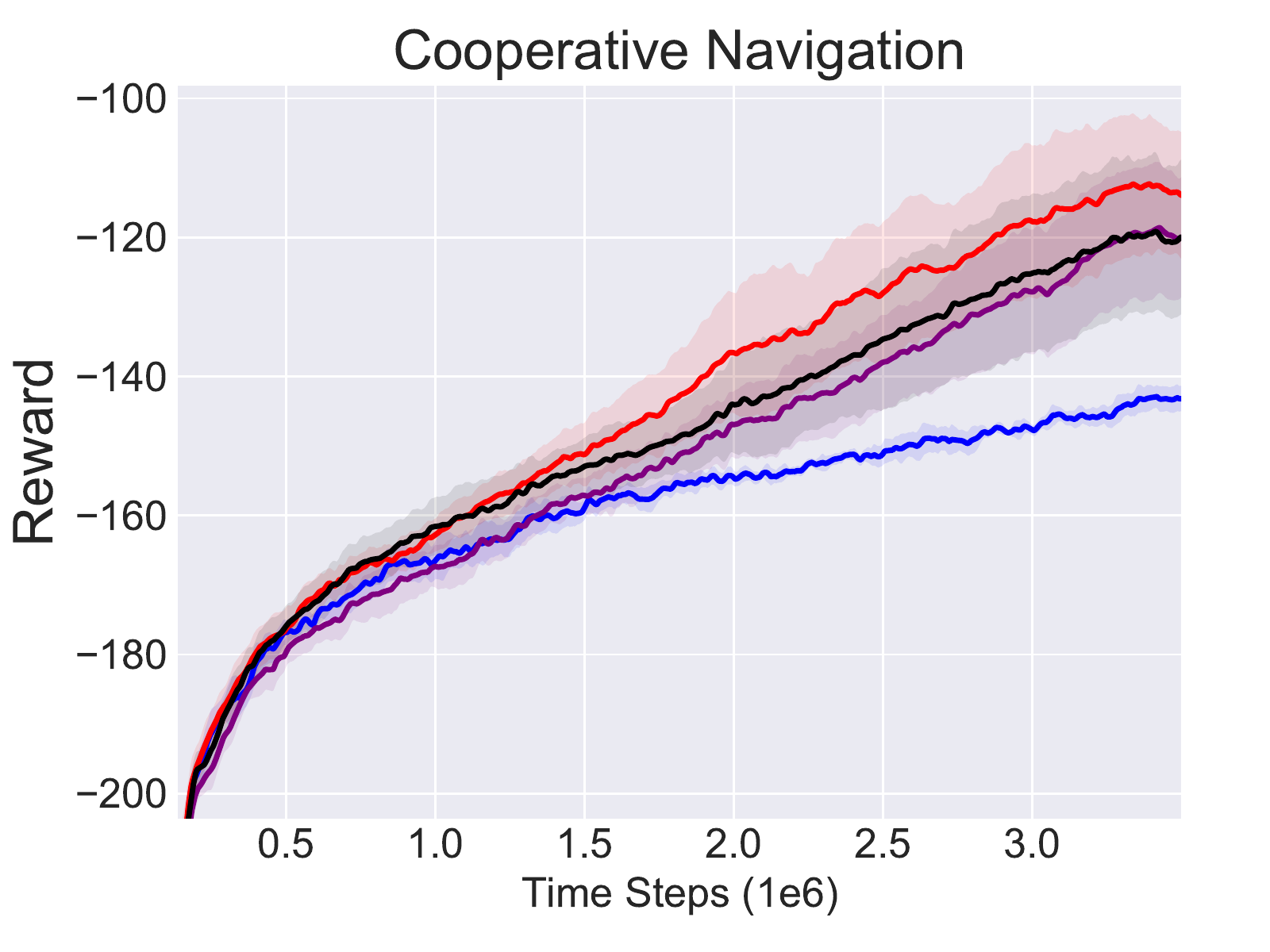}
\caption{Ablation on MI maximization \& minimization.}
\label{abd on max min I}
\end{figure}

\begin{figure}[t]
\setlength{\belowcaptionskip}{-0.15cm}
\centering
\includegraphics[width=0.49\linewidth]{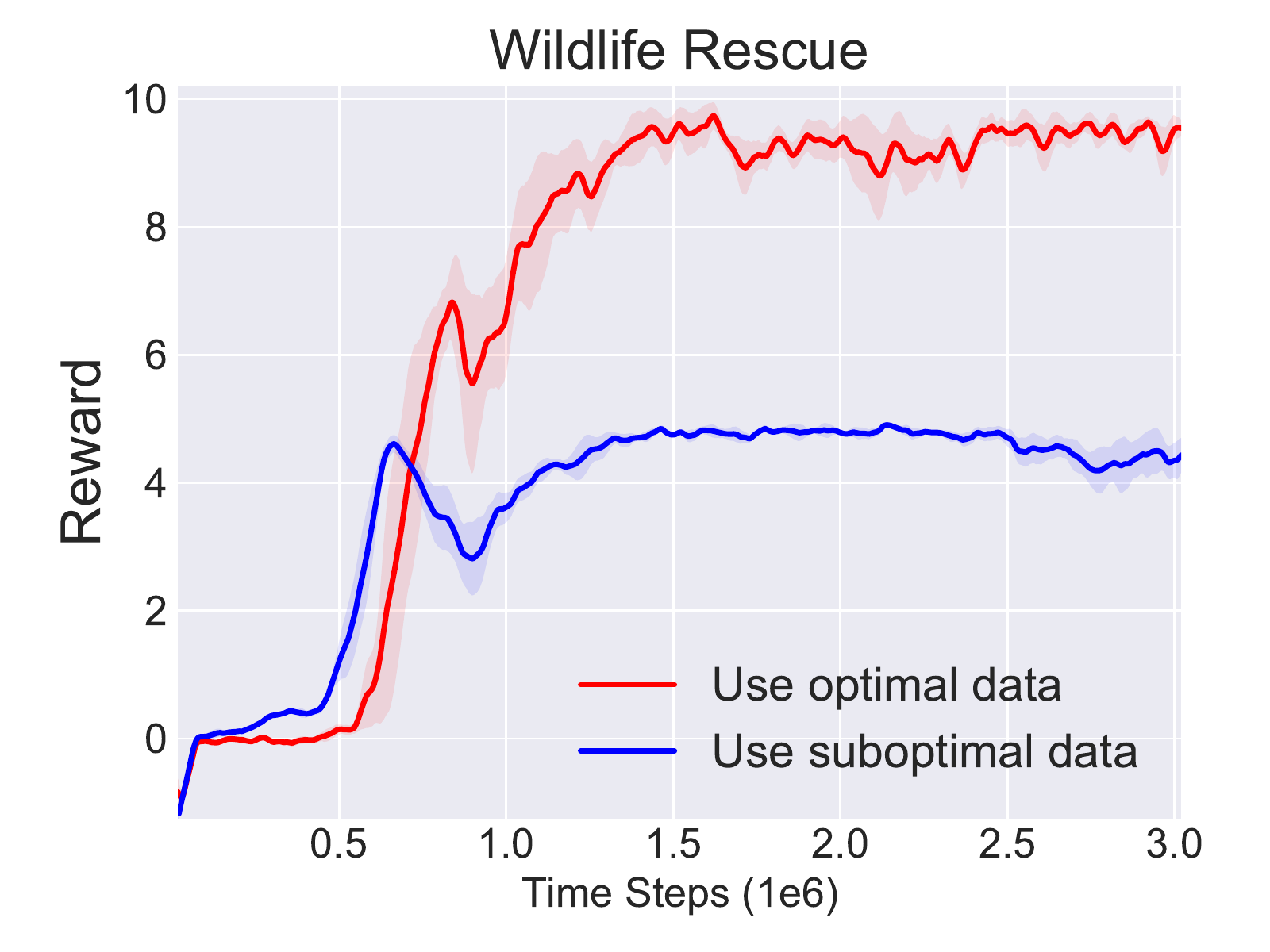}
\includegraphics[width=0.49\linewidth]{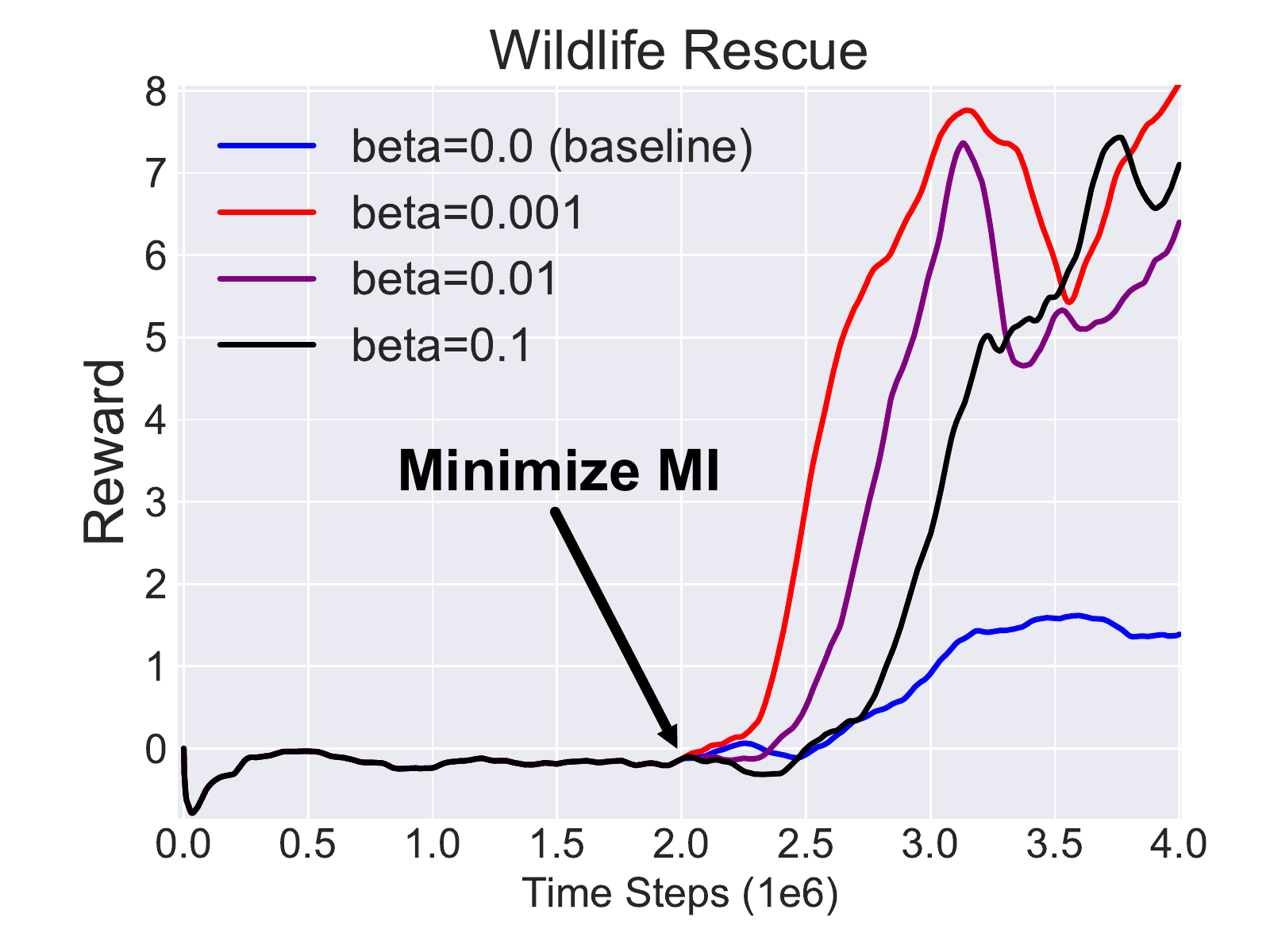}
\caption{\textit{Left}: maximizing $I(s;u)$ associated with optimal and suboptimal data to guide agents. \textit{Right}:  minimizing $I(s;u)$ to break inferior behaviors .}
\label{effect on max min I}
\end{figure}


\begin{figure}[t]
\setlength{\belowcaptionskip}{-0.15cm}
\centering
\includegraphics[width=0.49\linewidth]{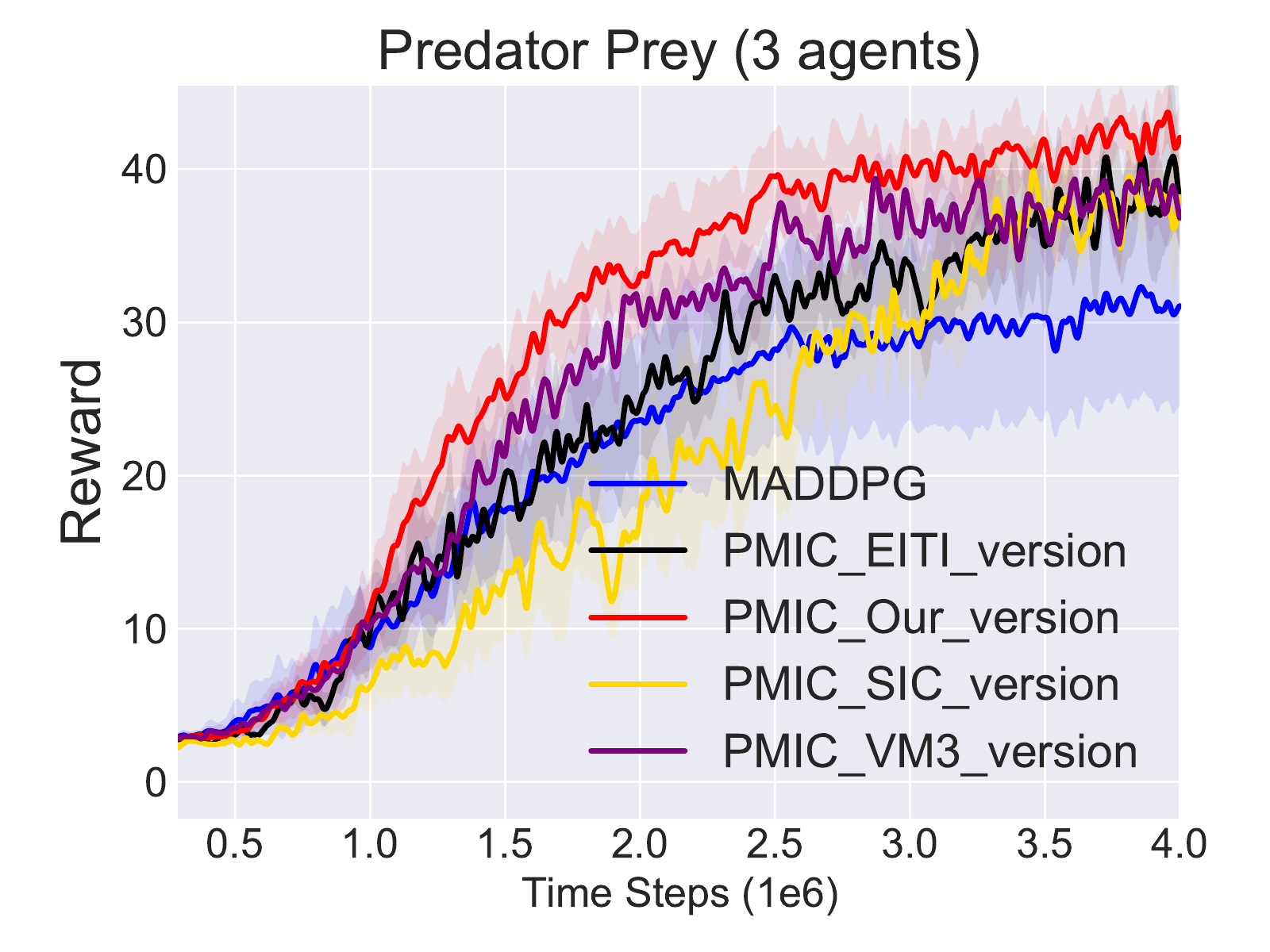}
\includegraphics[width=0.49\linewidth]{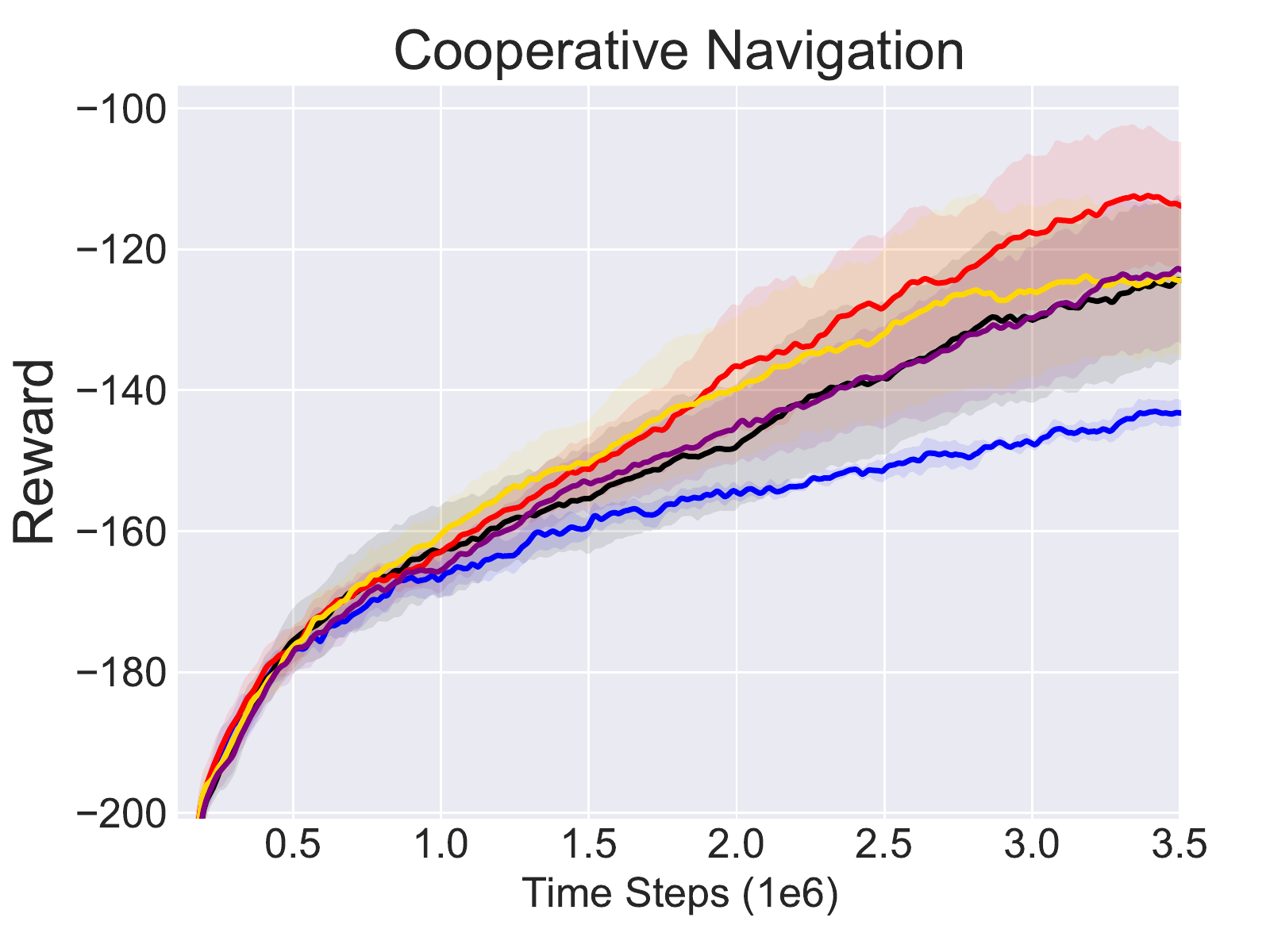}
\caption{Comparisons of different MI under \alg. EITI version measures the MI of agents' current actions/states and other agents' next states. 
SIC version measures the MI of the shared latent variables and agents' joint policy 
VM3 version measures the MI of any two agents' actions.}
\label{compare MI under PMIC. }
\end{figure}

To answer RQ2, we first provide an ablation study on maximizing and minimizing $I(s;u)$ to investigate the effect of the two mechanisms. The results in Figure~\ref{abd on max min I} show that maximizing and minimizing $I(s;u)$ can achieve faster convergence and higher performance, relative to only maximizing or only minimizing $I(s;u)$.

To verify whether the MINE estimator that maximizes $I(s;u)$ can guide agents correctly, we collect optimal (i.e., two agents both catch A) and sub-optimal (i.e., two agents both catch C) trajectories to train MINE respectively, then use the trained MINE to guide an initialized MADDPG from the beginning. As shown in the left of Figure~\ref{effect on max min I}, training MINE with optimal trajectories can quickly guide the algorithm to learn the optimal joint behavior, which indicates that maximizing $I(s;u)$ on superior trajectories can guide agents correctly and quickly. However, if MINE operates on sub-optimal trajectories, the policy converges to the sub-optima. This indicates that only maximizing MI without distinguishing the quality of trajectories can make agents fall into inferior ones. This further verifies our assumption in the motivation section and also indicates the necessity of designing \dnb. 


Next, we examine whether the CLUB estimator that minimizes $I(s;u)$ can escape sub-optima and help agents achieve high-quality collaboration. We test on Wildlife Rescue and find a policy in a sub-optima (e.g., rescuing animals with lower rewards). After 2 million time steps, we begin minimizing $I(s;u)$ for the policy with different $\beta$. Figure~\ref{effect on max min I} (right) shows that the training curve gradually increases and achieves higher rewards than baseline. This indicates that minimizing $I(s;u)$ can help the algorithm escape from sub-optima and discover better ones. 

\begin{figure}[htb]
\centering
\includegraphics[width=0.49\linewidth]{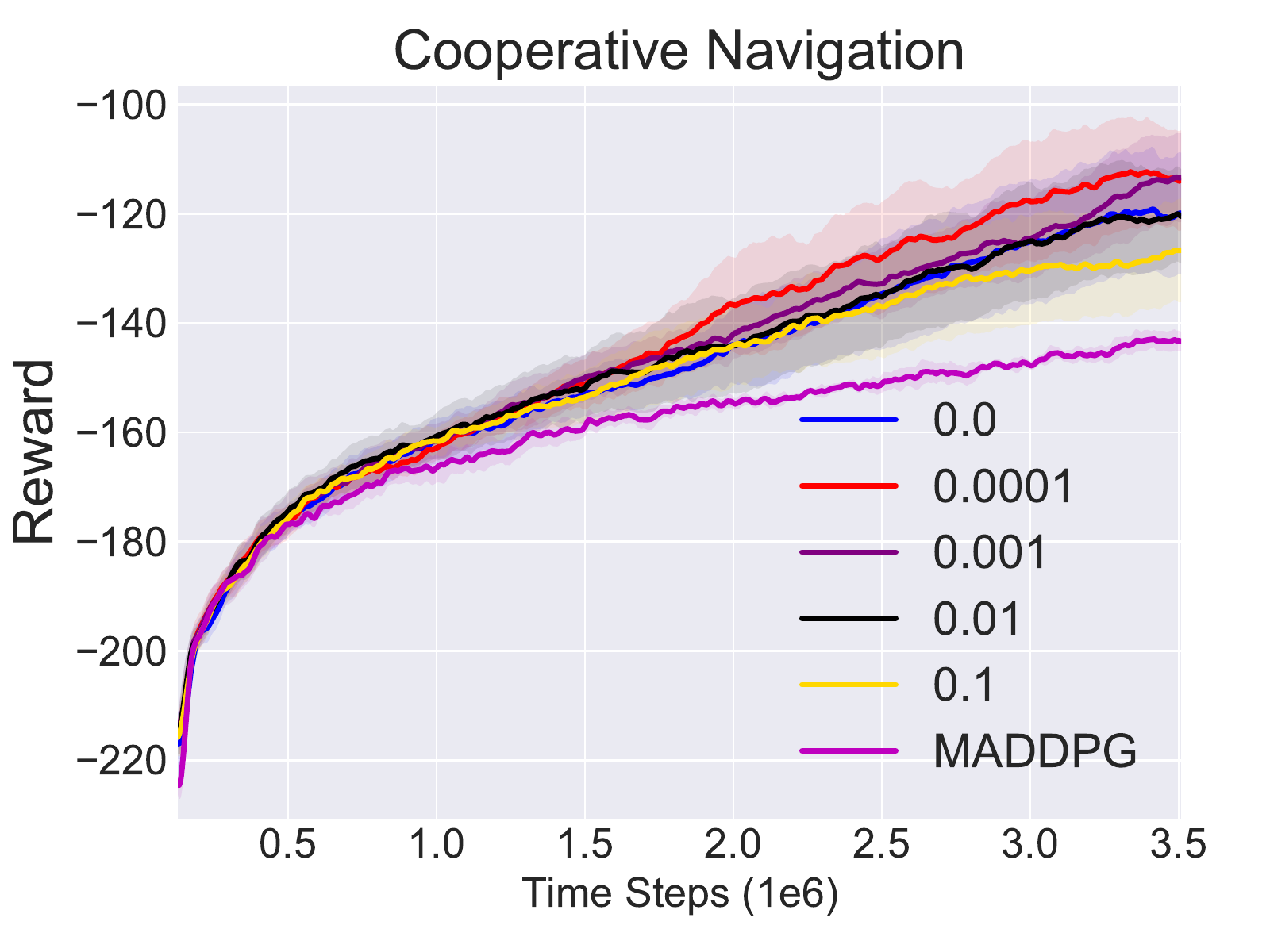}
\includegraphics[width=0.49\linewidth]{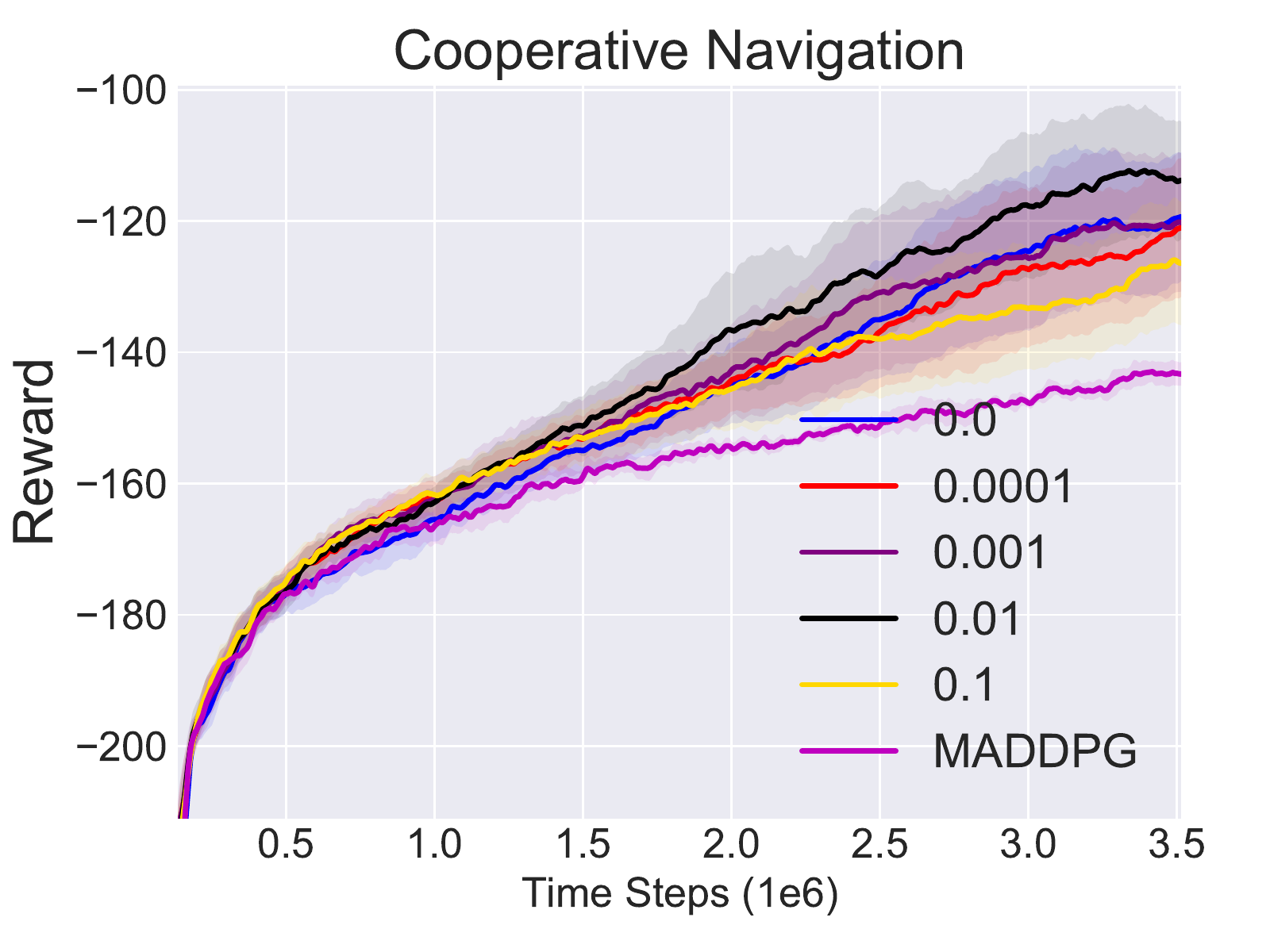}
\caption{Parameter analysis on $\alpha$ (left) and $\beta$ (right).}
\label{ablation of a and b }
\end{figure}

We further provide the parameter analysis on $\alpha$ and $\beta$. 
The results are shown in Figure~\ref{ablation of a and b }. The experiment proves that appropriately sized values of $\alpha$ and $\beta$ are critical for algorithm improvement. The performance loss occurs if $\beta$ or $\alpha$ is too large or too small. Thus, for different environments, we need to adjust $\alpha$ and $\beta$ to achieve the best performances.
Intuitively, if the environment has
multiple kinds of collaboration (e.g., Wildlife Rescue), one
may scale up $\beta$ to prevent falling into the inferior ones; if
the environment has a smooth underlying path from inferior
to superior collaboration, one may scale up $\alpha$ to enhance
guidance and accelerate learning.

To finish answering RQ2, we compare different MI measurements under \alg. Results in
Figure \ref{compare MI under PMIC. } show that our proposed measure is more effective than other alternatives.
Besides, we test the sensitivity of different parameters such as the size of \dnb, $\alpha$, and $\beta$.
Due to the space limitation, these ablation experiments are put in Appendix \ref{more exp}. 



\subsection{Necessity of \dnb (RQ3)}

\begin{figure}[t]
\setlength{\abovecaptionskip}{0.1cm}
\setlength{\belowcaptionskip}{-0.15cm}
\centering
\includegraphics[width=0.49\linewidth]{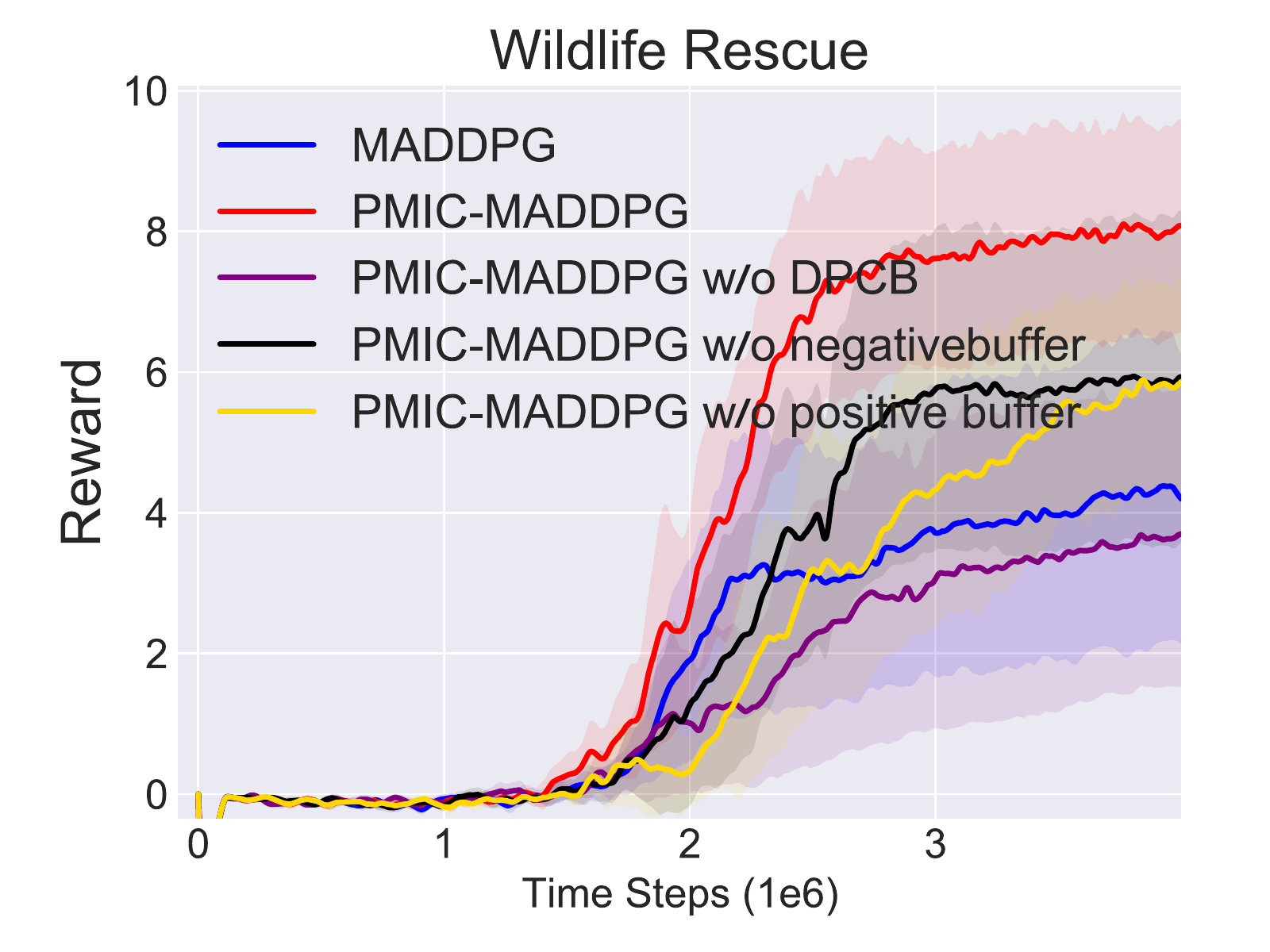}
\includegraphics[width=0.49\linewidth]{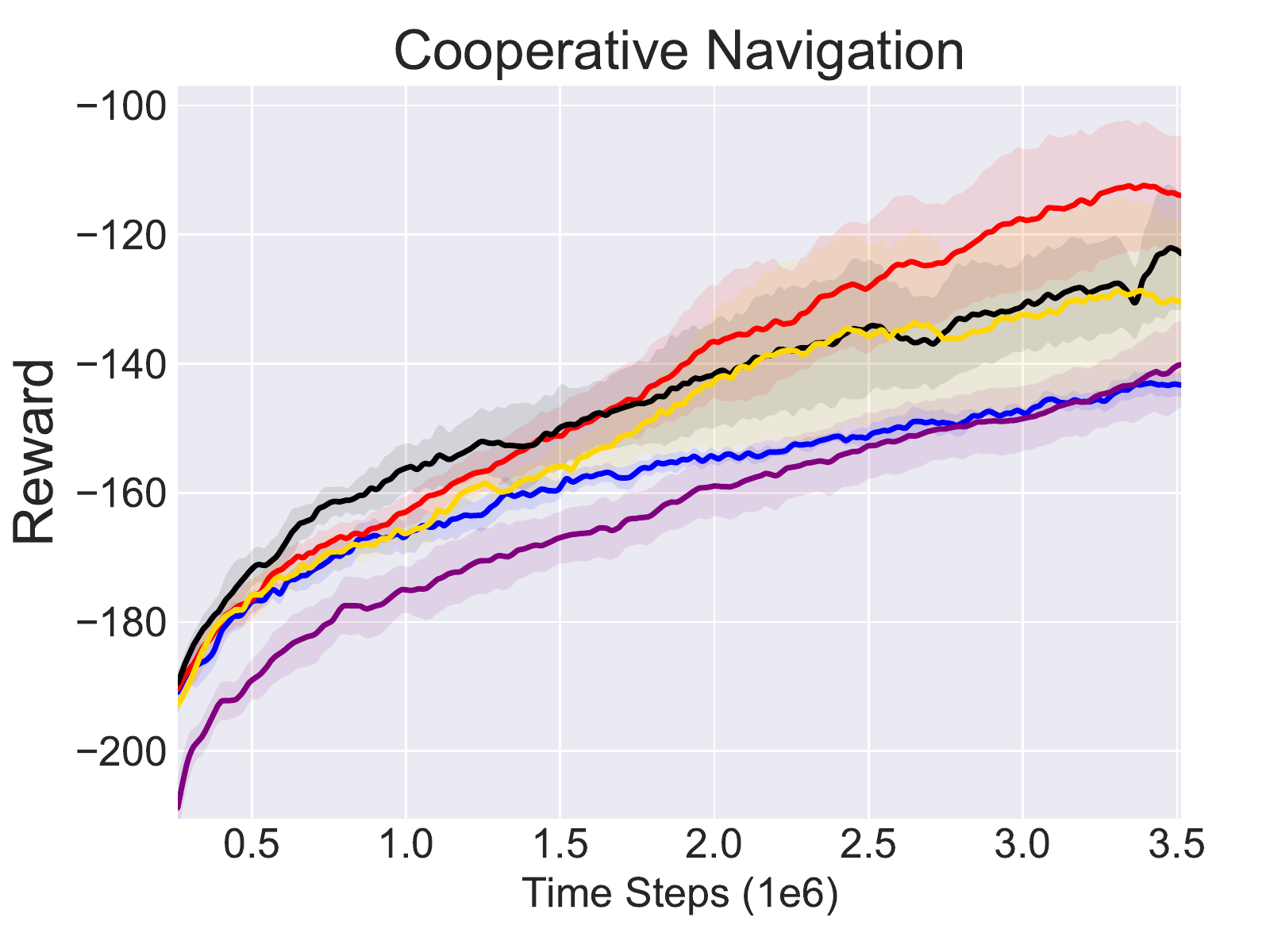}
\caption{Ablation experiments on \dnb. "PMIC-MADDPG w/o negative buffer” means
we use a regular buffer to train CLUB, i.e., without data
filtering as introduced in Sec.~\ref{subsection:progressive_mi_estimation}}
\label{Ablation dnb}
\vspace{-0.3cm}
\end{figure}
To address RQ3, we consider whether the positive and negative buffers in \dnb can be replaced with a normal replay buffer. Results in Figure~\ref{Ablation dnb} show that \dnb is indeed more effective. If we put all these trajectories into one normal buffer, MINE and CLUB cannot distinguish what behaviors are more favorable, in a result MINE and CLUB cannot provide useful guidance for agents. 


\section{Conclusion \& Future work}

To address the potentially detrimental effect of only maximizing mutual information,
we propose the \alg framework, consisting of a newly proposed collaboration criterion measured by the MI between global states and joint actions;
\dnb to progressively maintain superior and inferior trajectories;
and \mie to maintain two MI estimates of our new criterion from samples in \dnb.
In addition to maximizing the expected discounted return,
PMIC-MARL maximizes the MI estimates associated with superior collaboration to guide agents to facilitate better collaboration, while minimizing the MI estimates associated with inferior collaborations, effectively encouraging exploration while avoiding sub-optimal collaborations.
In our experiments, we evaluate several implementations of PMIC-MARL in a wide range of cooperative environments with both continuous action space and discrete action space.  The results demonstrate the effectiveness and generalization of PMIC.

For limitations, 
firstly our work is empirical proof of the effectiveness of the \alg idea and we provide no theory on optimality, convergence, and complexity.
Secondly, the choice of hyperparameters $\alpha$ and $\beta$ has a significant effect on the performances. Thirdly, the current approach to distinguishing data is relatively simple and straightforward. In the future, we would like to improve \alg to solve the above limitations by theoretical support, automatic hyperparameter tuning technology, and more accurate methods to distinguish data. In addition, we would like to further develop \alg by applying the idea to other fields, such as communication in MARL~\cite{DBLP:conf/aaai/SunZHML20} and hierarchical reinforcement learning (HRL)~\cite{tang2018hierarchical}. For example, in HRL, the maximization-minimization of MI could help lower-level agents achieve effective execution based on a target provided by an upper-level policy. On the other hand, extending PMIC with advanced techniques, like hybrid action space~\cite{DBLP:journals/corr/abs-2109-05490}, evolutionary RL~\cite{EMOGI}, or prior human knowledge~\cite{DBLP:conf/ijcai/ZhangHWTMDZ20}, is worth further study.

\section*{Acknowledgements}
The work is supported by the National Natural Science Foundation of China (Grant Nos.: 62106172, U1836214) and the new Generation of Artificial Intelligence Science and Technology Major Project of Tianjin under grant: 19ZXZNGX00010 and the Science and Technology on Information Systems Engineering Laboratory (Grant No. WDZC20205250407).


\bibliography{example_paper}

\begin{thebibliography}{40}
\providecommand{\natexlab}[1]{#1}
\providecommand{\url}[1]{\texttt{#1}}
\expandafter\ifx\csname urlstyle\endcsname\relax
  \providecommand{\doi}[1]{doi: #1}\else
  \providecommand{\doi}{doi: \begingroup \urlstyle{rm}\Url}\fi

\bibitem[Belghazi et~al.(2018)Belghazi, Baratin, Rajeshwar, Ozair, Bengio,
  Courville, and Hjelm]{belghazi2018mutual}
Belghazi, M.~I., Baratin, A., Rajeshwar, S., Ozair, S., Bengio, Y., Courville,
  A., and Hjelm, D.
\newblock Mutual information neural estimation.
\newblock In \emph{International Conference on Machine Learning}, pp.\
  531--540, 2018.

\bibitem[Chen et~al.(2021)Chen, Guo, Du, Fang, Zhang, Zhang, and
  Yu]{chen2019signal}
Chen, L., Guo, H., Du, Y., Fang, F., Zhang, H., Zhang, W., and Yu, Y.
\newblock Signal instructed coordination in cooperative multi-agent
  reinforcement learning.
\newblock In \emph{Distributed Artificial Intelligence}, volume 13170, pp.\
  185--205, 2021.

\bibitem[Cheng et~al.(2020)Cheng, Hao, Dai, Liu, Gan, and Carin]{cheng2020club}
Cheng, P., Hao, W., Dai, S., Liu, J., Gan, Z., and Carin, L.
\newblock Club: A contrastive log-ratio upper bound of mutual information.
\newblock In \emph{International Conference on Machine Learning}, pp.\
  1779--1788. PMLR, 2020.

\bibitem[de~Witt et~al.(2020)de~Witt, Peng, Kamienny, Torr, B{\"o}hmer, and
  Whiteson]{de2020deep}
de~Witt, C.~S., Peng, B., Kamienny, P.-A., Torr, P., B{\"o}hmer, W., and
  Whiteson, S.
\newblock Deep multi-agent reinforcement learning for decentralized continuous
  cooperative control.
\newblock \emph{arXiv preprint arXiv:2003.06709}, 2020.

\bibitem[Guo et~al.(2018)Guo, Oh, Singh, and Lee]{guo2018generative}
Guo, Y., Oh, J., Singh, S., and Lee, H.
\newblock Generative adversarial self-imitation learning.
\newblock \emph{arXiv preprint arXiv:1812.00950}, 2018.

\bibitem[Hernandez-Leal et~al.(2019)Hernandez-Leal, Kartal, and
  Taylor]{2019jaamas}
Hernandez-Leal, P., Kartal, B., and Taylor, M.~E.
\newblock A survey and critique of multiagent deep reinforcement learning.
\newblock \emph{Journal of Autonomous Agents and Multiagent Systems},
  33:\penalty0 750--797, October 2019.
\newblock \doi{https://doi.org/10.1007/s10458-019-09421-1}.

\bibitem[Hu et~al.(2021)Hu, Jiang, Harding, Wu, and Liao]{hu2021rethinking}
Hu, J., Jiang, S., Harding, S.~A., Wu, H., and Liao, S.-w.
\newblock Rethinking the implementation tricks and monotonicity constraint in
  cooperative multi-agent reinforcement learning.
\newblock \emph{arXiv e-prints}, pp.\  arXiv--2102, 2021.

\bibitem[Iqbal \& Sha(2019)Iqbal and Sha]{iqbal2019actor}
Iqbal, S. and Sha, F.
\newblock Actor-attention-critic for multi-agent reinforcement learning.
\newblock In \emph{International Conference on Machine Learning}, pp.\
  2961--2970. PMLR, 2019.

\bibitem[Jaques et~al.(2018)Jaques, Lazaridou, Hughes, G{\"{u}}l{\c{c}}ehre,
  Ortega, Strouse, Leibo, and de~Freitas]{08647}
Jaques, N., Lazaridou, A., Hughes, E., G{\"{u}}l{\c{c}}ehre, {\c{C}}., Ortega,
  P.~A., Strouse, D., Leibo, J.~Z., and de~Freitas, N.
\newblock Intrinsic social motivation via causal influence in multi-agent {RL}.
\newblock \emph{arXiv preprint arXiv:1810.08647}, 2018.

\bibitem[Jaques et~al.(2019)Jaques, Lazaridou, Hughes, Gulcehre, Ortega,
  Strouse, Leibo, and De~Freitas]{jaques2019social}
Jaques, N., Lazaridou, A., Hughes, E., Gulcehre, C., Ortega, P., Strouse, D.,
  Leibo, J.~Z., and De~Freitas, N.
\newblock Social influence as intrinsic motivation for multi-agent deep
  reinforcement learning.
\newblock In \emph{International Conference on Machine Learning}, pp.\
  3040--3049. PMLR, 2019.

\bibitem[Kim et~al.(2020)Kim, Jung, Cho, and Sung]{kim2020maximum}
Kim, W., Jung, W., Cho, M., and Sung, Y.
\newblock A maximum mutual information framework for multi-agent reinforcement
  learning.
\newblock \emph{arXiv preprint arXiv:2006.02732}, 2020.

\bibitem[Lanctot et~al.(2017)Lanctot, Zambaldi, Gruslys, Lazaridou, Tuyls,
  P{\'e}rolat, Silver, and Graepel]{lanctot2017unified}
Lanctot, M., Zambaldi, V., Gruslys, A., Lazaridou, A., Tuyls, K., P{\'e}rolat,
  J., Silver, D., and Graepel, T.
\newblock A unified game-theoretic approach to multiagent reinforcement
  learning.
\newblock \emph{arXiv preprint arXiv:1711.00832}, 2017.

\bibitem[Li et~al.(2021)Li, Tang, Zheng, Hao, Li, Wang, Meng, and
  Wang]{DBLP:journals/corr/abs-2109-05490}
Li, B., Tang, H., Zheng, Y., Hao, J., Li, P., Wang, Z., Meng, Z., and Wang, L.
\newblock Hyar: Addressing discrete-continuous action reinforcement learning
  via hybrid action representation.
\newblock \emph{CoRR}, abs/2109.05490, 2021.

\bibitem[Li et~al.(2019)Li, Qin, Jiao, Yang, Wang, Wang, Wu, and
  Ye]{li2019efficient}
Li, M., Qin, Z., Jiao, Y., Yang, Y., Wang, J., Wang, C., Wu, G., and Ye, J.
\newblock Efficient ridesharing order dispatching with mean field multi-agent
  reinforcement learning.
\newblock In \emph{The World Wide Web Conference}, pp.\  983--994, 2019.

\bibitem[Liu et~al.(2020)Liu, Zhou, Zhang, Zhuang, Wang, Liu, and
  Yu]{liu2020multi}
Liu, M., Zhou, M., Zhang, W., Zhuang, Y., Wang, J., Liu, W., and Yu, Y.
\newblock Multi-agent interactions modeling with correlated policies.
\newblock \emph{arXiv preprint arXiv:2001.03415}, 2020.

\bibitem[Lowe et~al.(2017)Lowe, Wu, Tamar, Harb, Abbeel, and
  Mordatch]{lowe2017multi}
Lowe, R., Wu, Y.~I., Tamar, A., Harb, J., Abbeel, O.~P., and Mordatch, I.
\newblock Multi-agent actor-critic for mixed cooperative-competitive
  environments.
\newblock In \emph{Advances in neural information processing systems}, pp.\
  6379--6390, 2017.

\bibitem[Lyu et~al.(2022)Lyu, Baisero, Xiao, and Amato]{lyu2022deeper}
Lyu, X., Baisero, A., Xiao, Y., and Amato, C.
\newblock A deeper understanding of state-based critics in multi-agent
  reinforcement learning.
\newblock \emph{arXiv preprint arXiv:2201.01221}, 2022.

\bibitem[Mahajan et~al.(2019)Mahajan, Rashid, Samvelyan, and
  Whiteson]{mahajan2019maven}
Mahajan, A., Rashid, T., Samvelyan, M., and Whiteson, S.
\newblock Maven: Multi-agent variational exploration.
\newblock In \emph{Advances in Neural Information Processing Systems}, pp.\
  7613--7624, 2019.

\bibitem[Matignon et~al.(2012)Matignon, Jeanpierre, and
  Mouaddib]{matignon2012coordinated}
Matignon, L., Jeanpierre, L., and Mouaddib, A.
\newblock Coordinated multi-robot exploration under communication constraints
  using decentralized markov decision processes.
\newblock In \emph{Proceedings of the Twenty-Sixth {AAAI} Conference on
  Artificial Intelligence}, 2012.

\bibitem[Merhej \& Chetouani(2021)Merhej and Chetouani]{merhej2021lief}
Merhej, R. and Chetouani, M.
\newblock Lief: Learning to influence through evaluative feedback.
\newblock In \emph{Adaptive and Learning Agents Workshop (AAMAS 2021)}, 2021.

\bibitem[Oliehoek \& Amato(2016)Oliehoek and Amato]{oliehoek2016concise}
Oliehoek, F.~A. and Amato, C.
\newblock \emph{A concise introduction to decentralized POMDPs}.
\newblock Springer, 2016.

\bibitem[Peng et~al.(2017)Peng, Wen, Yang, Yuan, Tang, Long, and
  Wang]{peng2017multiagent}
Peng, P., Wen, Y., Yang, Y., Yuan, Q., Tang, Z., Long, H., and Wang, J.
\newblock Multiagent bidirectionally-coordinated nets: Emergence of human-level
  coordination in learning to play starcraft combat games.
\newblock \emph{arXiv preprint arXiv:1703.10069}, 2017.

\bibitem[Rashid et~al.(2018)Rashid, Samvelyan, De~Witt, Farquhar, Foerster, and
  Whiteson]{rashid2018qmix}
Rashid, T., Samvelyan, M., De~Witt, C.~S., Farquhar, G., Foerster, J., and
  Whiteson, S.
\newblock Qmix: Monotonic value function factorisation for deep multi-agent
  reinforcement learning.
\newblock \emph{arXiv preprint arXiv:1803.11485}, 2018.

\bibitem[Samvelyan et~al.(2019)Samvelyan, Rashid, de~Witt, Farquhar, Nardelli,
  Rudner, Hung, Torr, Foerster, and Whiteson]{samvelyan2019starcraft}
Samvelyan, M., Rashid, T., de~Witt, C.~S., Farquhar, G., Nardelli, N., Rudner,
  T. G.~J., Hung, C., Torr, P. H.~S., Foerster, J.~N., and Whiteson, S.
\newblock The starcraft multi-agent challenge.
\newblock In \emph{International Conference on Autonomous Agents and MultiAgent
  Systems}, pp.\  2186--2188, 2019.

\bibitem[Shen et~al.(2020)Shen, Zheng, Hao, Meng, Chen, Fan, and Liu]{EMOGI}
Shen, R., Zheng, Y., Hao, J., Meng, Z., Chen, Y., Fan, C., and Liu, Y.
\newblock Generating behavior-diverse game ais with evolutionary
  multi-objective deep reinforcement learning.
\newblock In Bessiere, C. (ed.), \emph{Proceedings of the Twenty-Ninth
  International Joint Conference on Artificial Intelligence, {IJCAI} 2020},
  pp.\  3371--3377. ijcai.org, 2020.

\bibitem[Sun et~al.(2020)Sun, Zheng, Hao, Meng, and
  Liu]{DBLP:conf/aaai/SunZHML20}
Sun, J., Zheng, Y., Hao, J., Meng, Z., and Liu, Y.
\newblock Continuous multiagent control using collective behavior entropy for
  large-scale home energy management.
\newblock In \emph{The Thirty-Fourth {AAAI} Conference on Artificial
  Intelligence, {AAAI} 2020, The Thirty-Second Innovative Applications of
  Artificial Intelligence Conference, {IAAI} 2020, The Tenth {AAAI} Symposium
  on Educational Advances in Artificial Intelligence, {EAAI} 2020, New York,
  NY, USA, February 7-12, 2020}, pp.\  922--929. {AAAI} Press, 2020.

\bibitem[Sunehag et~al.(2017)Sunehag, Lever, Gruslys, Czarnecki, Zambaldi,
  Jaderberg, Lanctot, Sonnerat, Leibo, Tuyls, et~al.]{sunehag2017value}
Sunehag, P., Lever, G., Gruslys, A., Czarnecki, W.~M., Zambaldi, V., Jaderberg,
  M., Lanctot, M., Sonnerat, N., Leibo, J.~Z., Tuyls, K., et~al.
\newblock Value-decomposition networks for cooperative multi-agent learning.
\newblock \emph{arXiv preprint arXiv:1706.05296}, 2017.

\bibitem[Tang et~al.(2018)Tang, Hao, Lv, Chen, Zhang, Jia, Ren, Zheng, Meng,
  Fan, et~al.]{tang2018hierarchical}
Tang, H., Hao, J., Lv, T., Chen, Y., Zhang, Z., Jia, H., Ren, C., Zheng, Y.,
  Meng, Z., Fan, C., et~al.
\newblock Hierarchical deep multiagent reinforcement learning with temporal
  abstraction.
\newblock \emph{arXiv preprint arXiv:1809.09332}, 2018.

\bibitem[Wang et~al.(2020{\natexlab{a}})Wang, Gupta, Mahajan, Peng, Whiteson,
  and Zhang]{wang2020rode}
Wang, T., Gupta, T., Mahajan, A., Peng, B., Whiteson, S., and Zhang, C.
\newblock Rode: Learning roles to decompose multi-agent tasks.
\newblock \emph{arXiv preprint arXiv:2010.01523}, 2020{\natexlab{a}}.

\bibitem[Wang et~al.(2020{\natexlab{b}})Wang, Wang, Wu, and
  Zhang]{wang2019influencebased}
Wang, T., Wang, J., Wu, Y., and Zhang, C.
\newblock Influence-based multi-agent exploration.
\newblock In \emph{International Conference on Learning Representations},
  2020{\natexlab{b}}.

\bibitem[Wang et~al.(2020{\natexlab{c}})Wang, Yang, Liu, Hao, Hao, Hu, Chen,
  Fan, and Gao]{wang2019action}
Wang, W., Yang, T., Liu, Y., Hao, J., Hao, X., Hu, Y., Chen, Y., Fan, C., and
  Gao, Y.
\newblock Action semantics network: Considering the effects of actions in
  multiagent systems.
\newblock In \emph{Proceedings of the 8th International Conference on Learning
  Representations}, 2020{\natexlab{c}}.

\bibitem[Wen et~al.(2019)Wen, Yang, Luo, Wang, and Pan]{wen2019probabilistic}
Wen, Y., Yang, Y., Luo, R., Wang, J., and Pan, W.
\newblock Probabilistic recursive reasoning for multi-agent reinforcement
  learning.
\newblock \emph{arXiv preprint arXiv:1901.09207}, 2019.

\bibitem[Xie et~al.(2020)Xie, Losey, Tolsma, Finn, and Sadigh]{xie2020learning}
Xie, A., Losey, D.~P., Tolsma, R., Finn, C., and Sadigh, D.
\newblock Learning latent representations to influence multi-agent interaction.
\newblock \emph{arXiv preprint arXiv:2011.06619}, 2020.

\bibitem[Yang et~al.(2021{\natexlab{a}})Yang, Tang, Bai, Liu, Hao, Meng, and
  Liu]{DBLP:journals/corr/abs-2109-06668}
Yang, T., Tang, H., Bai, C., Liu, J., Hao, J., Meng, Z., and Liu, P.
\newblock Exploration in deep reinforcement learning: {A} comprehensive survey.
\newblock \emph{CoRR}, abs/2109.06668, 2021{\natexlab{a}}.

\bibitem[Yang et~al.(2021{\natexlab{b}})Yang, Wang, Tang, Hao, Meng, Mao, Li,
  Liu, Chen, Hu, Fan, and Zhang]{DBLP:conf/nips/YangWTHMMLLCHFZ21}
Yang, T., Wang, W., Tang, H., Hao, J., Meng, Z., Mao, H., Li, D., Liu, W.,
  Chen, Y., Hu, Y., Fan, C., and Zhang, C.
\newblock An efficient transfer learning framework for multiagent reinforcement
  learning.
\newblock In Ranzato, M., Beygelzimer, A., Dauphin, Y.~N., Liang, P., and
  Vaughan, J.~W. (eds.), \emph{Advances in Neural Information Processing
  Systems 34}, pp.\  17037--17048, 2021{\natexlab{b}}.

\bibitem[Zhang et~al.(2020)Zhang, Hao, Wang, Tang, Ma, Duan, and
  Zheng]{DBLP:conf/ijcai/ZhangHWTMDZ20}
Zhang, P., Hao, J., Wang, W., Tang, H., Ma, Y., Duan, Y., and Zheng, Y.
\newblock Kogun: Accelerating deep reinforcement learning via integrating human
  suboptimal knowledge.
\newblock In Bessiere, C. (ed.), \emph{Proceedings of the Twenty-Ninth
  International Joint Conference on Artificial Intelligence, {IJCAI} 2020},
  pp.\  2291--2297, 2020.

\bibitem[Zhang et~al.(2019)Zhang, Zhang, and Lin]{zhang2019efficient}
Zhang, S.~Q., Zhang, Q., and Lin, J.
\newblock Efficient communication in multi-agent reinforcement learning via
  variance based control.
\newblock \emph{arXiv preprint arXiv:1909.02682}, 2019.

\bibitem[Zheng et~al.(2018{\natexlab{a}})Zheng, Meng, Hao, and
  Zhang]{DBLP:conf/pricai/ZhengMHZ18}
Zheng, Y., Meng, Z., Hao, J., and Zhang, Z.
\newblock Weighted double deep multiagent reinforcement learning in stochastic
  cooperative environments.
\newblock In Geng, X. and Kang, B. (eds.), \emph{{PRICAI} 2018: Trends in
  Artificial Intelligence - 15th Pacific Rim International Conference on
  Artificial Intelligence, Nanjing, China, August 28-31, 2018, Proceedings,
  Part {II}}, volume 11013 of \emph{Lecture Notes in Computer Science}, pp.\
  421--429. Springer, 2018{\natexlab{a}}.

\bibitem[Zheng et~al.(2018{\natexlab{b}})Zheng, Meng, Hao, Zhang, Yang, and
  Fan]{DBLP:conf/nips/ZhengMHZYF18}
Zheng, Y., Meng, Z., Hao, J., Zhang, Z., Yang, T., and Fan, C.
\newblock A deep bayesian policy reuse approach against non-stationary agents.
\newblock In Bengio, S., Wallach, H.~M., Larochelle, H., Grauman, K.,
  Cesa{-}Bianchi, N., and Garnett, R. (eds.), \emph{Advances in Neural
  Information Processing Systems 31: Annual Conference on Neural Information
  Processing Systems 2018, NeurIPS 2018, December 3-8, 2018, Montr{\'{e}}al,
  Canada}, pp.\  962--972, 2018{\natexlab{b}}.

\bibitem[Zheng et~al.(2020)Zheng, Hao, Zhang, Meng, and
  Hao]{DBLP:journals/jcst/ZhengHZMH20}
Zheng, Y., Hao, J., Zhang, Z., Meng, Z., and Hao, X.
\newblock Efficient multiagent policy optimization based on weighted estimators
  in stochastic cooperative environments.
\newblock \emph{J. Comput. Sci. Technol.}, 35\penalty0 (2):\penalty0 268--280,
  2020.

\end{thebibliography}
\bibliographystyle{icml2022}

\newpage
\appendix
\onecolumn
\newpage

\section{Environment Details}\label{env details}

\begin{figure}[h]
\centering
\subfloat[Predator-Prey]{
\includegraphics[width=0.32\linewidth]{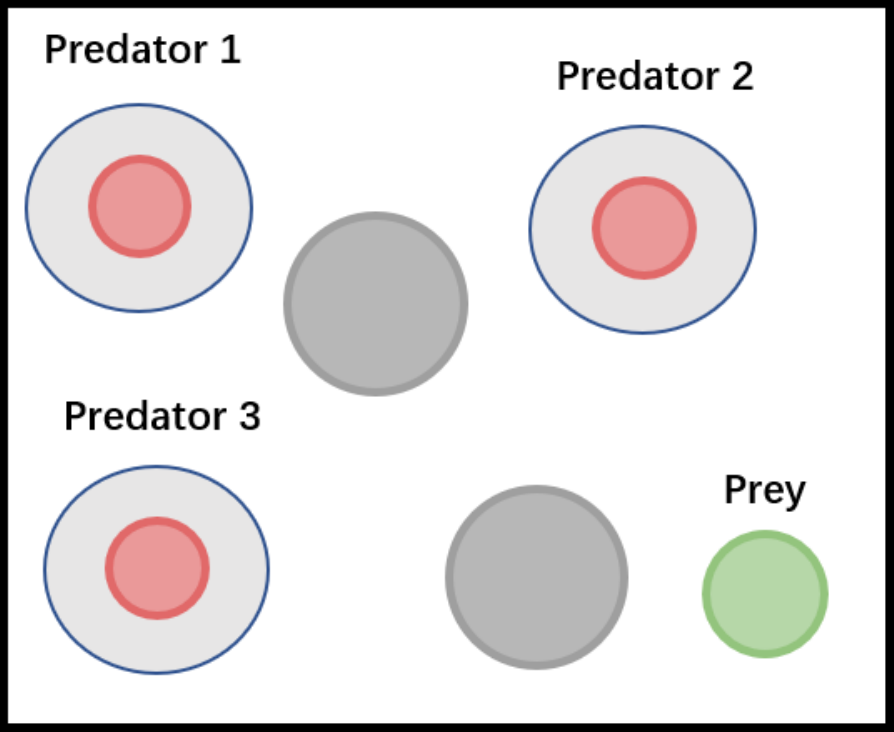}}
\subfloat[Cooperative Navigation]{
\includegraphics[width=0.32\linewidth]{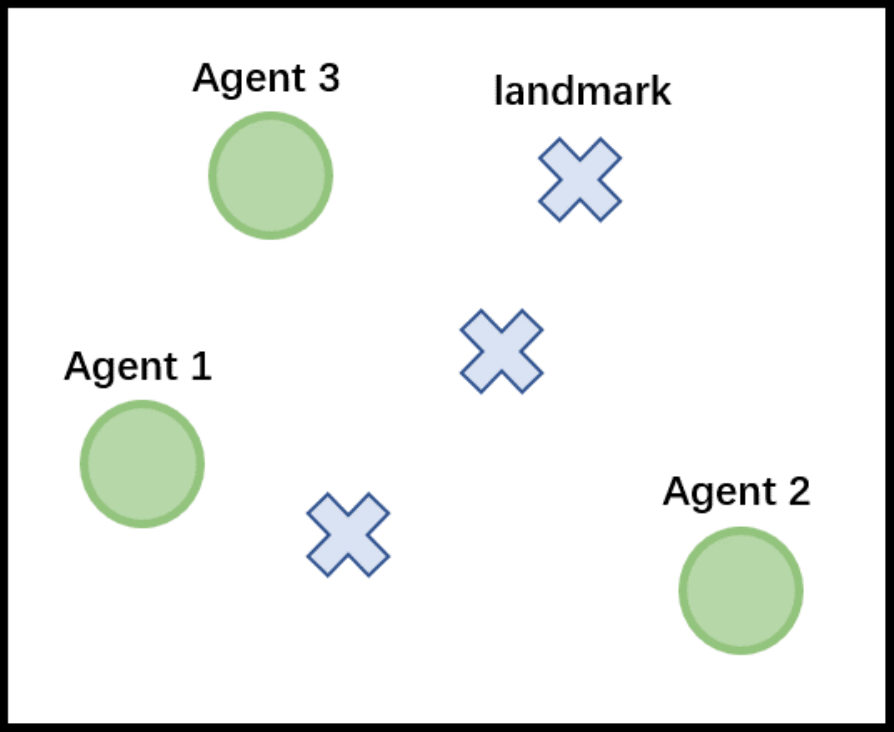}}
\subfloat[Wildlife Rescue]{
\includegraphics[width=0.32\linewidth]{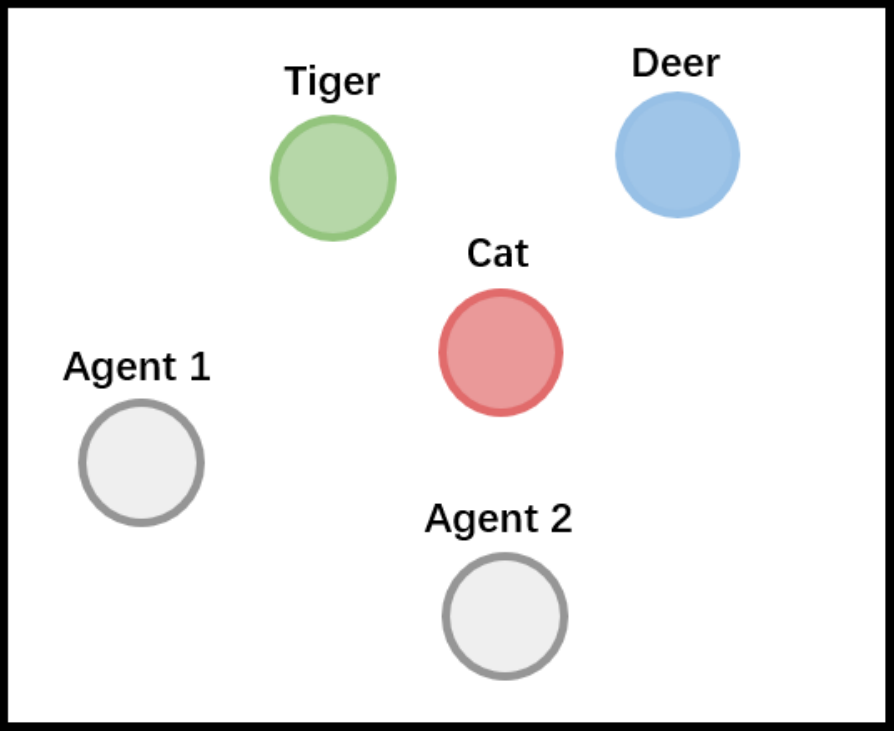}}
\caption{Multi-Agent Particle Environment.}
\label{MPE env}
\end{figure}


\textbf{Predator Prey}: N slower cooperating agents chase the faster adversary around a randomly generated environment with L large landmarks impeding the way. In our setting, agents control the predators to chase the prey, the policy of prey is fixed.
The agents are partially observable, the observation radius of each predator is 0.25. Only when the predator captures the prey, it can get the reward 10. we set N to 3, 6, 12, 24 separately. Each game has 25 steps.

\textbf{Cooperative Navigation}: Agents must cooperate through physical actions to reach a set of L landmarks. In our setting, agents receive a shared reward which is the sum of the minimum distance
of the landmarks from any agent, and the agents who collide with each other receive negative reward -1. Besides, all agents receive 1 if all landmarks are occupied. Each game has 25 steps.

\textbf{Wildlife Rescue (the motivating example)}: N agents must cooperate to rescue M wildlife with different risks and rewards. We provide a control time $T_c$. When an agent catches up with an animal, the agent can control $T_c$ seconds to wait for other agents to arrive. Each game has $T$ steps. In the motivating example and experiment, we leverage the same setting.
We set $T_c$ to 8, $T$ to 60. The specific reward at the end of each episode is set in Table ~\ref{wildlife reward}.

\begin{table}[!htp]
\caption{The reward matrix of the rescue agents at the end of each episode. Both agents receive the same reward.}
\centering
\begin{tabular}{|c|c|c|c|c|}
\hline
\diagbox{agent 1}{reward}{agent 2}&tiger&deer&cat&on the road\\ 
\hline
tiger&11&-30&0&-30\\
\hline
deer&-30&7&6&-30\\
\hline
cat&0&6&5&0\\
\hline
on the road&-30&-10&0&0\\
\hline
\end{tabular}
\label{wildlife reward}
\end{table}


\begin{wrapfigure}{r}{0.4\textwidth} 
\centering
\includegraphics[width=0.85\linewidth]{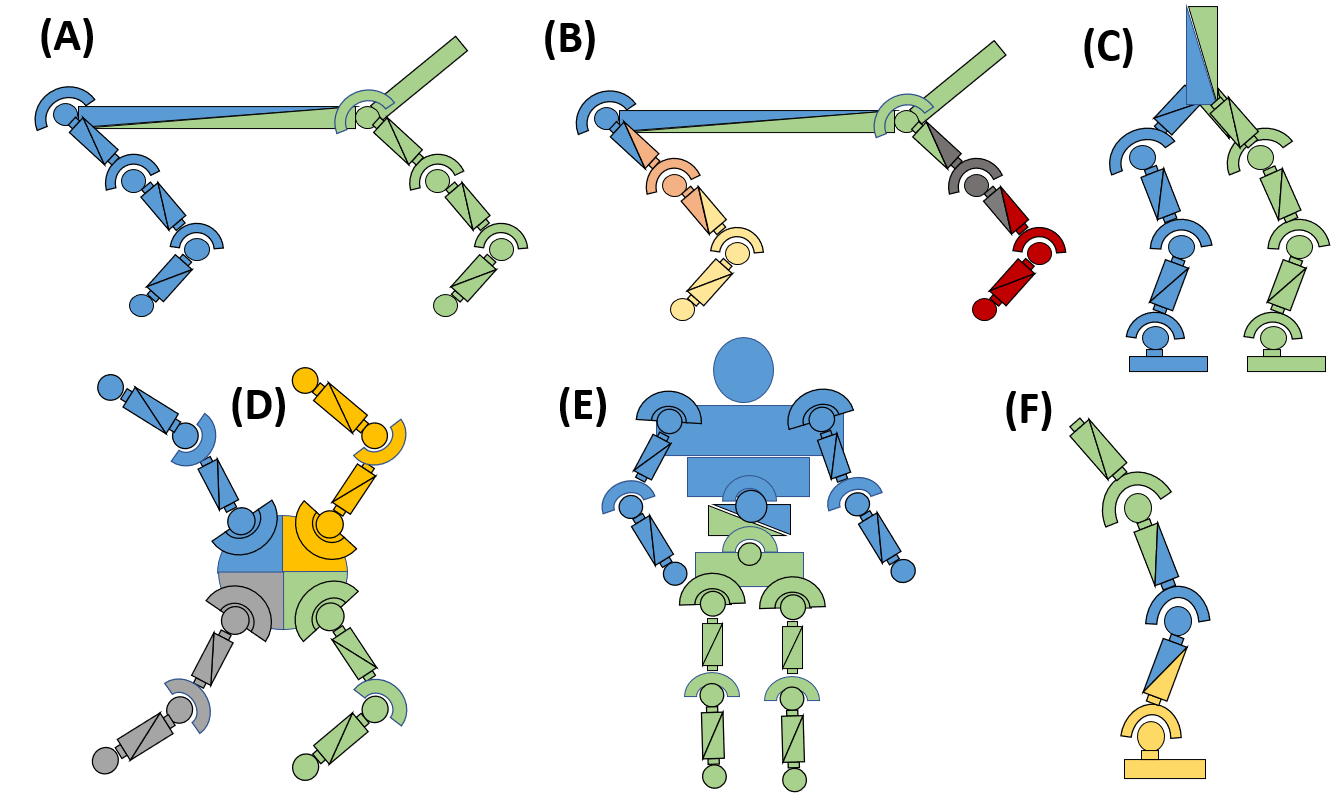}
\caption{The structure of Multi-Agent MuJoCo.}
\label{mujoco}
\end{wrapfigure}

Multi-Agent MuJoCo~\citep{de2020deep} is a range of challenging continuous multi-agent control tasks of Multi-Agent MuJoCo benchmark suite. Multi-Agent MuJoCo is designed for decentralized cooperative continuous multi-agent robotic control. The structure is shown in Figure~\ref{mujoco}.
For evaluation, we use the reward setting of the original paper, but we set $k$ to zero for all environments. $k$ controls how much information each agent can observe from its adjacent agents. When $k$ is zero, it means each agent can only observe information about its own joints, which is the hardest setting for Multi-Agent MuJoCo. 

For SMAC, we use the latest version 2.4.10, and default settings for all maps.

\section{Process of training MINE and CLUB}\label{MINE CLUB}

In this section, we detail the process of training MINE and CLUB.

First, we introduce the process of training MINE.
We train MINE based on the positive buffer of \dnb.
Figure~\ref{321} illustrates the detailed process of estimating the lower bound. We design the network $T$ with parameters $\omega_1$ as a state encoder $E_{\omega_s}$ and an action encoder $E_{\omega_u}$. MINE takes a batch of $(s_t, u_t) \sim \mathbb{P}_{\mathcal{S} \mathcal{U}}$ and $(s_t, u_k) \sim \mathbb{P}_{\mathcal{S}}\otimes\mathbb{P}_{\mathcal{U}}$ as inputs and encodes the data as vectors $E_{\omega_s}(s_t)$, $E_{\omega_u}(u_t)$, $E_{\omega_u}(u_k)$ with the same dimension. Then we take the inner product of $E_{\omega_s}(s_t)$ and $E_{\omega_u}(u_t)$, $E_{\omega_s}(s_t)$ and $E_{\omega_u}(u_k)$ to get $T_{\omega_1}(s_t, u_t)$ and $T_{\omega_1}(s_t, u_k)$ (e.q., scores) separately. Finally we do the subtraction operation on the scores to get the estimate of $I_{\text{MINE}}(s,u)$ according to Equation~(\ref{e7}). In summary, MINE is trained by sampling data from the positive buffer of \dnb to minimize $\mathcal{L}(\omega_1)$ in Equation~(\ref{e7}). 

Then we details the process of training CLUB.
We train CLUB based on the negative buffer of \dnb.
Figure~\ref{321} illustrates the detailed process of estimating the upper bound. We design the network $T$ with parameters $\omega_2$ as action prediction network.
The train process of CLUB is different from MINE,
we only need to sample data from the joint distribution $\mathbb{P}_{\mathcal{S} \mathcal{U}}$. Then we directly minimize the loss $\mathcal{L}(\omega_2)$ based on the samples to optimize CLUB.

\section{Integrate \alg with RODE }\label{inter RODE}
RODE is a role-based algorithm which decomposes the joint action space into a restricted role action space to reduce the primitive action-observation spaces. 
In RODE, each task can be decomposed into a sub-task which has a smaller action-observation space, and each sub-task is associated with a role $\rho$.
To demonstrate the generalization and effectiveness of \alg, we design \alg-RODE (i.e., integrate \alg with RODE). 
To integrate with \alg, we take the joint role $\rho$ into consideration. Thus we measure the MI between the joint actions and global states with the joint role $\rho$ (e.q., $I(u;\rho,s)$). Since the selection of actions for the agents needs to consider the selected role and the observations.
$I(u;\rho,s)$ has a similar effort with $I(u;s)$ in \alg-MADDPG where the decisions are only based on the observations. 
Different from \alg-MADDPG, we set $ \ r_{t}^{\text{\alg}} = \alpha I_{\text{MINE}}(s_{t};\rho_t,u_{t}) -\beta I_{\text{CLUB}}(s_t;\rho_t,u_t))$. $ \ r_{t}^{\text{\alg}}$ is used in the same way as in \alg-MADDPG.
To achieve the goal in Equation~\ref{etotal}, we only need to modify the $Q_{tot}$ as follows:
\begin{equation}
\mathcal{L}=\mathbb{E}_{s_t,u_t,r_t,s_{t+1} \sim \mathcal{D}}\left[\left(y_{rode}-Q_{t o t}(s_t, u_t)\right)^{2}\right],
\label{rode q}
\end{equation}
where $y_{rode} = r_t +  \ r_t^{\text{\alg}} + \gamma \max _{a_{t+1}} \bar{Q}_{t o t}\left(s_{t+1}, \pi_{\theta^{\prime}}(s_{t+1})\right)$ and $Q_{t o t}$ is a QMIX-style~\citep{rashid2018qmix} mixing network to estimate the total value of the state and actions and $\bar{Q}_{t o t}$ is the target network of $Q_{tot}$. \mie is updated according to Equation~\ref{e11}. Since RODE is QMIX-based method, credit assignment is also 
applied to the MI signals. The MI signals over superior trajectories and inferior trajectories can reward and punish the agents based on their contribution, which can help agents to find the optimal joint behaviors.

\alg can be easily combined with existing MARL algorithms. 
For example, to integrate \alg with RODE, we only need to initialize the networks of RODE in line 2 of Algorithm~\ref{algo2} and retain other components. All other processes remain the same except changing $I(u;s)$ to $I(u;\rho,s)$. Finally, we update the value network according to Equation~\ref{rode q} and others are the same with RODE to replace the line 15 in Algorithm~\ref{algo2}. The same applies in combination with other algorithms.



\section{Experimental Settings}
\label{Experiment setting}
For experiments on MPE and SMAC, we use episodic rewards as the criterion for adding data to \dnb. 

On Multi-Agent MuJoCo, we can more accurately collect superior trajectories. Since the episode reward of one trajectory is high, there is no guarantee that there are no poor sub-trajectories that affect \dnb. Thus we leverage a more fine-grained filtering method to filter the trajectories on Multi-Agent MuJoCo. We leverage sub-trajectories to replace the complete trajectory to accurately filter the superior and inferior trajectories. In particular, we set the length of sub-trajectories to 50 steps and use the total reward of each sub-trajectory as the criterion for adding data to \dnb.
The update frequency of MINE and CLUB is 1 on all tasks.

The experiments on MPE are carried out on Intel(R) Xeon(R) Gold 5117 CPU @ 2.00GHz.
The experiments on Multi-Agent MuJoCo are carried out on NVIDIA GTX 2080 Ti GPU with Intel(R) Xeon(R) CPU E5-2680 v4 @ 2.40GHz. 
The experiments on SMAC are carried out on Intel(R) Xeon(R) CPU E5-2680 v4 @ 2.40GHz and Intel(R) Xeon(R) CPU E5-2679 v4 @ 2.50GHz.

\section{Additional Experiments}
\label{more exp}
In this section, we present more experiments to help better understand our framework.
Six questions are raised and more experiments about \alg-MASAC and \alg-QMIX are provided:

\textbf{RQ1}: Is \dnb buffer size sensitive?


\textbf{RQ2}: Is only maximizing $I(s;\pi(\cdot|s))$ better than other MI forms?

\textbf{RQ3}: What is the time consumption of \alg-MADDPG?

\textbf{RQ4}: Can MINE be replaced by normal MI estimator?

\textbf{RQ5}: Can MINE trained with the positive buffer of \dnb provide a good guide?

\textbf{RQ6}: How to choose the $k$ (Update frequency of MI
estimators)?

\begin{figure}[htb]
\centering
\includegraphics[width=0.49\linewidth]{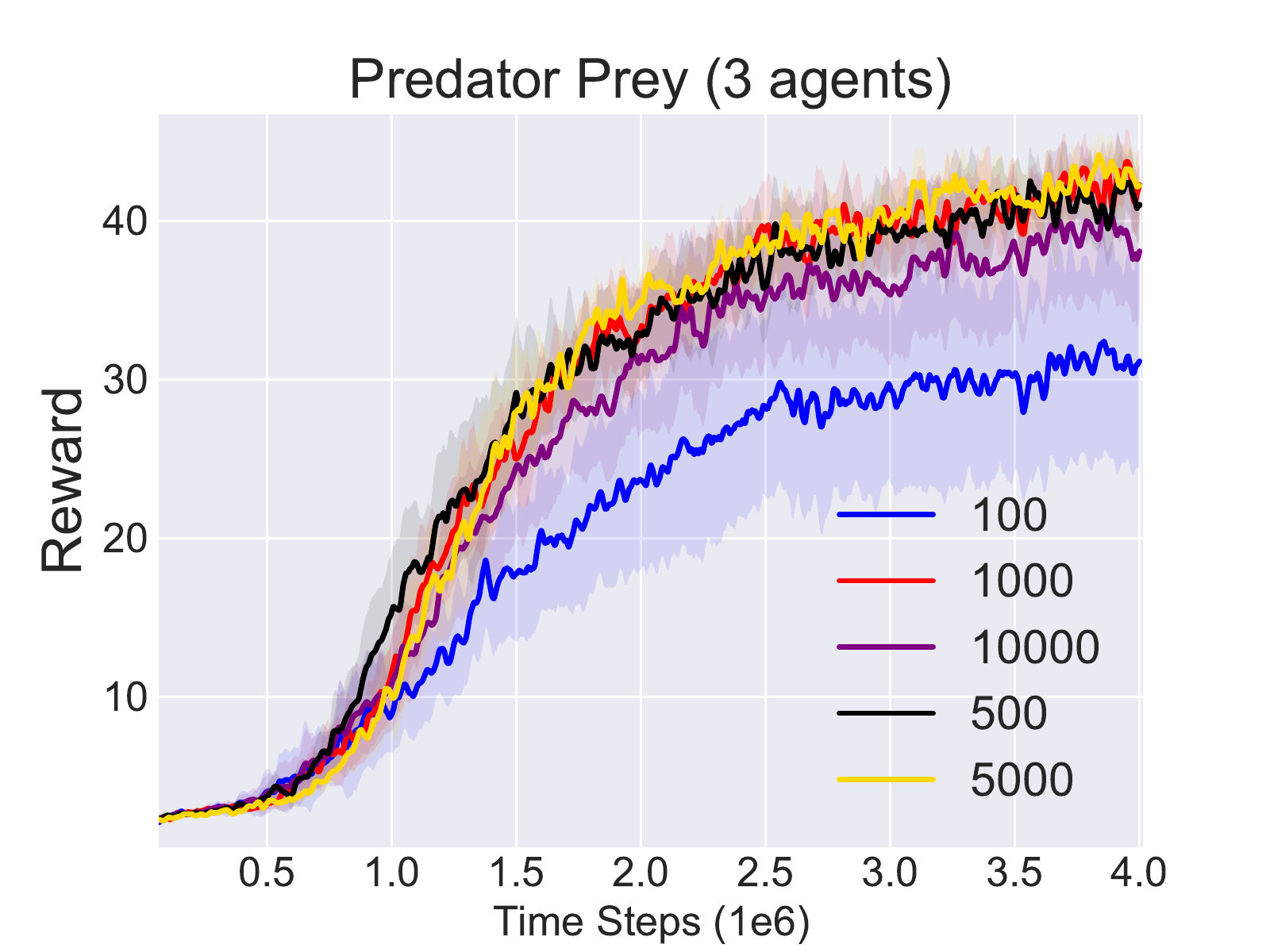}
\includegraphics[width=0.49\linewidth]{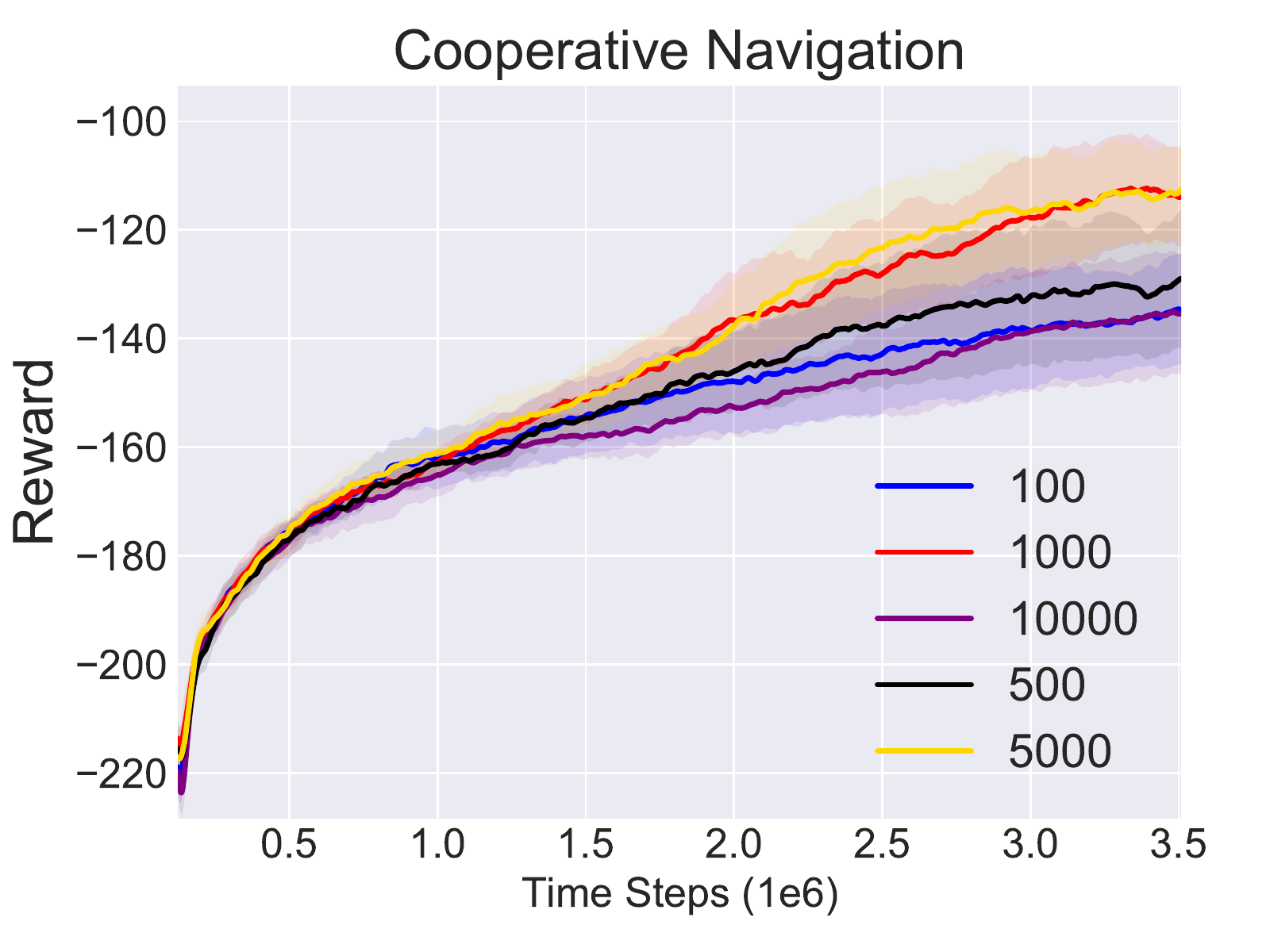}
\caption{Inﬂuence of \dnb size.}
\label{buffer size}
\end{figure}

\textbf{To answer RQ1}, we design experiments on MPE and adjust the size of \dnb. Specifically, we test 5 different buffer sizes (100, 500, 1000, 5000, 10000) for positive buffer and negative buffer of \dnb.
The results shown in Figure~\ref{buffer size} indicate that both too large size and too small size are harmful to the performance. 
A buffer size that is too small prevents MINE and CLUB from characterizing trajectories well, while too large size hinders the optimality of \dnb which makes it impossible for MINE and CLUB to adapt quickly to the new joint behavior. 
Therefore, an appropriate size is important.



\begin{figure}[htb]
\centering
\includegraphics[width=0.49\linewidth]{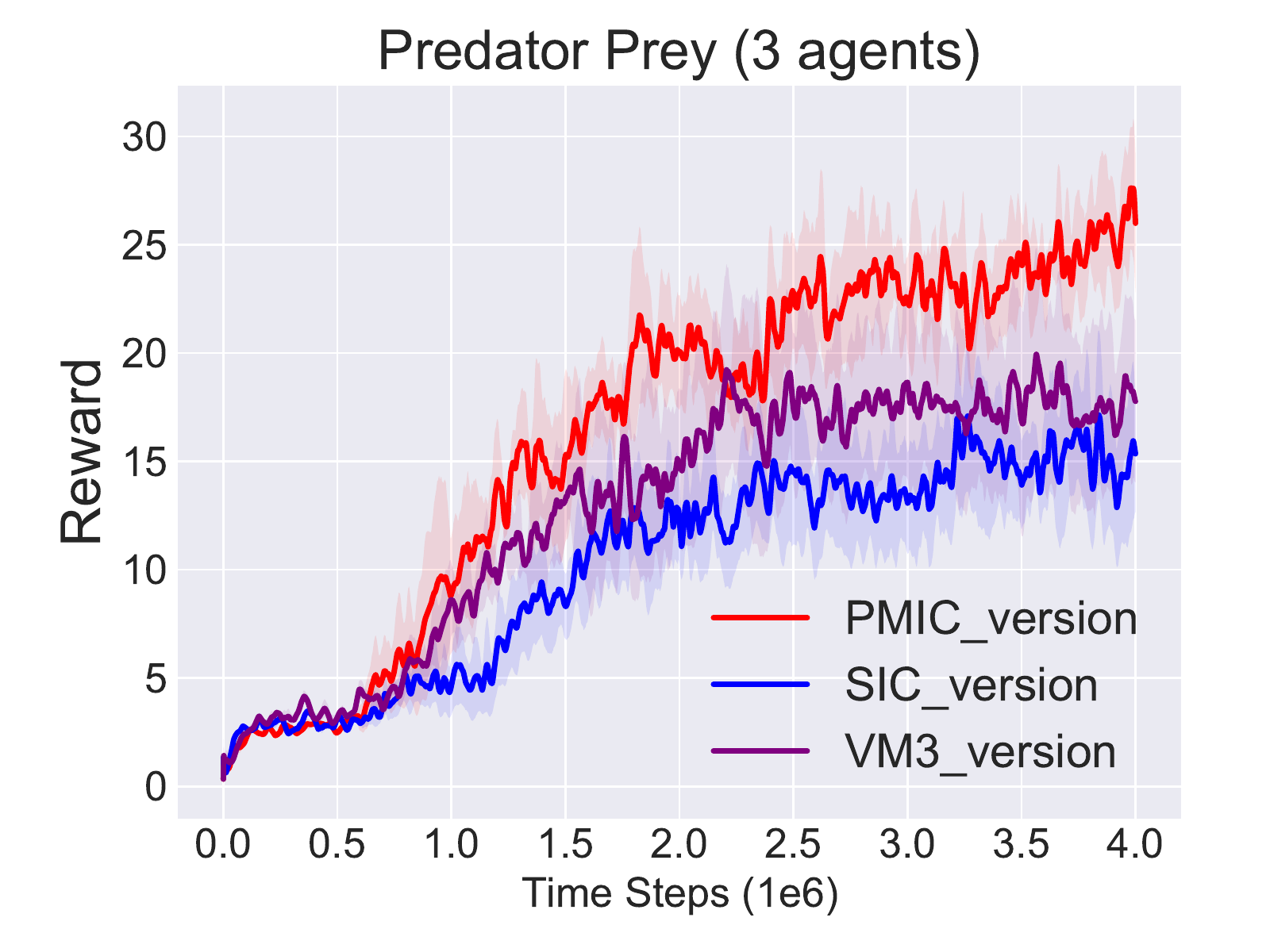}
\includegraphics[width=0.49\linewidth]{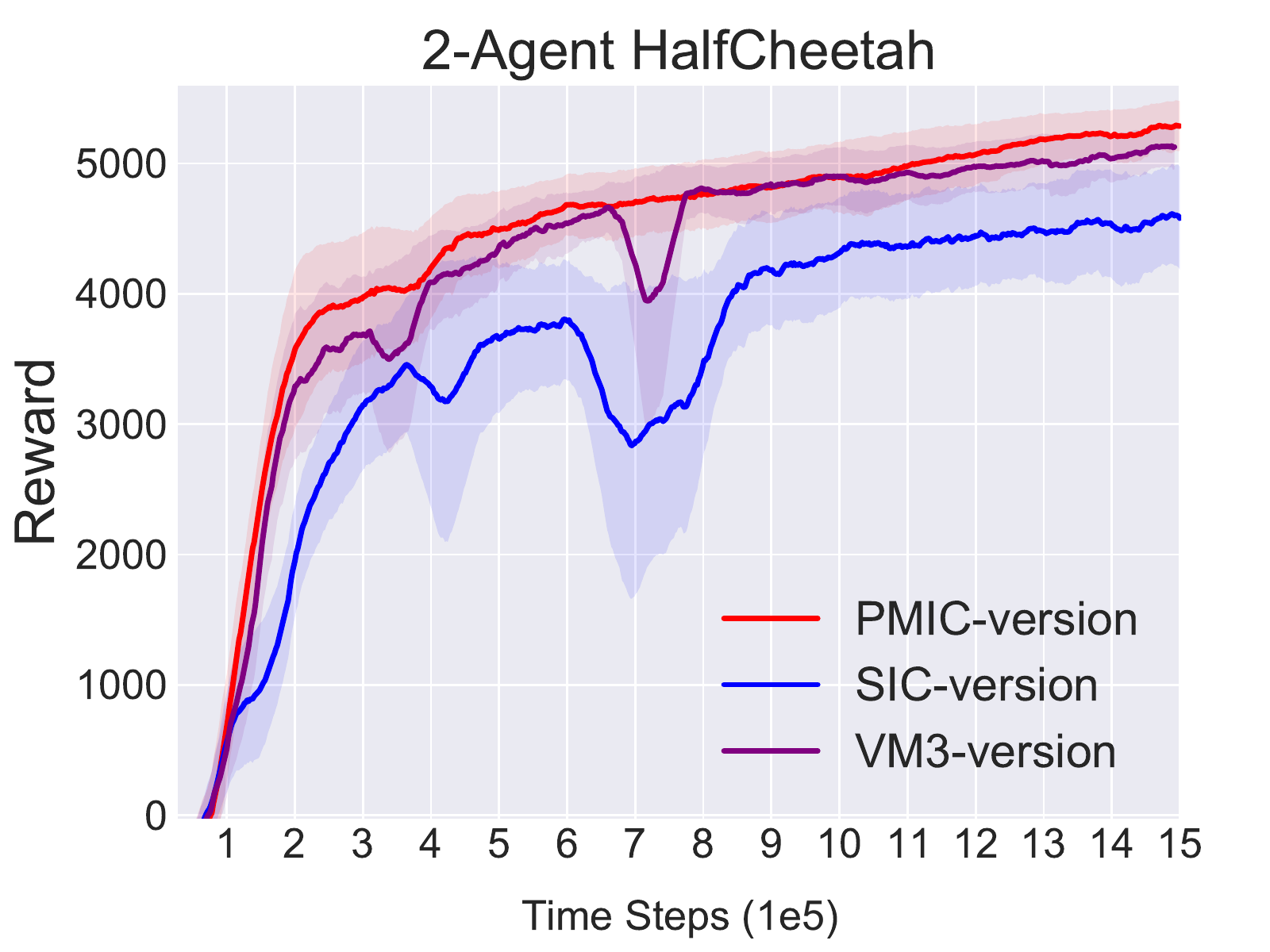}
\caption{Comparison of only maximizing mutual information with different forms. \alg-version leverages MI of global states and joint actions. VM3-version leverages the MI of any two agents' actions among agents. SIC-version leverages the MI of $z$ and the joint policy.}
\label{MI compare}
\end{figure}

\textbf{To answer RQ2}, we only maximize MI with different forms to give a comparison. 
To avoid differences caused by some mechanisms (e.g., double Q) or structures (e.g., MINE), we apply different forms of MI to MADDPG and leverage MINE as the MI estimator and other settings remain the same. The hyperparameters have been fine-tuned. The results are shown in Figure~\ref{MI compare}. Maximize MI of the global states and the joint actions is better than other MI forms in terms of the final performance.
This demonstrates that our proposed method of maximizing the MI of global states and joint actions is more effective than other MI forms. Besides, we also find that maximizing MI in environments with sparse reward brings performance degradation to the original algorithm. The reason is:
In environments with sparse reward, the MI signals play a greater role in guiding than reward signals. Beside, maximizing MI has the problem of easily making agents fall into sub-optimal collaborations, thus agents quickly fall into sub-optimal collaborations.

\textbf{To answer RQ3}, we evaluate the time consumption of different algorithms on 6-Agent HalfCheetah. The experiment is carried out on NVIDIA GTX 2080 Ti GPU with Intel(R) Xeon(R) CPU E5-2680 v4 @ 2.40GHz. We evaluate the time consumption of each algorithm individually with no additional programs running on the device. Each result is the average of 10 time-consuming calculations. The time consumption includes the time consumption of the execution phase and the time consumption of the centralized training phase.
The results are shown in Table~\ref{tab:time}. \alg-MADDPG brings less time consuming than other methods.

\begin{table*}[h]
		\centering
		\caption{Time consumption of different algorithms on 6-Agent HalfCheetah every 1000 time steps.}
		
		\begin{tabular}{|l|c|c|c|c|}\hline
			algorithm&\alg-MADDPG & MADDPG & SIC-MADDPG & VM3-AC \\
			\hline
			seconds & 30.96 & 24.43 & 31.01 & 182.58\\ 
			\hline
			algorithm& COMIX & Fac-MADDPG & MASAC & \\
			\hline
			seconds & 34.43 & 87.13 & 41.68 & \\
			\hline
		\end{tabular}
		\label{tab:time}
	\end{table*}

\begin{figure}[htb]
\centering
\includegraphics[width=0.49\linewidth]{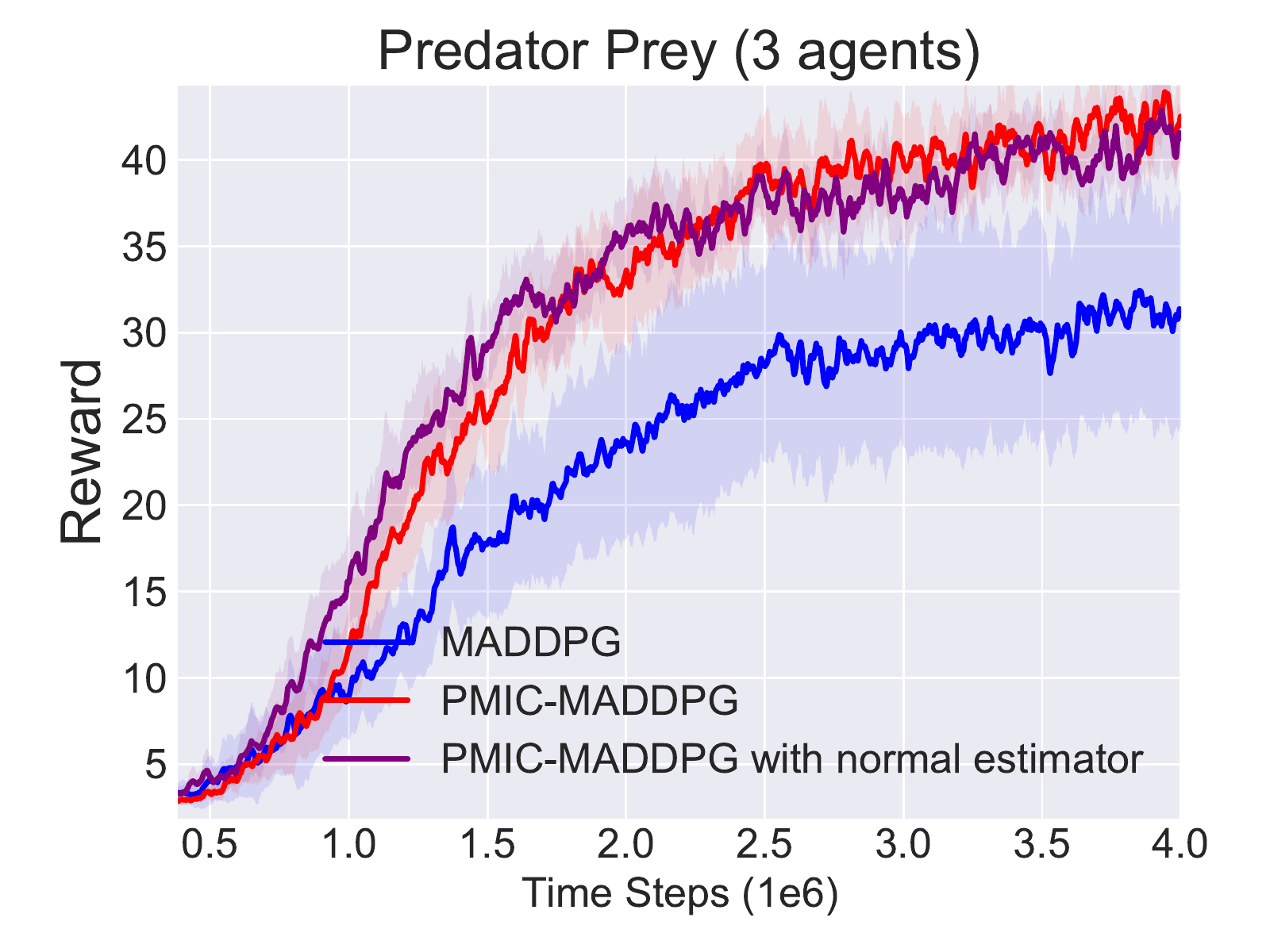}
\includegraphics[width=0.49\linewidth]{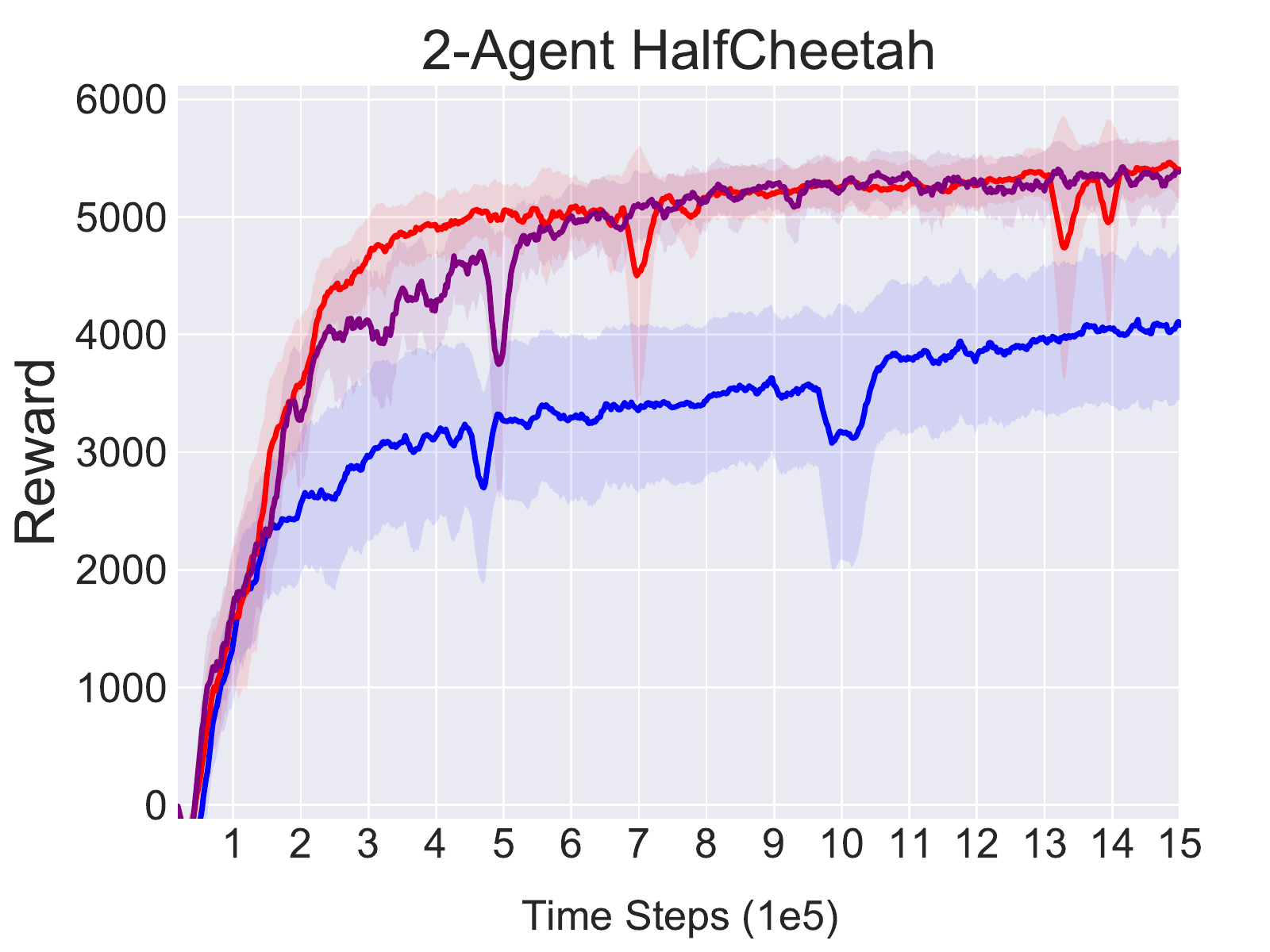}
\caption{Comparison of PMIC with MINE and PMIC with normal estimator.}
\label{MINE and normal }
\end{figure}

\textbf{To answer RQ4}, we evaluate the impact of MINE on the performance of the algorithm. Because we use MINE for the first time in the field of multi-agent reinforcement learning, thus we need to provide a comparison of MINE and the normal mutual information estimation method. The results are shown in Figure~\ref{MINE and normal }. We can see that the performance of both methods is similar, which demonstrates that the effectiveness of PMIC is introduced by the mechanism of the maximization-minimization MI and new collaboration criterion, rather than MINE.

\begin{figure}[t]
\centering
\includegraphics[width=0.45\linewidth]{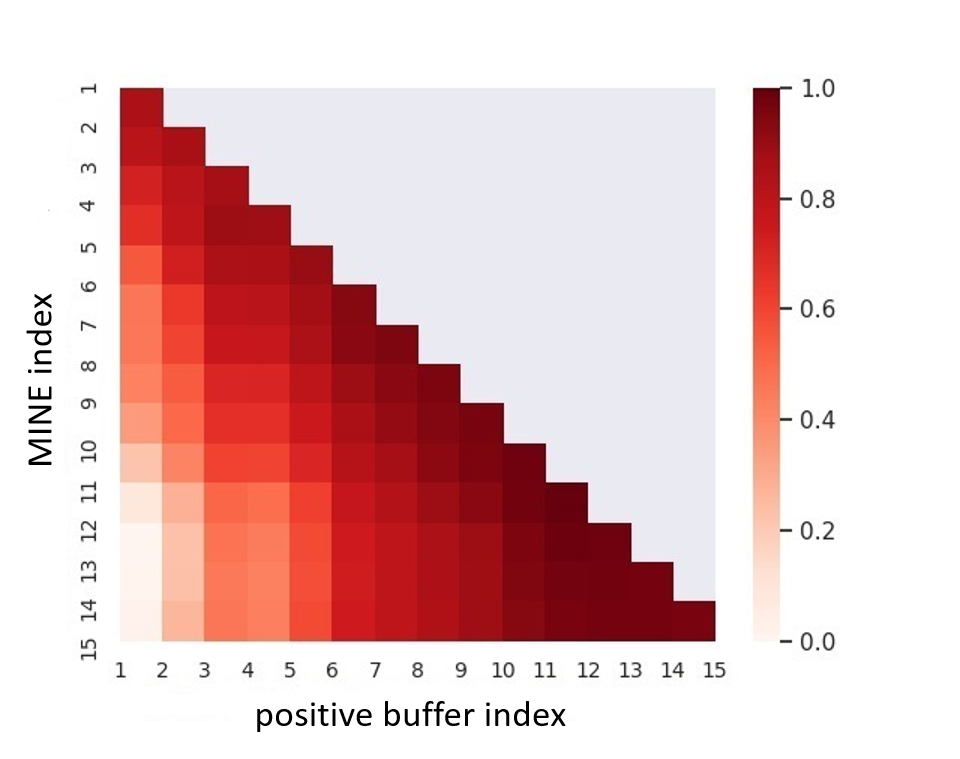}
\includegraphics[width=0.45\linewidth]{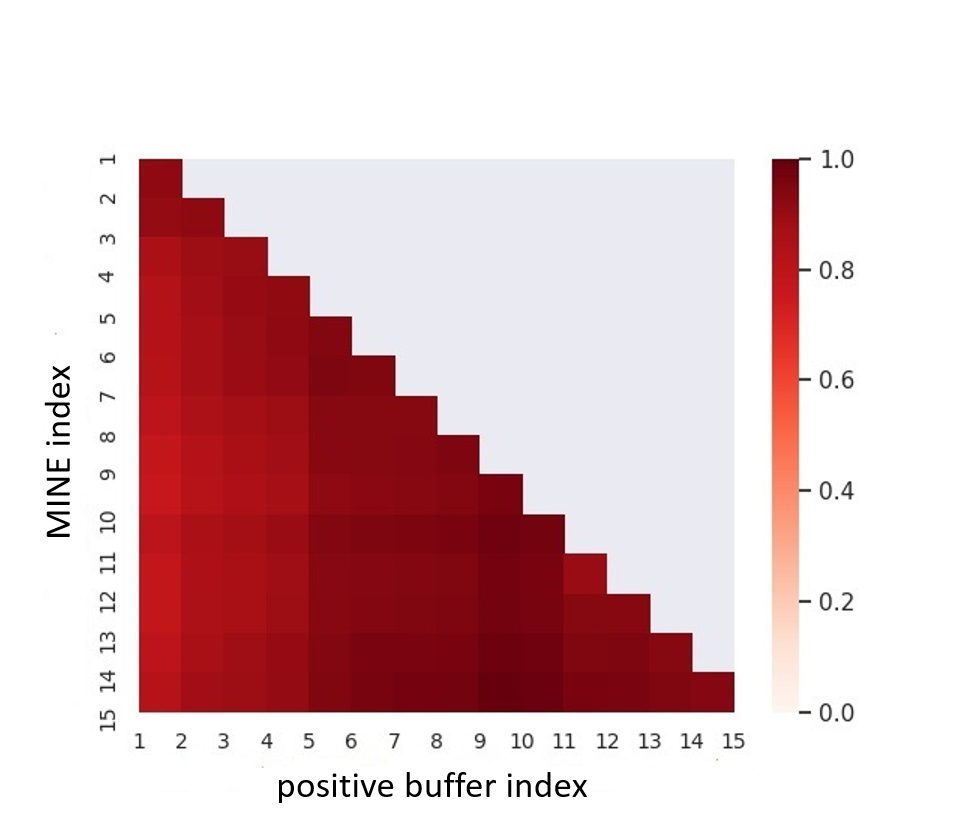}
\caption{Visualization results of the MI estimated by different MINE for different positive buffer (The left uses our method, the right uses the regular method). X-axis denotes the index of positive buffer. Y-axis denotes the index of MINE. The value in coordinates $(x,y)$ represents the MI estimated by the MINE with index y for the buffer with index x. }
\label{analyze pcb}
\end{figure}

\textbf{To answer RQ5}, we further analyze whether MINE trained with the positive buffer in \dnb could act as a good guide, which requires the MINE to estimate a large value over superior trajectories and a small value over inferior trajectories. We train MADDPG in 2-Agent Walker. During training, we save the positive buffer every 100000 steps, which ensures that the trajectories saved each time are better than the previous time. Then we save the MINE every 100000 steps and use it to estimate MI of trajectories saved in the positive buffer. The estimated MI of different positive buffer by different MINE are plotted on Figure~\ref{analyze pcb}. The larger the index, the newer the MINE and buffer. When we use \dnb to train MINE, the color gradually darkens from left to right in each line, indicating that MINE has a higher estimation value for the better trajectories. However, when training MINE in the same way with VM3-AC \citep{kim2020maximum}, MINE estimates the same value of MI for both inferior and superior trajectories. From the perspective of visualization, the MI of superior coordination estimated by MINE with \dnb is large so that such coordination can be more encouraged, which can give accurate guidance to superior joint behavior.

\begin{figure}[htb]
\centering
\includegraphics[width=0.4\linewidth]{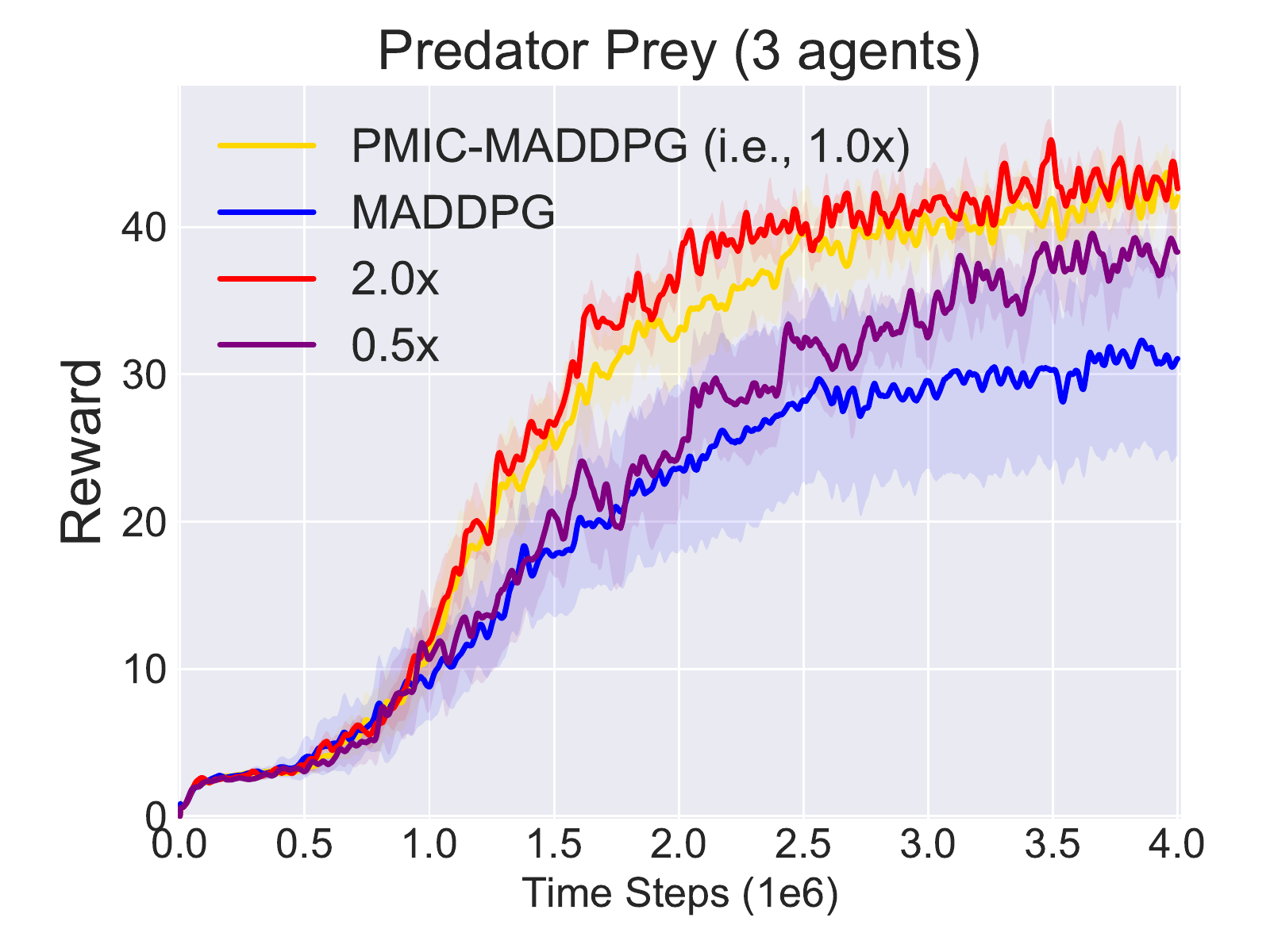}
\caption{Hyperparameter analysis of $k$ on \textit{Predator Prey (3 agents)}.}
\label{abd:k}
\end{figure}
We train Du-MIE at the same frequency as the critic in all our experiments and it works well. \textbf{To answer RQ6}, we complete the hyperparameter analysis on $k$ by changing this frequency to 0.5x and 2x of the critic (with other settings unchanged) and evaluating in \textit{Predator Prey (3 agents)}. As shown in Fig.\ref{abd:k},
PMIC learns faster and achieves slightly better performance when increasing the update frequency of Du-MIE to 2x (red). Increasing the frequency of training Du-MIE provides more accurate estimates, but it also brings additional time consumption, which needs to be traded off.



\begin{figure}[htb]
\centering
\includegraphics[width=0.325\linewidth]{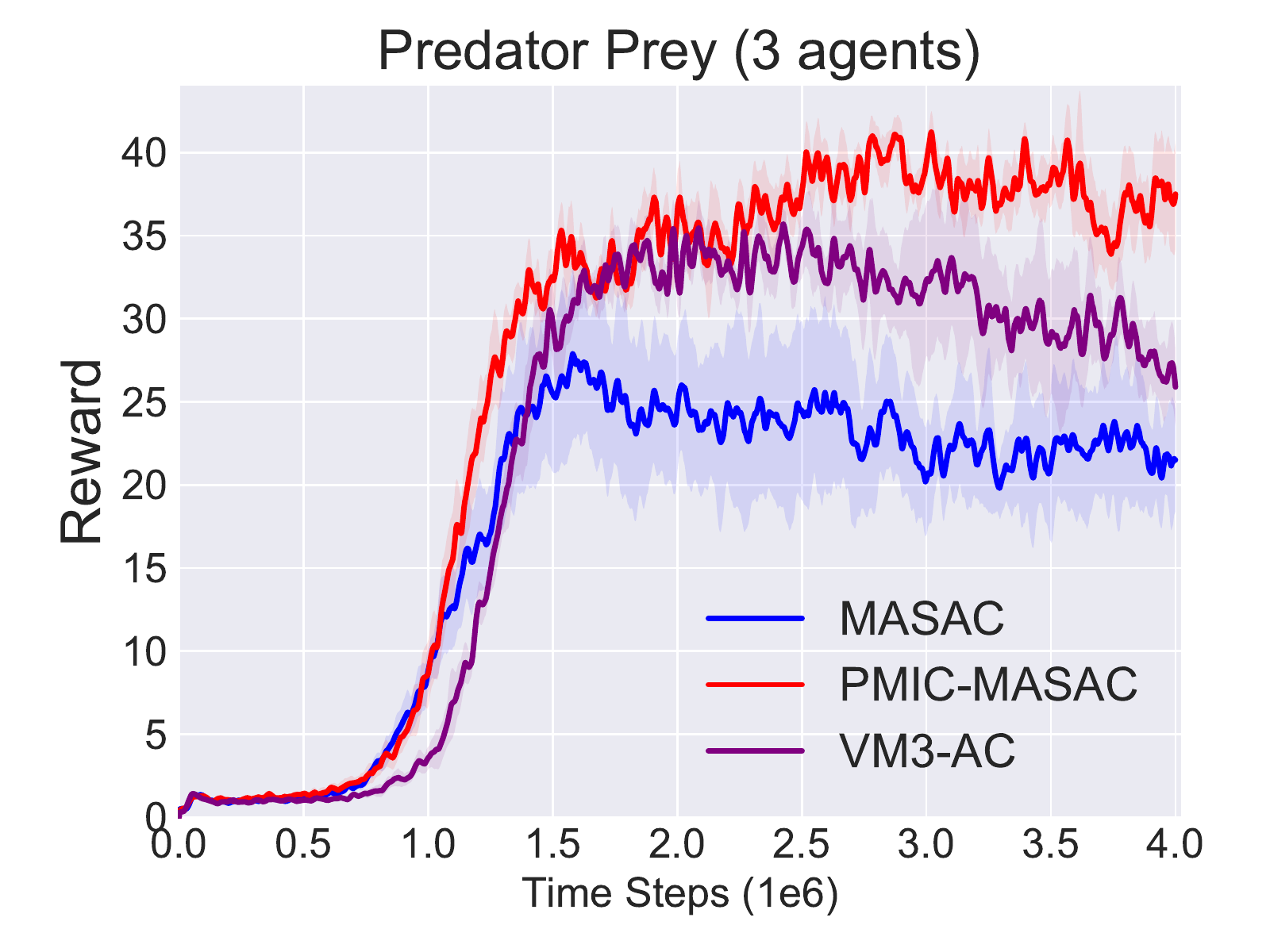}
\includegraphics[width=0.325\linewidth]{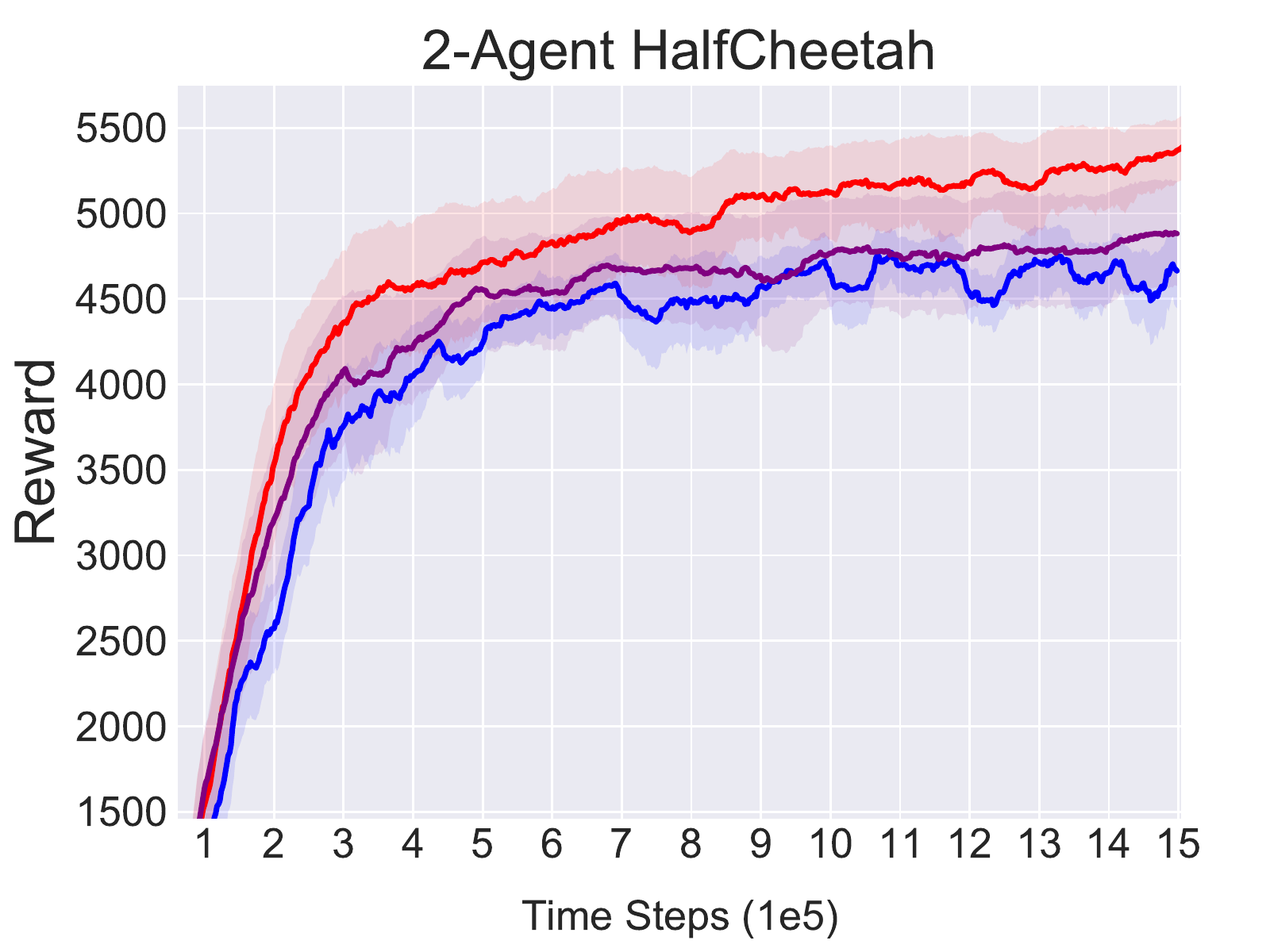}
\includegraphics[width=0.325\linewidth]{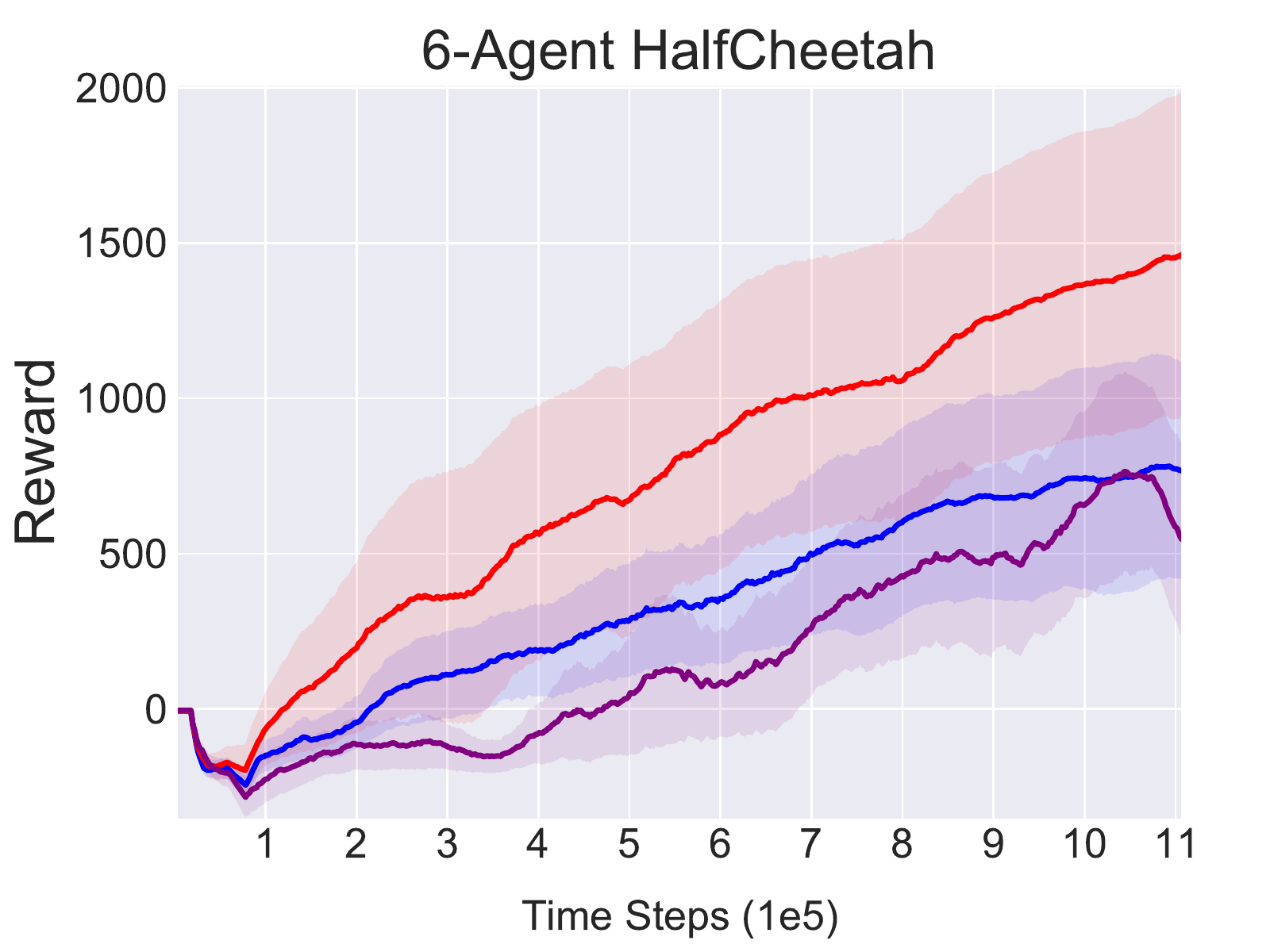}
\caption{Comparisons of MASAC-based methods on MPE and Multi-Agent MuJoCo.}
\label{PMIC-MASAC methods}
\end{figure}

\textbf{Integrate \alg with MASAC:}
we integrate PMIC with MASAC to further verify the generalisation and effectiveness of PMIC. The results are shown in Figure~\ref{PMIC-MASAC methods}. \alg also has significant improvement on MASAC. Beside, \alg-MASAC is better than the other MASAC-based method, VM3-AC. The effectiveness and generalisation of PMIC in continuous action space is more convincing by the above experiments.

\begin{figure}[htb]
\centering
\includegraphics[width=0.325\linewidth]{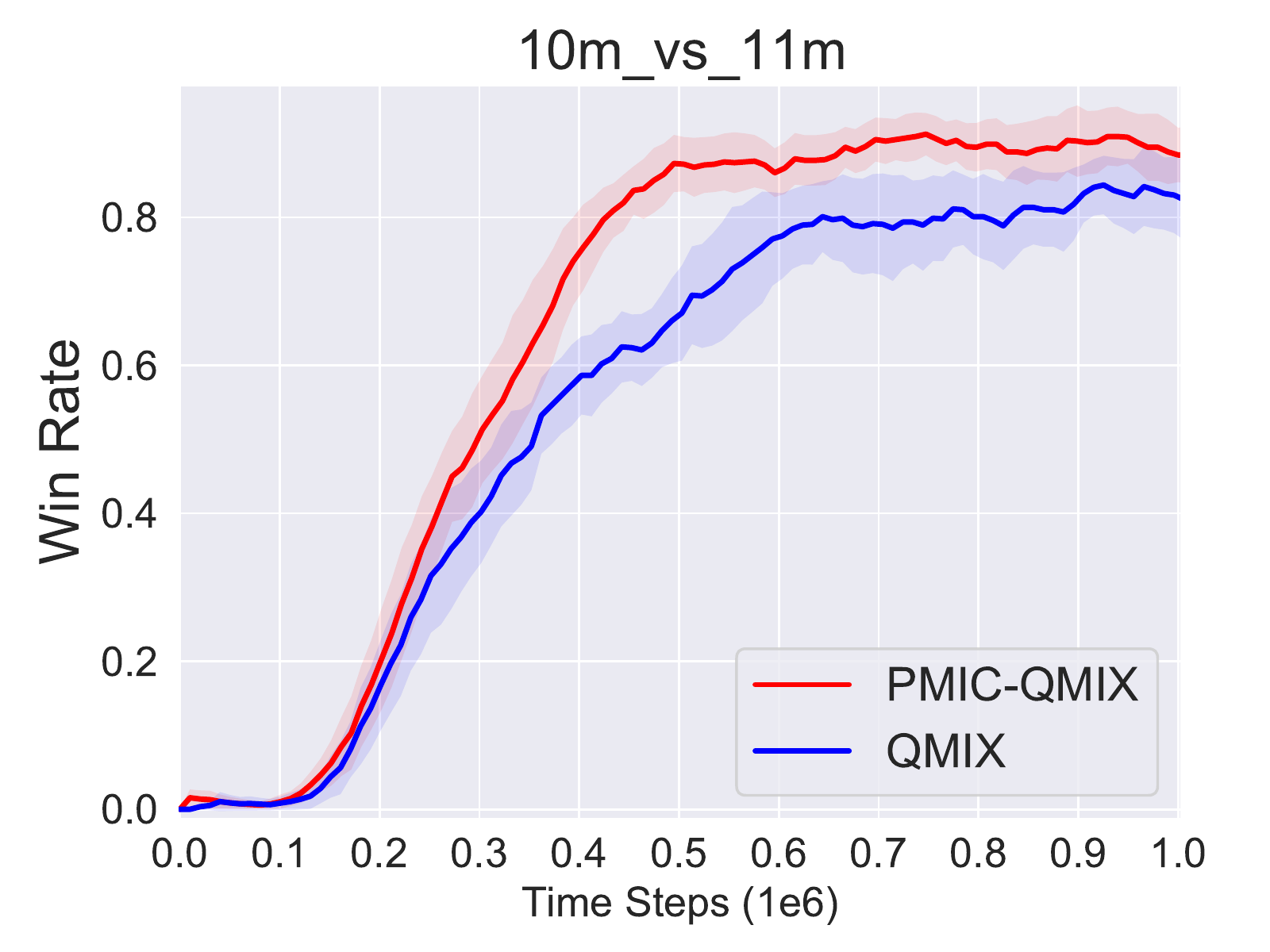}
\includegraphics[width=0.325\linewidth]{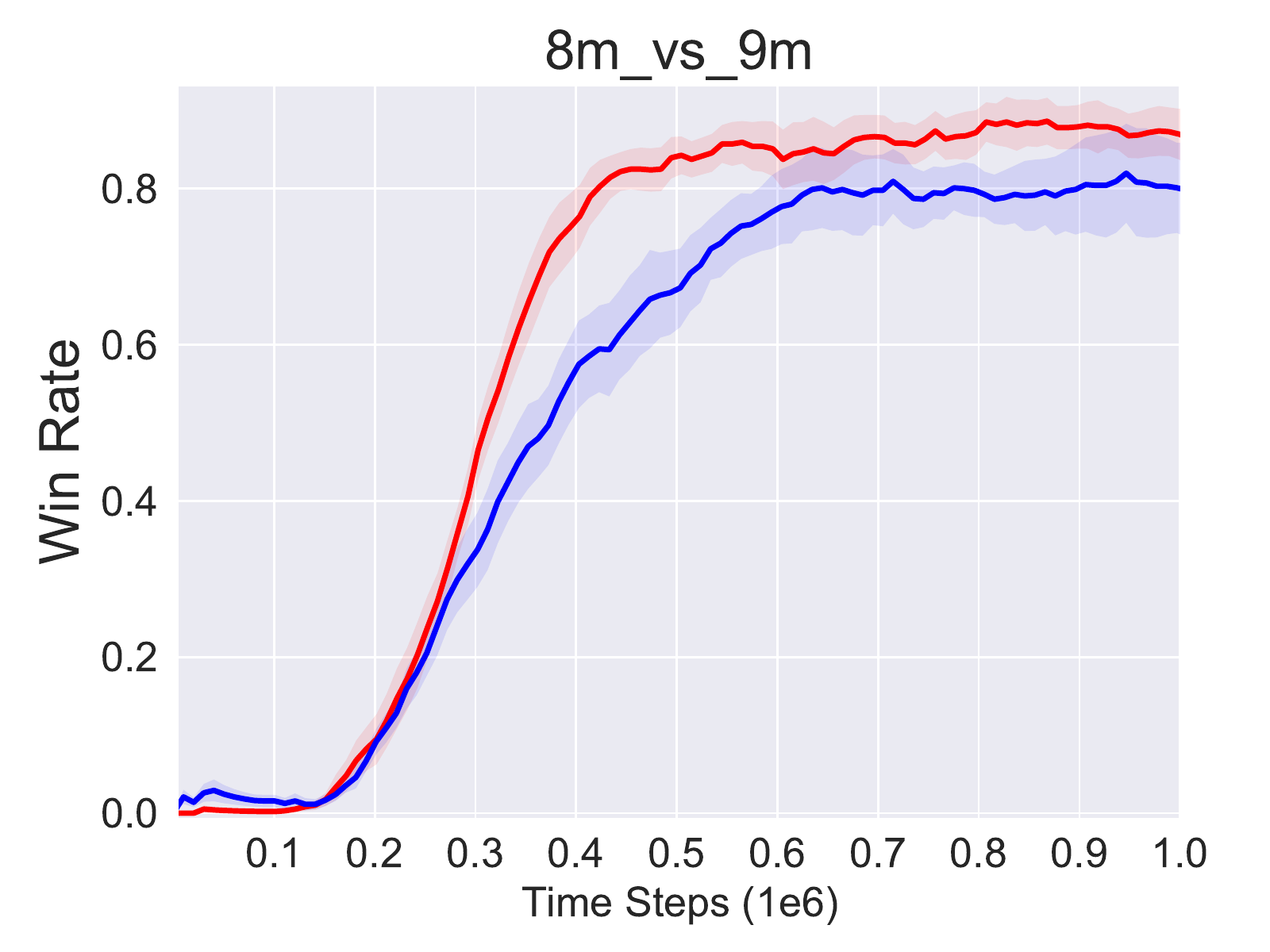}
\includegraphics[width=0.325\linewidth]{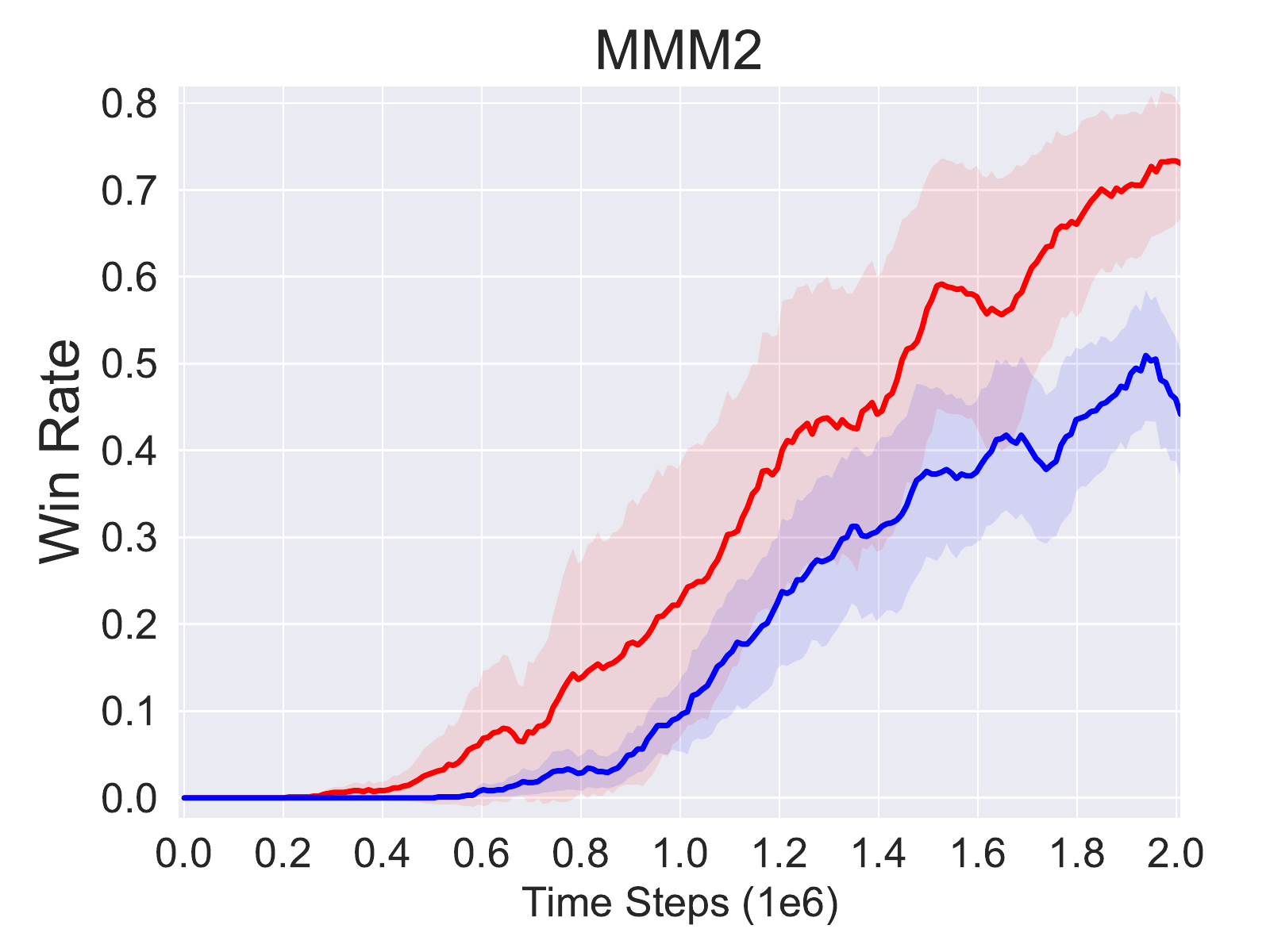}
\caption{Comparisons of \alg-QMIX and QMIX on SMAC}
\label{PMIC-QMIX methods}
\end{figure}

\textbf{Integrate \alg with QMIX:}
To better evaluate the generalization ability, we integrate PMIC with QMIX based on PyMARL2~\citep{hu2021rethinking} and evaluate PMIC-QMIX and QMIX on 3 maps. We run QMIX and \alg-QMIX both in serial mode, and other settings remain the same as in the original paper.
Fig.\ref{PMIC-QMIX methods} shows that \alg can also improve QMIX. This further verifies the generalization and effectiveness of \alg on the discrete action space.

\section{Comparison with Other Related Algorithms}
\label{Comp with Dis}

\begin{figure}[htb]
\centering
\includegraphics[width=0.49\linewidth]{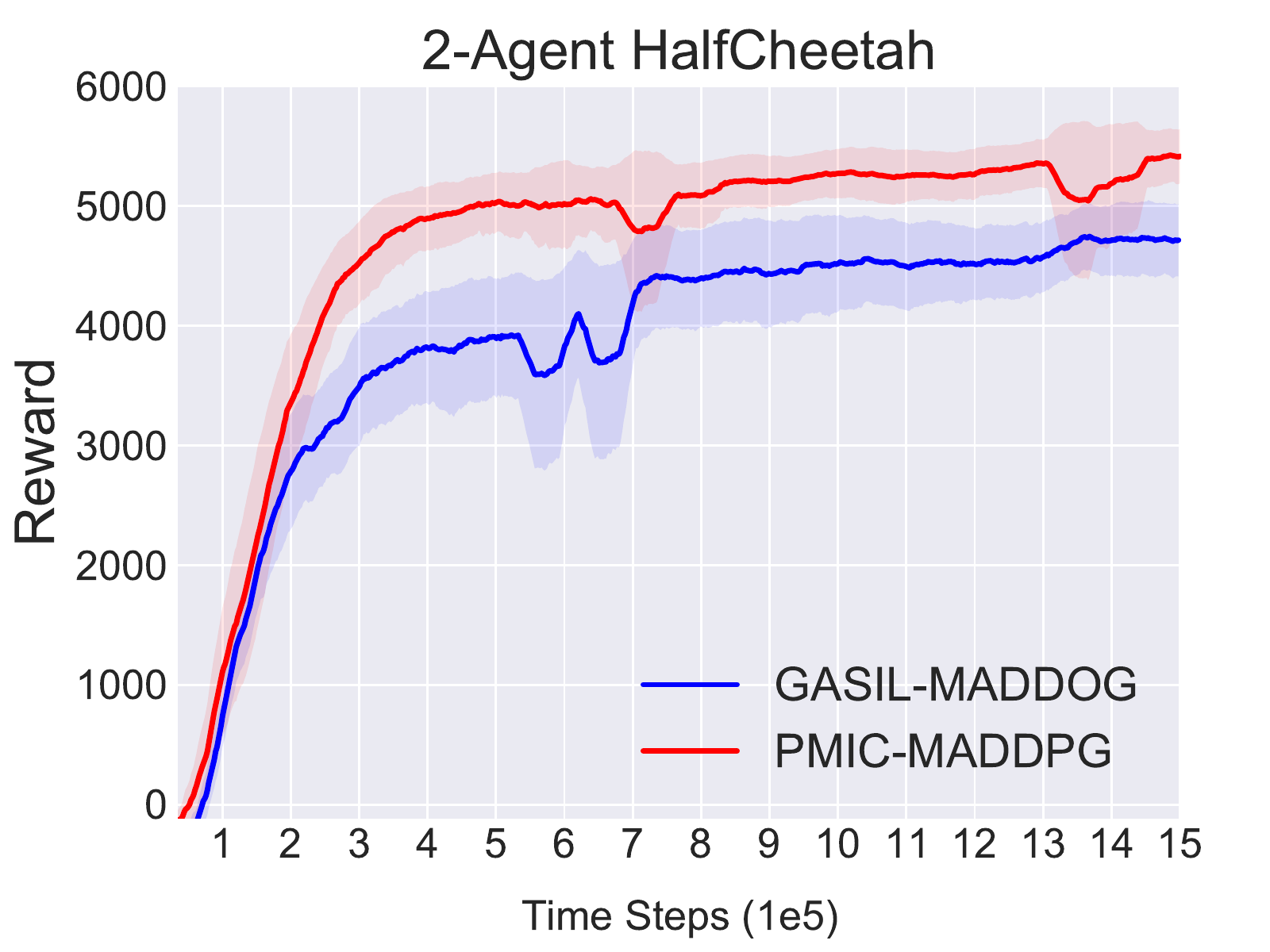}
\includegraphics[width=0.49\linewidth]{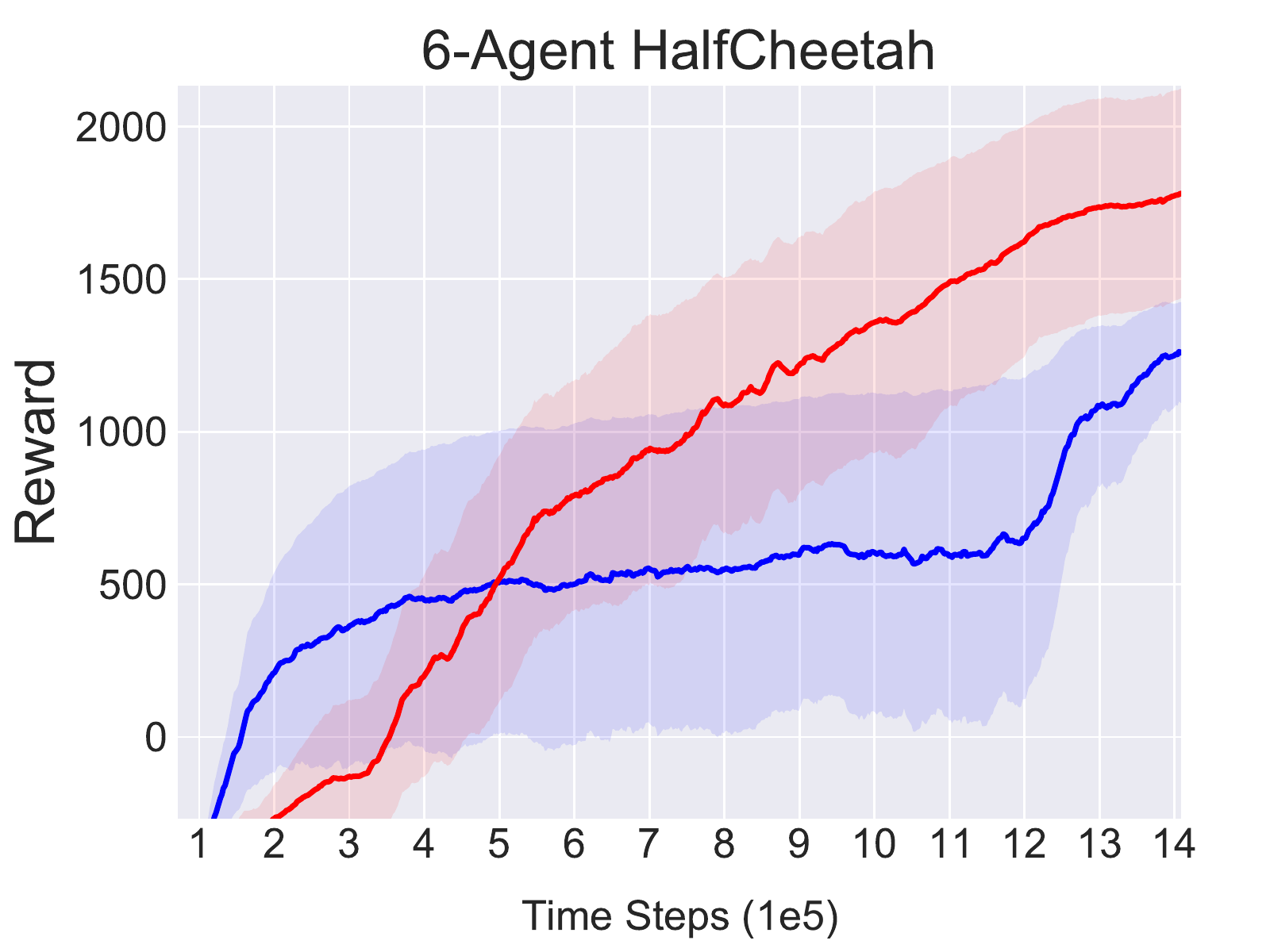}
\caption{Comparisons between leveraging \alg and GASIL.}
\label{compre dis}
\end{figure}
In this section, we give a new question: Whether \alg can be replaced by a discriminator, which can also achieve the purpose of guiding agents to better collaboration. 
To answer the question, more experiments are carried out.
This idea of using a discriminator is similar to Generative Adversarial Self-Imitation Learning (GASIL) \citep{guo2018generative} where a discriminator is trained to discriminate between superior trajectories and inferior trajectories, while the policy learns to fool the discriminator by imitating superior trajectories.
Specifically, GASIL maintains a good trajectory experience. The discriminator scores the good trajectory to 1.0 and scores the 
trajectory generated by current policy to 0.0. The results are shown in Figure~\ref{compre dis} which indicates that our MI based architecture \alg is more effective than GASIL, which may be mainly due to the fact: 1) GASIL only has a guiding role based on the discriminator. The inferior collaboration data is not utilized to encourage agents exploration. 2) The discriminator is difficult to train and suffers from model collapse.

\section{Hyperparameters}\label{Hyperparameters}


The hyperparameters for network architecture:

1: On MPE, we use two fully connected layers comprised of units 64 with ReLU nonlinearity and a final layer with tanh to output actions as the policy network for each agent, the critic network adopts the same architecture as the policy network except tanh for the final layer.

2: For Multi-Agent MuJoCo, the policy networks and critic network use the same architecture as those on MPE, but with 200 and 100 units for two fully connected layers.

3: For SMAC, we use code provided by RODE, the parameters and details are the same as the original paper.


For a fair comparison, all algorithms' network architecture remain consistent. In addition, we do not use the attention mechanism in the critic for all algorithms. For MADDPG, we use a centralized critic network and $N$ policy networks. For MINE, we use two fully connected layers (100 units on Muti-Agent MuJoCo and 64 on MPE) with Leaky ReLU nonlinearity to encode global states and joint actions respectively, the results are obtained by a dot product of the embeddings of global states and joint actions.
For CLUB, we use two fully connected layers (50 units on Muti-Agent MuJoCo and 32 on MPE) with Leaky ReLU nonlinearity to encode global states.
For SMAC, we need to encode $\rho$, thus CLUB uses two fully connected layers (64 units) to encode $\rho$ and the global states, then the outputs are concatenated to form a vector of 128 dimensions, and use the vector to predict the mean and variance of the global actions. MINE uses two fully connected layers (32 units) to encode $\rho$ and the global states and uses a fully connected layer (64 units) to encode the joint action.
The results are obtained by a dot product of the embeddings of the new vector (e.q., combine vectors of the global states and the joint role $\rho$) and joint actions).
The architectures used to calculate MI in other MI-related algorithms (SIC-MADDPG and VM3-AC) use the same number of units and the activation function of MINE and the other parts are consistent with the setting in their original papers.

For SIC-MADDPG and VM3-AC, there are two parameters to adjust: the dimension of $z$ and $\alpha$. For SIC-MADDPG, we select $z$'s dimension from $[2,3,5,8,10,15,20]$ following the setting in original paper on MPE, $[2,3,4,5,8,10,15]$ on Multi-Agent MuJoCo and select $\alpha$ from $[0.1,0.01, 0.001,0.0001,0.00001]$ following the setting in original paper. For VM3-AC, we select $z$'s dimension from $[2,4,8]$ following the setting in original paper and select $\alpha$ from $[0.1,0.01,0.001,0.0001,0.00001]$. 
For MASAC, we adjust $\beta$ from $[1.0,0.1,0.01,0.001,0.0001],0.00001$ to control the entropy.
For \alg, we need to adjust $\alpha$ and $\beta$, we select from $[1.0,0.1,0.01,0.001,0.0001]$. The final choice of $\alpha$, $\beta$ and dimension of $z$ is shown in Table~\ref{tab:a}. For FacMADDPG and COMIX, we use the official code and the parameters of the original paper. 

For other hyperparameters on MPE and Multi-Agent MuJoCo, $1\times10^{-3}$ for the critic and $1\times10^{-4}$ for the actors on MPE except $1\times10^{-2}$ on Wildlife Rescue and use Adam optimizer with learning rate $1\times10^{-3}$ for the critic and $1\times10^{-4}$ for the actors on Multi-Agent MuJoCo. For MINE and CLUB, the learning rate is $1\times10^{-4}$ on all environments except $1\times10^{-3}$ on Wildlife Rescue.
The discounted factor $\gamma$ and $\tau$ are 0.99 and 0.002 on Multi-Agent MuJoCo and 0.95 and 0.001 on MPE. Replay buffer size is $1\times10^6$ on MPE and Multi-Agent MuJoCo except $3\times10^5$ on Wildlife Rescue. The batch size is 1024 on MPE and 100 on Multi-Agent MuJoCo. The size of positive buffer and negative buffer of \dnb is 1000 on MPE except 6000 on Wildlife Rescue.
5000 for positive buffer and negative buffer on Multi-Agent MuJoCo. 500 for the positive buffer and 3000 for negative buffer on SMAC.

For hyperparameters of RODE, all parameters remain the same as in the code provided in the original paper. RODE has two main adjusted hyperparameters: The number of role clusters and
role interval. Number of role clusters is used to control the number of role types. The role interval decides how frequently the action spaces change and may have a critical inﬂuence on the performance. 
All parameter settings are consistent with the original paper.
The selection of $\alpha$ and $\beta$ is shown in Table.~\ref{tab:smac}. The batch size to update MINE and CLUB is 128.
We apply the maximization and minimization of MI after 100000 timesteps on SMAC.

\begin{table*}[h]
		\centering
		\caption{Selection of $\alpha$, $\beta$ and dimension of $z$ for different algorithms on MPE and Multi-Agent MuJoCo.}
		\begin{tabular}{|l|c|c|c|c|c|}\hline
			&SIC-MADDPG&\alg-MADDPG&VM3-AC & MASAC\\
			\hline
			Env name & $\alpha | z$ dim & $\alpha | \beta$ & $\alpha | z$ dim & $\beta$\\
			\hline
			Predator Prey (3 Agents)&0.00001 $|$ 2&0.01 $|$ 0.1 & 0.1 $|$ 4 & 0.1\\
			Predator Prey (6 Agents)&0.0001  $|$ 8&0.01 $|$ 0.1 & 0.01 $|$ 8 & 0.1\\
			Predator Prey (12 Agents)&0.0001 $|$ 3& 0.1 $|$ 0.1 & 0.01 $|$ 4 & 0.1\\
			Predator Prey (24 Agents)&0.0001 $|$ 2&0.01 $|$ 0.1 & 0.01 $|$ 4 & 0.1\\
			Cooperative Navigation& 0.0001 $|$ 2  &0.0001 $ | $ 0.01 & 0.1 $|$ 4 & 0.01 \\
			Wildlife Rescue&0.001 $|$ 3        & 0.001 $|$ 0.1 & 0.001 $|$ 2 & 0.001\\
			2 agents HalfCheetah&0.0001 $|$ 3  &0.1 $|$ 0.0001 & 0.1 $|$ 8& 0.01\\
			6 agents HalfCheetah&0.0001 $|$ 2 & 0.1 $|$ 0.001&0.1 $|$ 2 & 0.01\\
			\hline
		\end{tabular}
		\label{tab:a}
	\end{table*}

\begin{table*}[h]
		\centering
		\caption{Selection of $\alpha$, $\beta$ on SMAC.}
		\begin{tabular}{|l|c|}\hline
			&\alg-RODE\\
			\hline
			Map name & $\alpha | \beta$ \\
			\hline
			MMM2 & 0.0001 $|$ 0.0001\\
			MMM & 0.0001 $|$ 0.001 \\
			2s3z & 0.001 $|$ 0.1\\
			3s\_vs\_5z & 0.001 $|$ 0.01\\
		    3s5z & 0.001 $|$ 0.01\\
			10\_vs\_11m & 0.01 $|$ 0.01\\
			\hline
		\end{tabular}
		\label{tab:smac}
	\end{table*}

\section{Experiments about MASAC-related Algorithms on Wildlife Rescue}\label{ver MASAC}
Through experiments, we find that the MASAC-related algorithms can not learn positive rewards on Wildlife Rescue environment. Thus we experiment MASAC on Wildlife Rescue environment to find the reason.

There are two differences between MASAC and MADDPG:  $\beta H(\pi)$ and double Q mechanism.
Firstly, we make adjustments to $\beta$ and find that no matter how we adjust $\beta$, we can not get positive rewards, even if $\beta$ is set to 0.0. Secondly, we change the network architecture of MASAC by removing the double Q mechanism, then find that the algorithm can get positive rewards. Thus we hypothesize that the main reason why MASAC-related algorithms can not learn to cooperate on Wildlife Rescue is caused by the double Q mechanism. 

\begin{wrapfigure}{r}{0.65\textwidth} 
\centering
\includegraphics[width=0.49\linewidth]{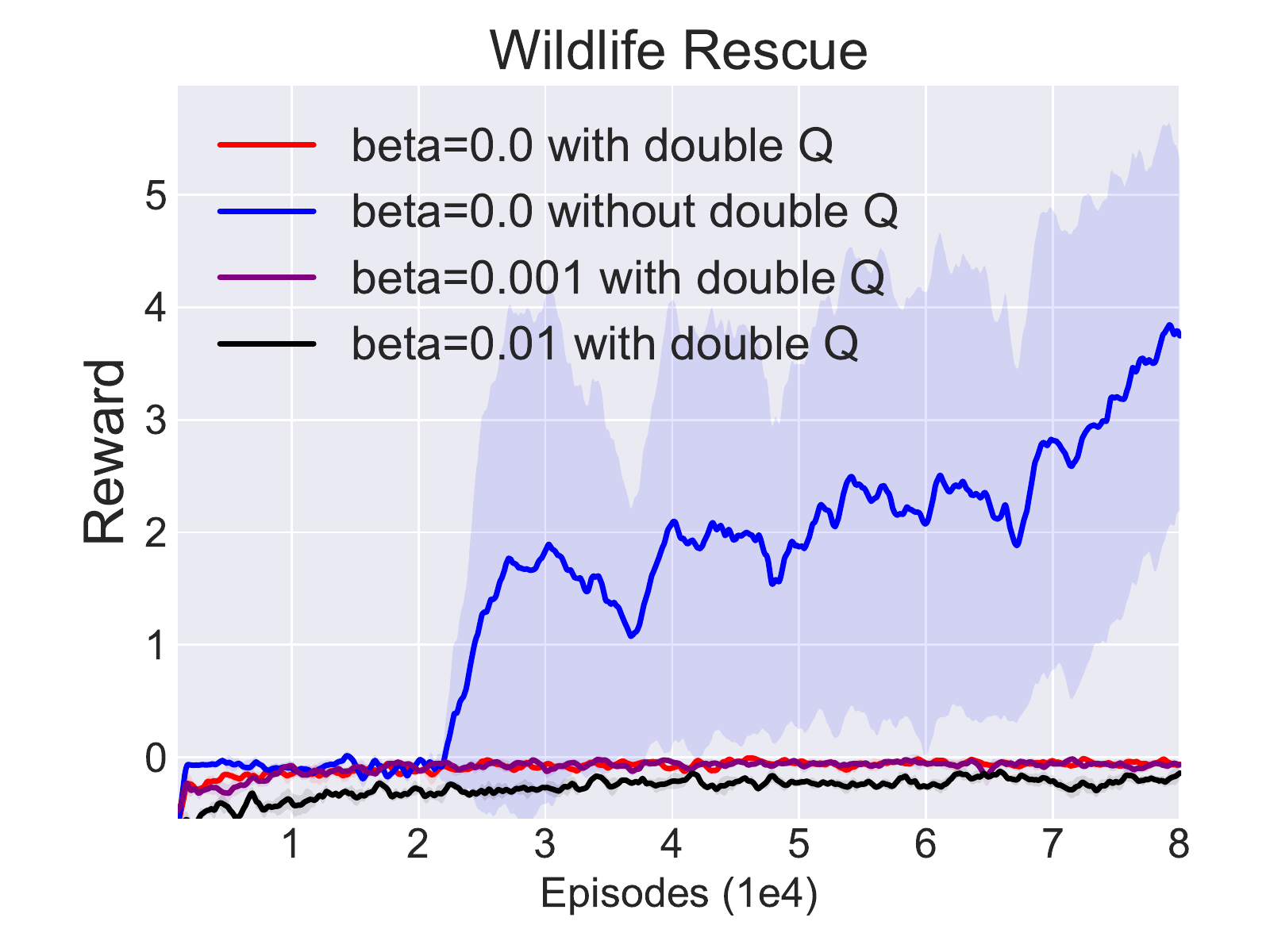}
\includegraphics[width=0.49\linewidth]{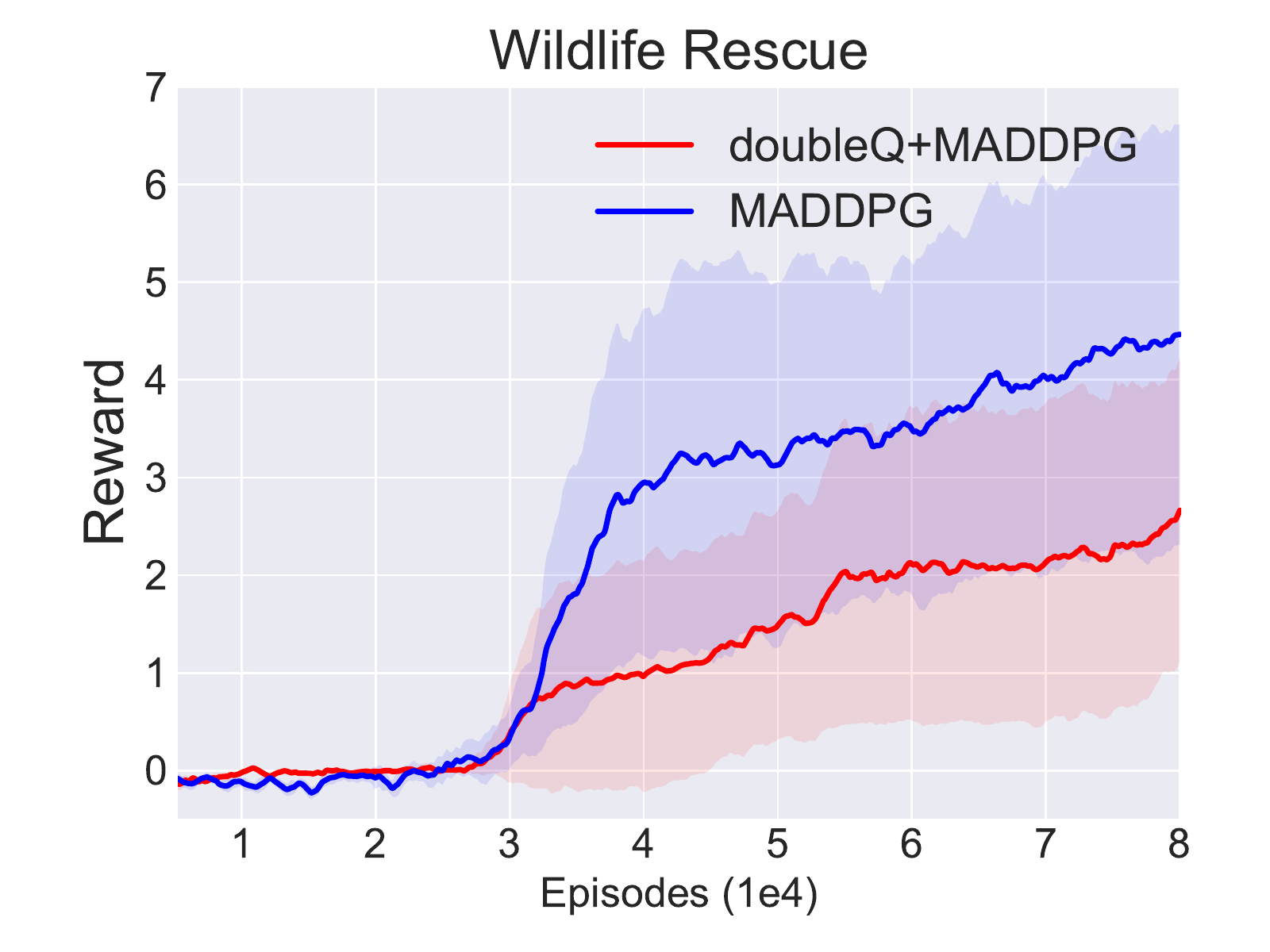}
\caption{Ablation experiments of $\beta H$ and double Q on Wildlife Rescue.}
\label{test MASAC}
\end{wrapfigure}

To further verify our hypothesis, we add double Q to MADDPG and find significant performance degradation and slower convergence, which proves that the double Q mechanism is the main factor.
We give some explanations, the Wildlife Rescue is characterized by high punishment for miss-coordination and low positive rewards. Double Q prevents overestimation but can lead to underestimation, thus using double Q might ignore less frequent positive rewards, leading to underestimation.


\end{document}